Accepted version for Engineering Optimization

# Overlap-free multi-material topology optimization for minimum compliance in two and three dimensions by level-set-based negative-mapping interpolation

Dong Wang [1] Qianglin Ran [1] Xuanliang Wang [1] Wei Xiang [1, 2]
Wenming Cheng [1, 2] Run Du [1, 2, 3, 4]

1. School of Mechanical Engineering, Southwest Jiaotong University, Chengdu 610031; 2. Technology and Equipment of Rail Transit Operation and Maintenance Key Laboratory of Sichuan Province, Southwest Jiaotong University, Chengdu 610031; 3. State Key Laboratory of Bridge Intelligent and Green Construction, Southwest Jiaotong University, Chengdu, Sichuan 611756, China; 4. School of Mechanical Engineering, University of Leeds, Leeds LS29JT, UK

**Corresponding author**: E-mail address: rdu@swjtu.edu.cn, r.du@leeds.ac.uk (R. Du).

**Abstract** Gray elements and material overlaps at the interfaces limit density-based multi-material topology optimization. This article extends the level set-based negative-mapping interpolation to the multi-material proportional topology optimization of macro-scale structures in two and three dimensions. An alternating active-phase algorithm decomposes $M$-phase problems into two-phase sub-problems described by level set functions; an evolutionary strategy removes the need for sentitivity analysis; and the negative-mapping interpolation eliminates the overlaps of the phases in the interface elements. Cantilever and half-MBB beams in two dimensions and a cantilever beam in three dimensions, with three and four material phases, give smooth boundaries, no gray elements and overlap-free designs. The compliance lies below the SIMP value in four of the eight two-dimensional cases and 0.3%, 0.4%, 4.8% and 12.7% above it in the other four. The influence of the material properties, of the interface treatment and of the algorithmic parameters is also examined.

## 1 Introduction

Aerospace and automotive structures place increasing demands on lightweight design and mechanical performance. Single-phase topology optimization struggles to meet such multi-objective requirements, whereas multi-material topology optimization can overcome the performance bottleneck of a single material by combining materials with different elastic moduli.

For thermoelastic metamaterials with extreme thermal expansion, Sigmund and Torquato (1997) proposed a material interpolation scheme within the Solid Isotropic Material with Penalization (SIMP) framework. Subsequently, Bendsøe and Sigmund (1999) extended the mixed model, assigning one design variable to each material phase. Gao and Zhang (2010) investigated multi-material structures under thermoelastic stress loads. Hvejsel and Lund (2011) derived multi-phase interpolation schemes within the variable density method. Yin and Ananthasuresh (2001) proposed a material interpolation model based on peak functions, where each material is described by a single design variable. Stegmann and Lund (2005) developed discrete material optimization for composite laminated shell structures. Gao and Zhang (2011) compared recursive and uniform multiphase materials interpolation (RMMI and UMMI) for mass-constrained problems. Zuo and Saitou (2017) developed an ordered SIMP interpolation. Chu et al. (2019) developed a density-based method for multi-material structures with graded interfaces. Building on previous work, Tavakoli and Mohseni (2014) proposed the alternating active-phase algorithm (AAPA), which decomposes the M-phase problem into M(M−1)/2 two-phase sub-problems. Wan et al. (2024) proposed a multi-SIMP model that converts the multi-material problem into several SIMP problems. Recent work has continued along these lines: Nguyen et al. (2025) coupled an ordered SIMP interpolation with a phase field model for thermoelastic multi-material structures, Wan et al. (2025) proposed a mapping-based interpolation within an isogeometric multi-material formulation, and Hu et al. (2026) extended the AAPA to three-dimensional structures under time-varying loads.

Topology optimization increasingly adopts the Level Set Method (LSM). Although the "Color" level set method proposed by Wang and Wang (2004) reduced computational costs, it requires handling redundant phase constraints. Luo et al. (2009) optimized piezoelectric materials with a piecewise constant level set model. Wang et al. (2015) constructed the MM-LS model, which uses $M$ level set functions to precisely describe $M$+1 phases. Based on the MM-LS model, Cui et al. (2016) introduced reaction-diffusion equations into the multi-material LSM. Liu and Ma (2018) enabled independent length-scale control for each phase. Sha et al. (2021) combined the MM-LS method (Wang et al. 2015) with the AAPA and extended it to three dimensions. Cui and co-workers further proposed an improved parameterized level set method for thermo-mechanically coupled problems (Wang et al. 2026) and an asymptotic concentration approach combined with isogeometric analysis (Cui et al. 2023). Two recent contributions act directly on the representation of the phases: Shen et al. (2025) described each phase on its own subdomain level set, which keeps the material regions disjoint in geometrically nonlinear problems, and Le et al. (2026) combined logical functions with the reaction–diffusion level set technique to control which phases remain active.

Proportional Topology Optimization (PTO) distributes materials based on compliance/stress proportions, avoiding the burden of gradient calculations. Biyikli and To (2015) first proposed the PTO method; subsequently, Cui et al. (2018) proposed a multi-phase material interpolation model, realizing multi-material proportional topology optimization. Nguyen et al. (2023) combined the PTO method with the AAPA. Rao et al. (2024) proposed a Proportional Topology Optimization method based on Level Set description and Evolutionary Strategy (PTO-LSES), which resolved the gray areas and sawtooth boundary issues prevalent in PTO methods, resulting in smooth and clear boundaries. Recently, the authors combined the PTO-LSES method with the AAPA (referred to there as the alternating active-phase strategy, AAPS) and introduced a negative-mapping interpolation that removes the overlap of the phases in the interface elements, for the design of multi-material microstructures with extreme bulk or shear modulus (Du et al. 2026). More recently, Gao et al. (2026) proposed a phased strategy in which an enhanced PTO is followed by a level set-based post-processing of the result. The PTO framework itself also continues to be refined, with new density filtering weight functions (Li et al. 2025) and enhanced proportional distribution schemes for volume-constrained compliance problems (Wang et al. 2025).

Beyond the density- and level set-based frameworks discussed above, topology optimization has recently been extended to isogeometric analysis (IGA) settings, e.g. multi-patch IGA for cellular structures using Nitsche's method (Gao et al. 2023), T-spline-based IGA for plate and shell structures with arbitrary geometries (Zhang et al. 2024) and for aerodynamic shell design (Zhang et al. 2025), as well as Lagrangian–Eulerian particle-flow formulations combined with the isogeometric material point method (Lin et al. 2025). Interfaces have also received attention at the multi-scale level, where the lattice microstructures assigned to different materials must remain connected across their common boundaries (Zhou and Kim 2026). These developments considerably broaden the geometric representations and the physics that topology optimization can handle, whereas the present work focuses on the treatment of the material interfaces in multi-material problems.

This article extends the level set-based negative-mapping interpolation framework of Du et al. (2026) from periodic microstructures to macro-scale load-bearing structures under prescribed loads and boundary conditions, with the structural compliance as objective, and to three-dimensional problems. It uses $M(M-1)/2$ level set functions to describe the $M$-phase topology, adopts the AAPA to decompose the $M$-phase problem into $M(M-1)/2$ two-phase sub-problems, applies the negative-mapping interpolation to eliminate interface overlaps, introduces compliance proportion filtering to improve boundary smoothness, and solves the resulting model by the optimality criteria method. With respect to Du et al. (2026), the new elements are the application to load-bearing structures, with a comparison against SIMP and against the AAPA without interface treatment, the extension to three dimensions, a quantitative measure of the interface overlap together with a final negative-mapping pass that makes the designs strictly overlap-free, and a systematic study of the convergence behaviour, of the material properties and of the algorithmic parameters.

## 2. Multi-Material Proportional Topology Optimization with level set description

### 2.1 Optimization and AAPA decomposition

The multi-material topology optimization model takes the minimum structural compliance as the objective function and volume as the constraint, described as follows:

$$\begin{cases} \text{minimize} & C(x) = \sum_{m=1}^{M}\sum_{e=1}^{n} u_e^T k_m^e u_e = \sum_{m=1}^{M}\sum_{e=1}^{n} E(x_m^e) u_e^T k_0^e u_e \\ \text{subject to} & F = KU \\ & V_m = \sum_{e=1}^{n} x_m^e v_e \le V_m^0 = f_m V_0 \\ & 0 < (x_{\min})_m \le x_m^e \le 1 \end{cases} \tag{1}$$

where $x = \{x_1, x_2, ..., x_M\}^T$ is an $M$-dimensional vector describing the density of the $M$-phase materials within the element; $C$ is the objective function (compliance); $M$ is the total number of material phases; $n$ is the total number of elements in the design domain; $u_e$ is the element displacement vector; $k_0^e$ is the element stiffness matrix; $K$ and $U$ are the global stiffness matrix and displacement vector of the structure, respectively; $F$ is the external load vector; $V_m$ is the volume of the $m$-th material after optimization; $f_m$ represents the volume fraction of the $m$-th material; and $(x_{\min})_m$ denotes the lower bound value of the element density, which is typically assigned a small positive value to prevent stiffness singularity. $E$ is the Young's modulus, which can be calculated by the following equation:

$$E(x_m^e) = E_{\min} + (E_m - E_{\min}) \cdot (x_m^e)^p, \quad x_m^e \in [0,1] \tag{2}$$

where $x_m^e$ is the design variable of the $m$-th phase material in the $e$-th element, $E_m$ represents the elastic modulus of the $m$-th phase material, and the sum of the relative densities of each element is 1, which can be expressed as:

$$\sum_{m=1}^{M} x_m^e = 1 \tag{3}$$

In the design domain $\Omega$ of multi-material topology optimization, the distribution of the $m$-th phase material is determined by its volume fraction $\alpha_m \left(m = 1,2, \quad ,M\right)$, which must satisfy the following relationships:

$$l_m \le \alpha_m \le u_m \tag{4}$$

$$\sum_{m=1}^{M} \alpha_m = 1 \tag{5}$$

where $l_m$ and $u_m$ represent the lower and upper bounds of the volume fraction, respectively, and their values must be between 0 and 1. Each phase also has a global volume constraint:

$$\int_{\Omega} \alpha_m dx = \Lambda_m \left|\Omega\right|, 0 \le \Lambda_m \le 1, \sum_{m=1}^{M} \Lambda_m = 1 \tag{6}$$

where $\Lambda_m$ is the prescribed volume constraint for each material phase; and the vector field $\alpha = \left\{\alpha_1, \alpha_2, \quad ,\alpha_M\right\}$ represents the set of design variables.

Let '$a$' and '$b$' denote the two material phases to be solved in the sub-problem of the AAPA (Tavakoli and Mohseni 2014). The other $M$−2 phases stay fixed, so the volume constraint $r_{ab}$ for the active phases $a$ and $b$ can be calculated:

$$r_{ab} = 1 - \sum_{\substack{m=1 \\ m \ne \{a,b\}}}^{M} \alpha_m \tag{7}$$

For each element *e*, the constraint in Eq. (3) is satisfied. If the volume fraction of phase *a* material is determined, the volume fraction of phase *b* material is expressed as:

$$r_b = r_{ab} - r_a \tag{8}$$

Combining the single-phase global material volume constraint of Eq. (6) and the active phase material volume constraint of Eq. (7), the temporary upper and lower limits for the volume constraint of phase *a* material in a single optimization sub-problem can be solved. The temporary lower limit remains $l_m$ unchanged, and the temporary upper limit is:

$$u_{a,temp} = \min\left(u_a, r_{ab}\right) \tag{9}$$

## *2.2 Compliance Proportion and filtering*

In the sub-problems of the AAPA, the compliance value used for distribution is that of the *m*-th phase. The elastic modulus interpolation of the *m*-th phase material in element *e* is calculated by Eq. (2), and its compliance is expressed as:

$$C_m^e = E_m^e u_e^T k_0^e u_e \tag{10}$$

The compliance proportion is the ratio of the compliance value of the *m*-th phase material in each element to the total compliance value of that phase across all elements within the design domain:

$$C_m^{ep} = \frac{1}{\sum_{e=1}^{n} C_m^e v_m^e} C_m^e \tag{11}$$

where $C_m^e$ represents the compliance value of the *m*-th phase material in element *e*; $v_m^e$ represents the volume of the *m*-th phase material in element *e*; *n* represents the total number of elements partitioned within the design domain; and $C_m^{ep}$ represents the compliance proportion value of the *m*-th phase material in element *e*.

After obtaining the compliance proportion values, a structural compliance proportion filtering scheme is adopted to reduce local intricate structures:

$$C_{mf}^{ep} = \frac{1}{\sum_{j=1}^{N_r} \gamma_{e,j}} \sum_{j=1}^{N_r} \gamma_{e,j} C_m^{jp} \tag{12}$$

where $C_m^{jp}$ represents the compliance proportion value of element *j* before executing the proportional filtering (within the filtering radius $r_f$ ); $C_{mf}^{ep}$ represents the compliance proportion of element *e* after executing the proportional filtering (involved in the finite element calculation); $N_r$ represents the total number of elements within the filtering radius; and $\gamma_{i,j}$ represents the weighting factor, calculated as:

$$\gamma_{e,j} = \begin{cases} \dfrac{r_{\mathrm{f}} - Dist(e,j)}{r_{\mathrm{f}}}, & \text{for } Dist(e,j) < r_{\mathrm{f}} \\ 0 \quad , & \text{for } Dist(e,j) \geq r_{\mathrm{f}} \end{cases} \tag{13}$$

where *Dist(e, j)* represents the distance between the geometric centers of element *e* and element *j*.

To obtain a topological configuration with clear and smooth boundaries, the LSF (Level Set Function) is employed to describe the boundaries of the structure. When PTO is adopted as the topology optimization framework, it is necessary to calculate the nodal compliance proportion. The nodal compliance proportion value $C_k^{NP}$ is the average of the compliance proportion values of the elements to which node *k* belongs, calculated as:

$$C_k^{NP} = \frac{1}{N_e} \sum_{e=1}^{N_e} C_{mf}^{ep} \tag{14}$$

where $N_e$ represents the total number of elements sharing node *k*. For bilinear four-node quadrilateral elements it takes the values 1, 2 and 4, for a corner node, a boundary node and an internal node respectively.

To improve the convergence performance of the algorithm, the compliance proportion of each node is further smoothed:

$$C_k^{NP(n)} \leftarrow \frac{C_k^{NP(n)} + C_k^{NP(n-1)}}{2} \tag{15}$$

where *n* represents the algorithm iteration; $C_k^{NP(n)}$ and $C_k^{NP(n-1)}$ represent the compliance proportion of node *k* at the current iteration and the previous one, respectively.

## *2.3 Level set description and density mapping*

In this work, *M*(*M*−1)/2 distinct LSFs are employed to describe the configuration of *M*-phase materials as in Du et al. (2026), one for each pair of phases (*m*, *n*), *m* < *n*, i.e. one for each two-phase sub-problem of the AAPA (Tavakoli and Mohseni 2014). The *M*−1 functions associated with the pairs (*m*, void) define the geometry of the solid phases and are the ones plotted in the following sections; the remaining functions are used only inside the corresponding two-phase sub-problems. The mathematical model is formulated as follows:

$$\begin{cases} \Phi_m(\eta,t) > 0, & \forall \eta \in \Omega_m \setminus \Gamma_m \\ \Phi_m(\eta,t) = 0, & \forall \eta \in \Gamma_m, \\ \Phi_m(\eta,t) < 0, & \forall \eta \in D \setminus (\Omega_m \cup \Gamma_m) \end{cases} \tag{16}$$

where *η* denotes a point within the design domain *D*; *Γ* represents the boundary of the solid domain Ω; and *t* indicates the pseudo-time, which describes the update iterations during the evolution of the LSFs.

To facilitate the Finite Element Analysis (FEA) of the structure, it is necessary to convert the LSFs into a physical model. A precise mapping function $M\left(\Phi_m\right)$ is employed to map the LSFs onto the surface of the design model. The mathematical expression is given by:

$$M(\Phi_m)=\begin{cases}1, & \text{if } \Phi_m \geq 0\\ 0, & \text{if } \Phi_m < 0\end{cases} \tag{17}$$

To enhance the computational efficiency of the finite element analysis, an ersatz material model is adopted here. By discretizing the structure into a fixed mesh during the optimization process, the need for remeshing or mesh modification is eliminated. The LSF-based ersatz material model is represented by a density distribution $x_m^e$ based on an exact Heaviside function $M(\Phi)$:

$$x_m^e=\frac{\int_{D^e} M(\Phi_m)d\Omega}{\int_{D^e} d\Omega} \tag{18}$$

where $D^e$ represents the element domain. In the practical implementation for 2D structural optimization, the design variables $x_m^e$ are obtained via the following equation:

$$x_m^e=\frac{A_m^s}{A_i} \tag{19}$$

where $A_m^s$ denotes the solid area of the *m*-th phase material in the 2D structure, and $A_i$ represents the total element area within the 2D structure. After calculating the element density $x_m^e$, the Young's modulus $E(x_m^e)$ is determined using Eq. (2). In the implementation, this ratio is evaluated numerically: 11×11 sample points are distributed in each element, the nodal values of the LSF are interpolated bilinearly at these points and the fraction of sample points with a non-negative LSF value is taken as the element density. The same procedure is used for three-dimensional problems with 11×11×11 sample points and trilinear interpolation of the eight nodal values, so that no modification of the method is required when passing from 2D to 3D.

# 3. Negative-mapping interpolation for interface reconstruction

## *3.1 Overlap detection*

The AAPA constrains only the volume fractions of the two active phases in every element (Eqs. (7)–(9)); it does not control which part of an interface element is occupied by which phase, so that the LSFs of two solid phases can be non-negative in the same region of an element, i.e. the phases overlap geometrically. To remove such overlaps, the negative-mapping interpolation introduced in Du et al. (2026) modifies the LSF of one of the two phases in the interface elements as follows:

$$\Phi_n(i,j)=-\Phi_m(i,j), \qquad \forall\, 0<x_m^e<1, 0<x_n^e<1, x_m^e+x_n^e=1 \tag{20}$$

where $\Phi_m$ and $\Phi_n$ are the level set functions of the (*m*, void) and (*n*, void) pairs of the two overlapping solid phases, *m* being the dominant phase (the phase with the larger elastic modulus) and *n* the suppressed phase, $(i,j)$ are the indices of a node of the fixed grid, and $x_m^e$ and $x_n^e$ are the densities of the two phases in the element *e* to which the node belongs.

The strategy is applied once the design has become nearly stationary (after about 5/7 of the maximum number of outer iterations). All elements of the design domain are then scanned and those in which the two active phases overlap are collected, as illustrated in Fig. 1: Fig. 1(a) shows such elements, Figs. 1(b) and 1(c) the material regions of the two phases, and Figs. 1(d) and 1(e) the LSFs $\Phi_m$ and $\Phi_n$ together with their zero level contours. The two material regions intersect, i.e. the phases overlap, in the region between the two zero contours.

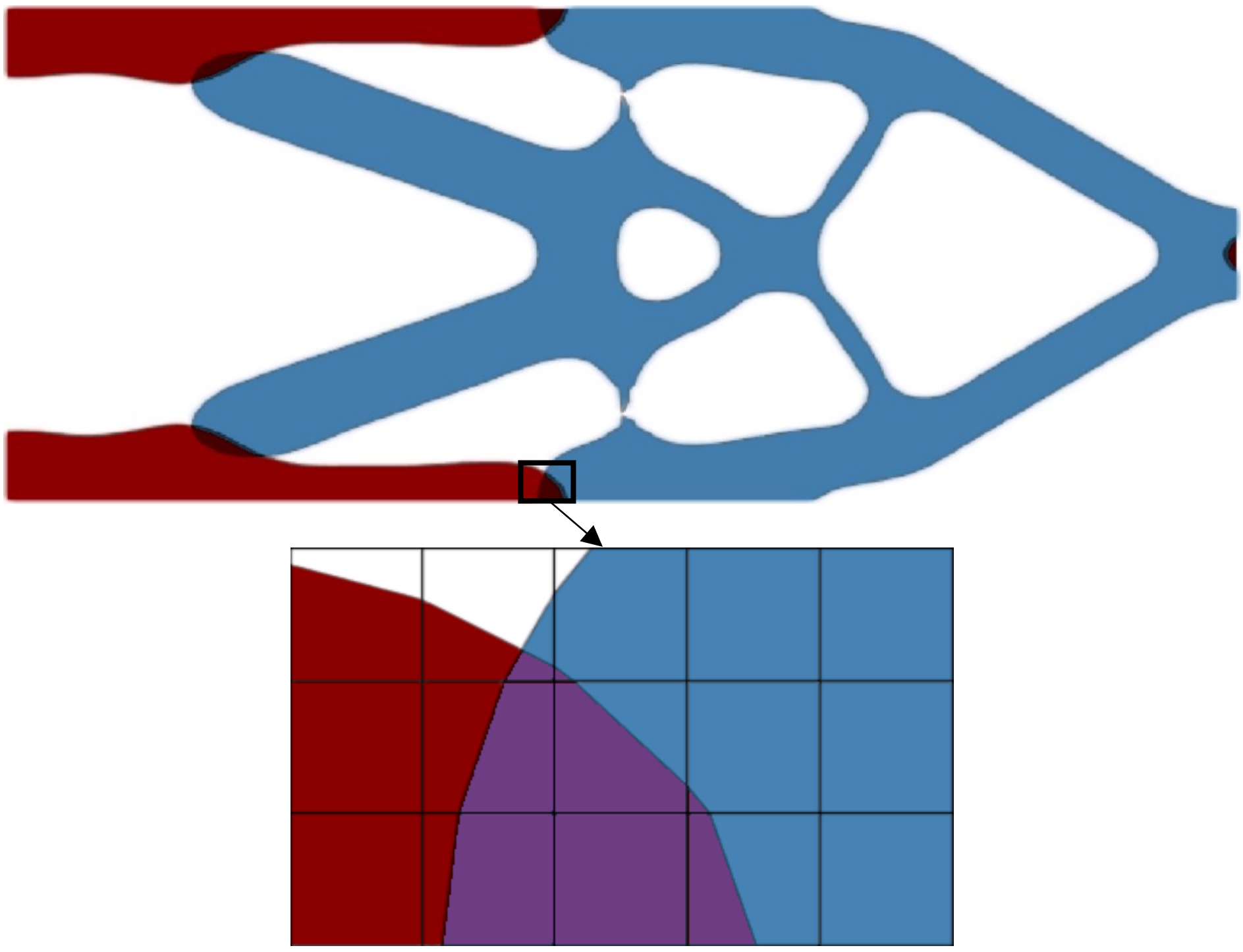

(a) Elements with Material Overlap During Iteration

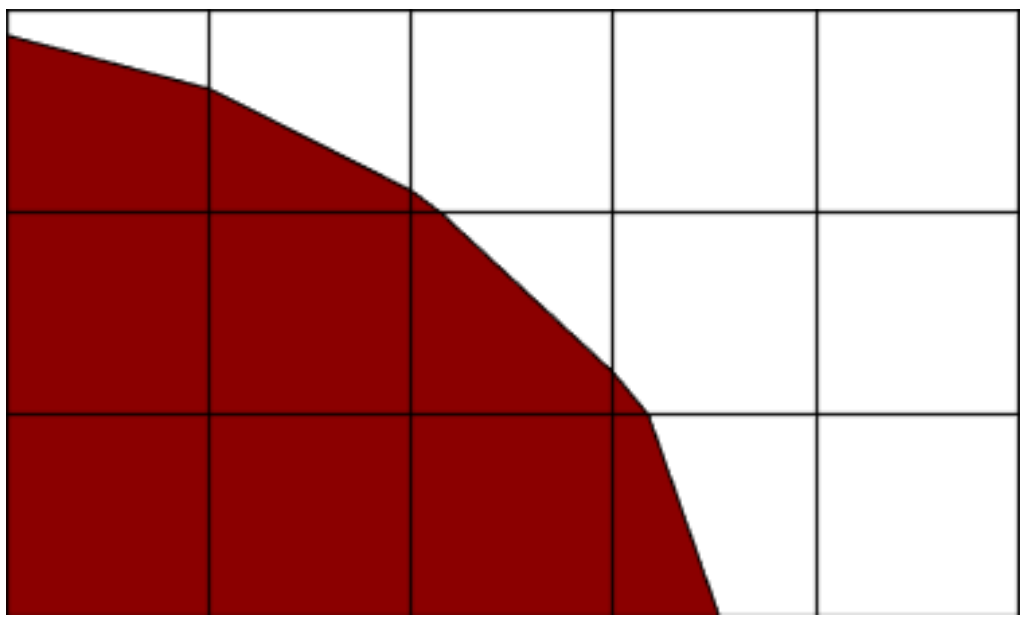

(b) Schematic of Material Distribution for $\Phi_m$

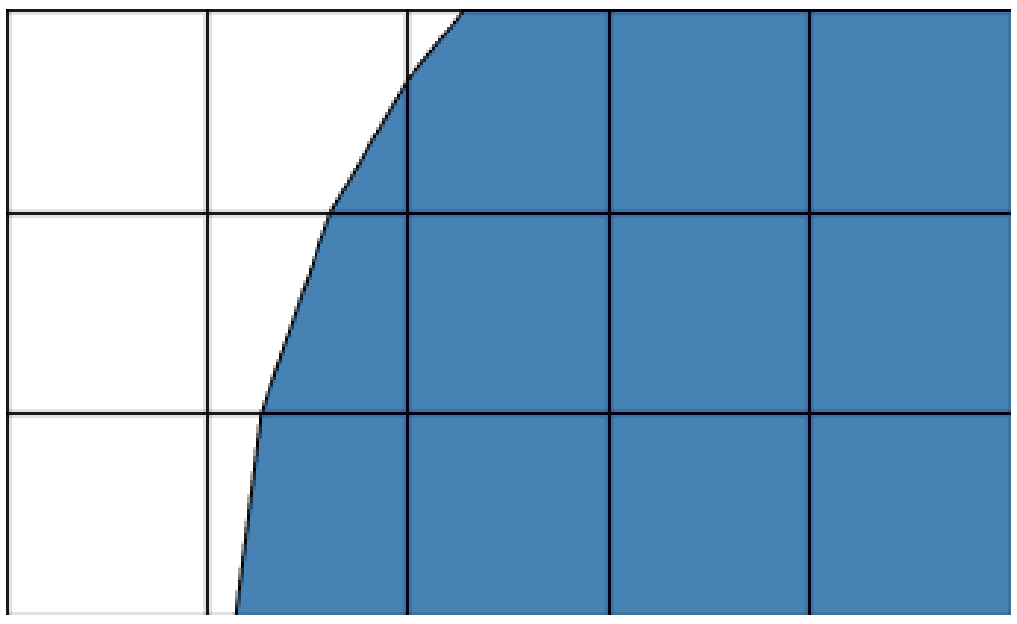

(c) Schematic of Material Distribution for $\Phi_n$

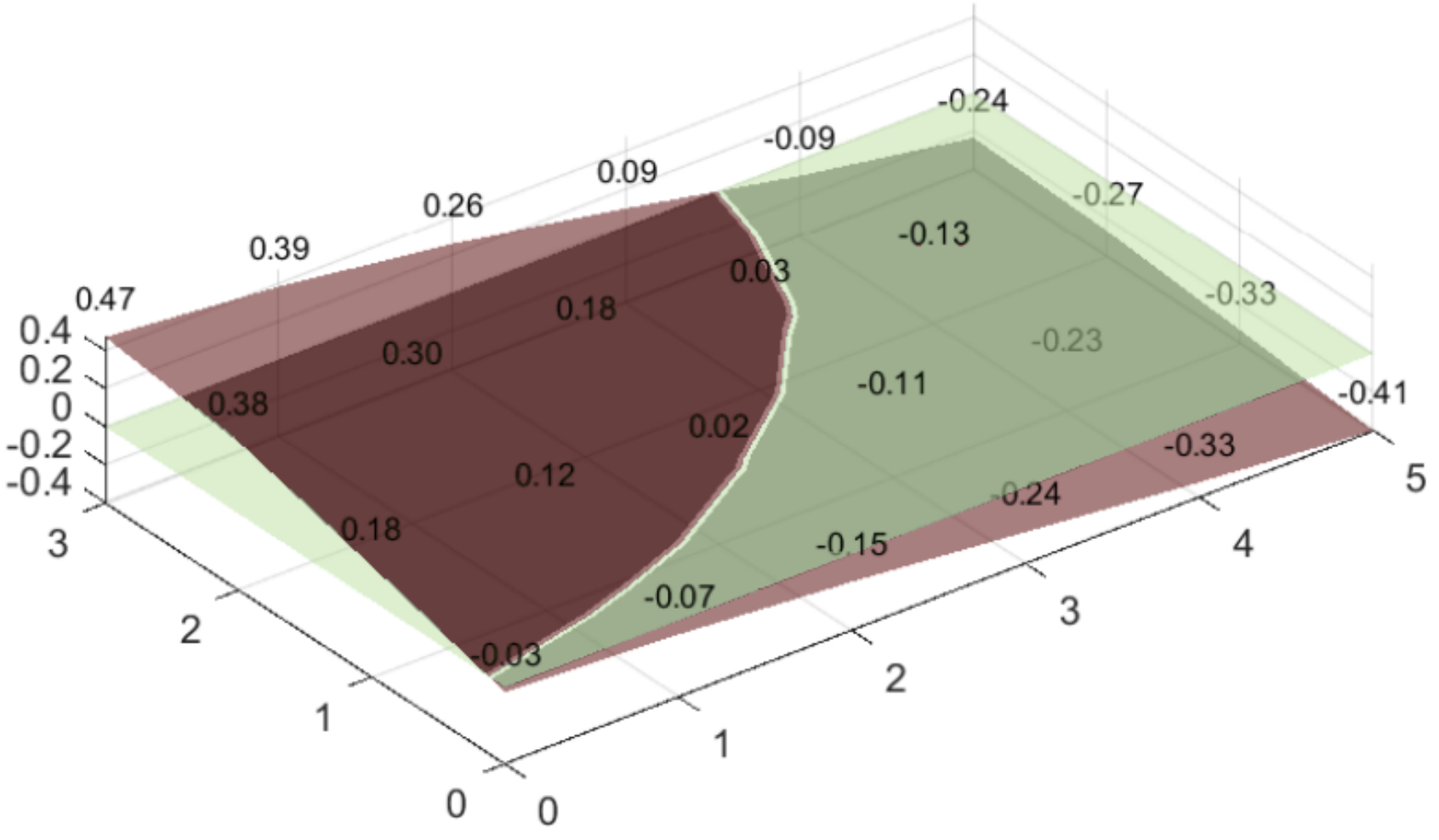

(d) Schematic Representation of Level-Set Function $\Phi_m$

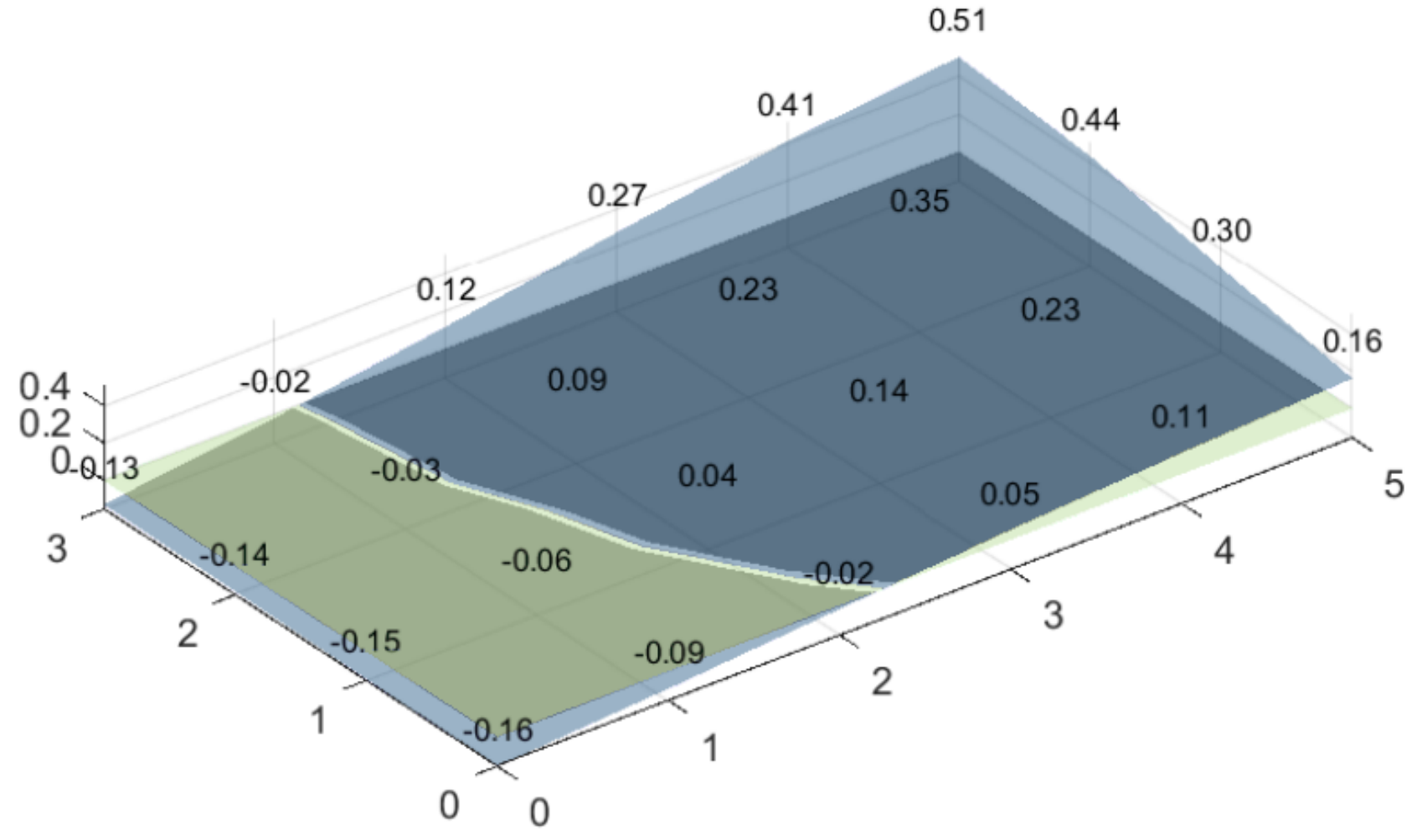

(e) Schematic Representation of Level-Set Function $\Phi_n$

Fig 1. Schematic Before Implementation of Negative-Mapping Interpolation

For every node of an overlap element the LSF of the dominant phase is kept unchanged (Figs. 2(b) and 2(d)), whereas that of the suppressed phase is replaced by its negative according to Eq. (20) (Figs. 2(c) and 2(e)).

After this operation the two zero contours coincide on the interface, the material regions of $\Phi_m$ and $\Phi_n$ are complementary within the element (Figs. 2(a)–(c)) and no overlap remains.

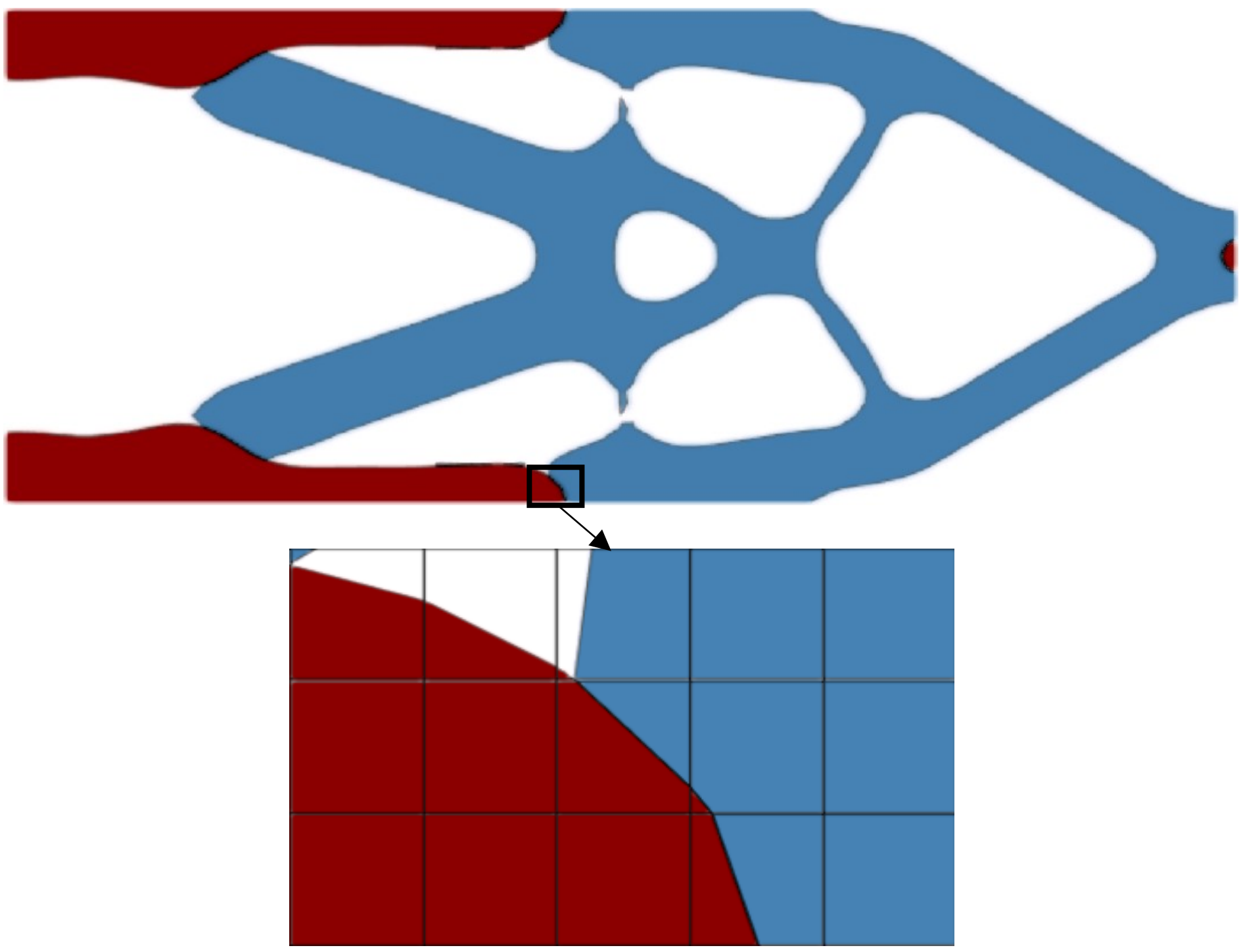

(a) Elements with Material Overlap During Iteration

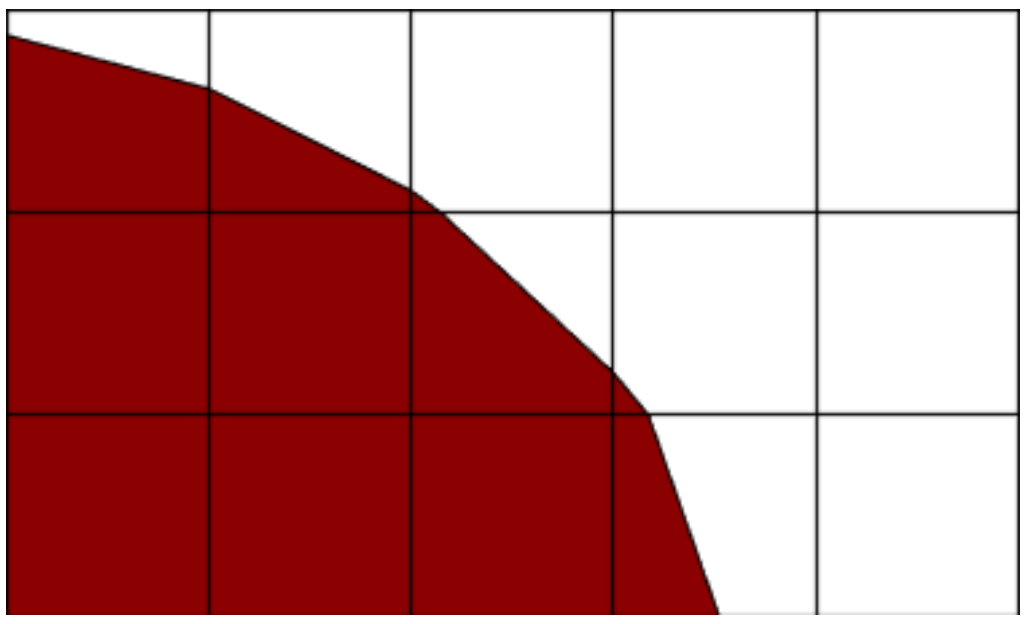

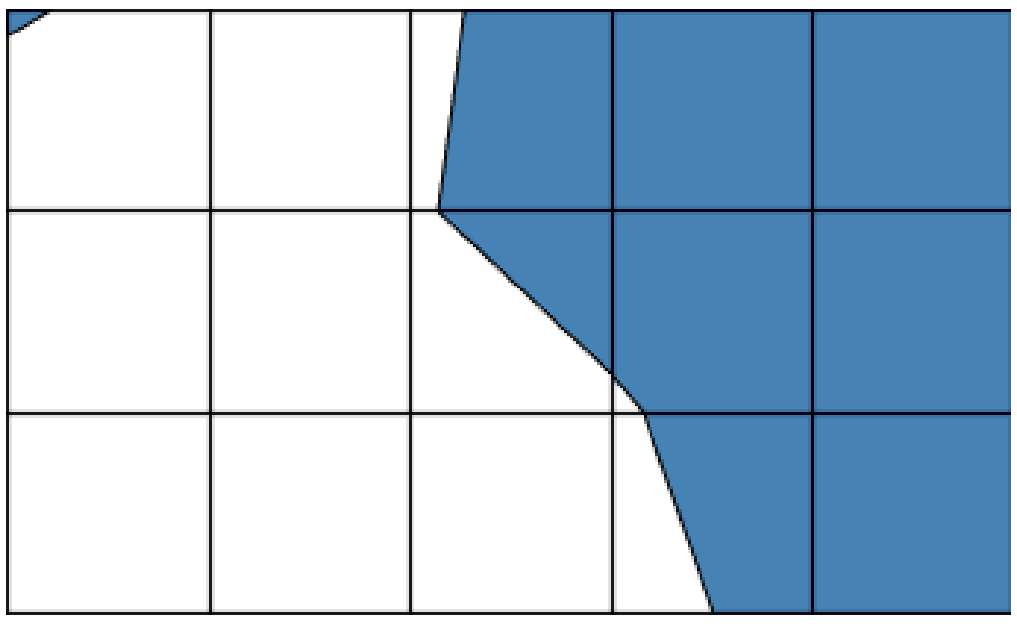

(b) Schematic of Material Distribution for $\Phi_m$ (c) Schematic of Material Distribution for $\Phi_n$

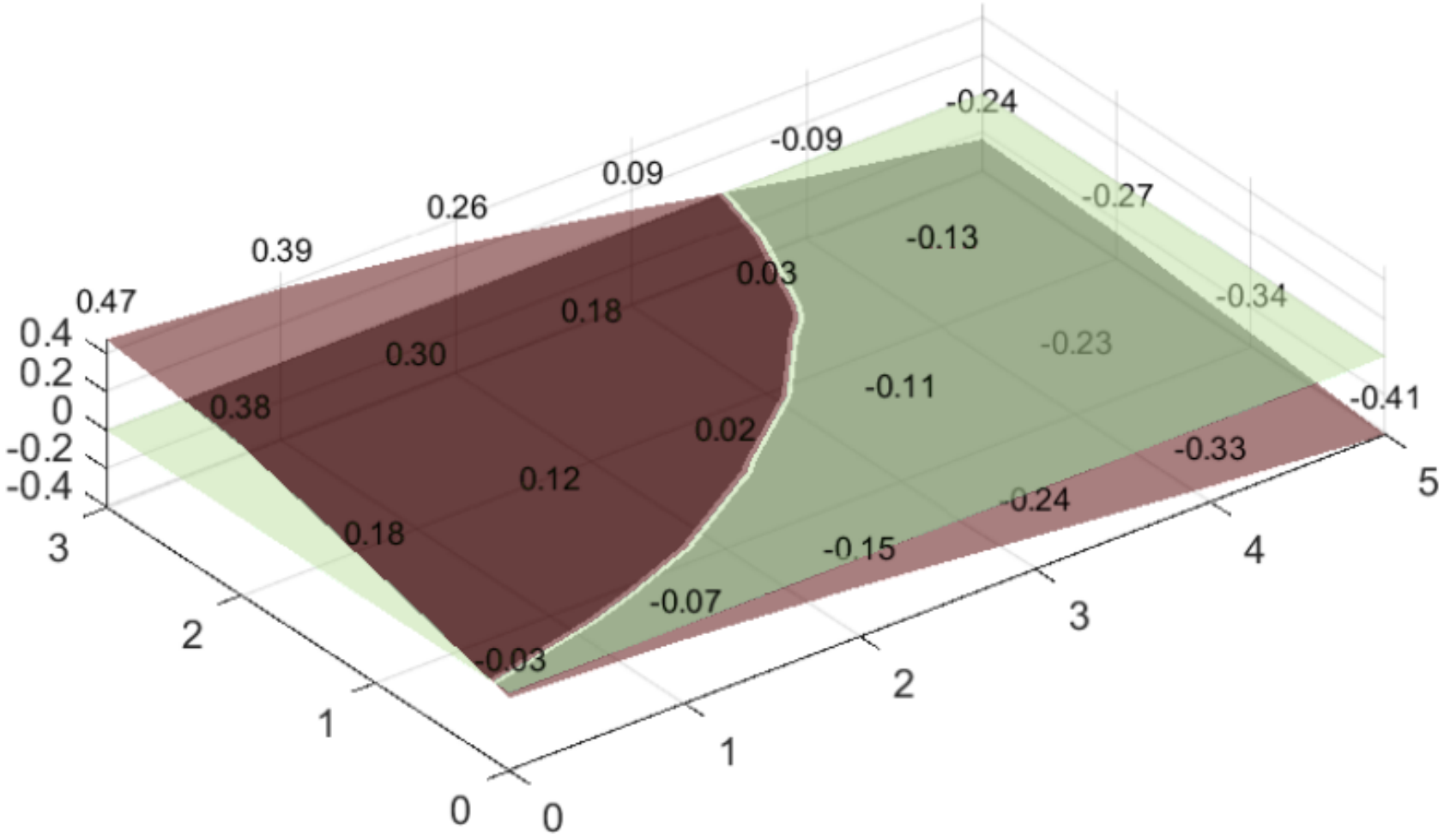

(d) Schematic Representation of Level-Set Function $\Phi_m$

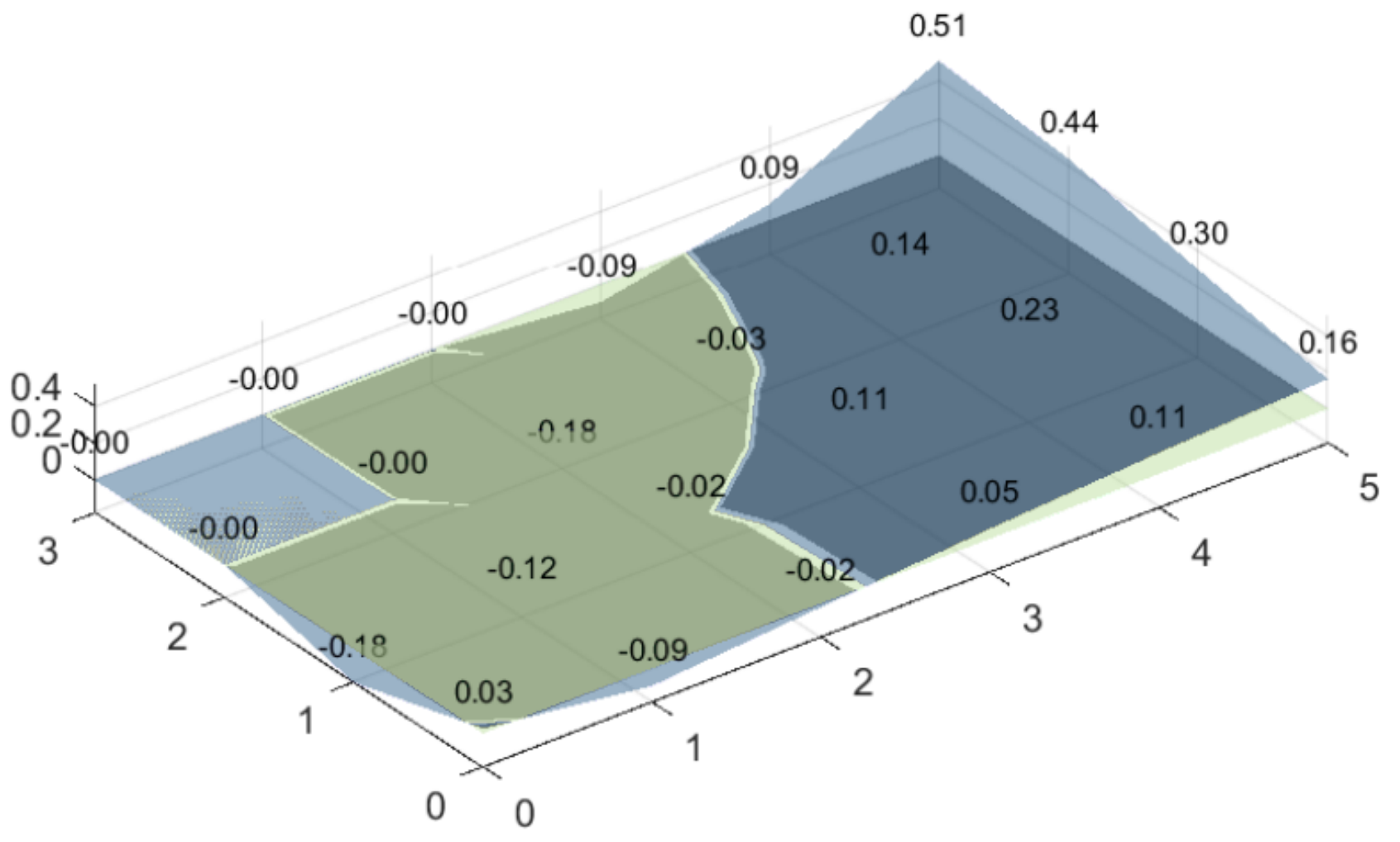

(e) Schematic Representation of Level-Set Function $\Phi_n$

Fig 2. Schematic After Implementation of Negative-Mapping Interpolation

Three points about the implementation. (i) Before the sign flip, a node whose neighbours are all occupied by a stronger phase is given a small negative value ($-10^{-9}$) of the suppressed LSF, so that isolated fragments cannot survive inside the dominant phase; with three solid phases the test covers every stronger phase. (ii) Once activated, the negative-mapping is applied in every subsequent inner iteration of the ($m$, void) sub-problems, because the level set functions of the other phases keep evolving and may generate new (small) overlaps. (iii) After the last iteration one additional pass is carried out on the converged level set functions: the overlap elements are detected from the nodal values (elements containing points where the LSFs of two solid phases are both non-negative) and, the solid phases being numbered in order of decreasing elastic modulus, the LSF of the weaker phase s is set at their nodes to $\Phi_s(i,j) = -\max_{t<s} \Phi_t(i,j)$, the natural generalization of Eq. (20) to elements in which more than two solid phases meet. Repeated until no overlap element is left (two to four sweeps in all examples), this pass does not change the topology and makes the design strictly overlap-free.

## *3.2 Negative-mapping interpolation*

In the notation of Du et al. (2026), let ($m$, $n$) be a pair of solid phases whose material regions overlap and $\Omega_{ov}$ the set of overlap elements detected from the nodal values of the level set functions $\Phi^{(m)}$ and $\Phi^{(n)}$. In every element $e \in \Omega_{ov}$ the phase with the larger elastic modulus, $u = \arg\max(E_e^{(m)}, E_e^{(n)})$, is taken as the dominant phase and the other one, $v = \arg\min(E_e^{(m)}, E_e^{(n)})$, as the suppressed phase.

For all nodes $k \in N(e)$ of such an element the negative-mapping replaces the level set value of the suppressed phase by the negative of that of the dominant phase,

$$\tilde{\phi}_v^k(i,j) = -\phi_u^k(i,j), \forall k \in N(e), 0 < x_v^e, x_v^e + x_u^e = 1 \tag{21}$$

Collecting these element-wise operations over all $e \in \Omega_{ov}$ defines the global negative-mapping operator

$$\mathcal{N}_{m\leftrightarrow n}\left[\phi^{(m)}, \phi^{(n)}; \rho^{(m)}, \rho^{(n)}\right] = \left(\tilde{\phi}^{(m)}, \tilde{\phi}^{(n)}\right),$$

which leaves the level set functions of the elements outside $\Omega_{ov}$ unchanged. It modifies only the level set function of the suppressed phase, which is the one updated in the current ($v$, void) sub-problem, and is applied after the update (26) and before the density mapping (18)–(19) of the same inner iteration, so that the densities entering the next finite element analysis are free of overlap.

## 4. Update Scheme Based on Evolutionary Strategy and Optimality Criteria

Since the standard level set method with an initial design of a specified material volume is often disadvantageous for structural topology changes—leading to instability in algorithm convergence or even structural collapse—this method adopts an initial full material design. The specified material volume (i.e. the volume constraint) is then reached gradually through an evolutionary strategy (Li et al. 1999). The expression is as follows:

$$V^{(n)} = \begin{cases} \max\left\{V_{\mathrm{pm}}, V^{(n-1)}(1-ER)\right\}, & \text{for } V_{\mathrm{pm}} < V_{\mathrm{igd}} \\ \min\left\{V_{\mathrm{pm}}, V^{(n-1)}(1+ER)\right\}, & \text{for } V_{\mathrm{pm}} \geq V_{\mathrm{igd}} \end{cases} \tag{22}$$

where $ER$ denotes the evolutionary rate, representing the volume of material added or removed in each iteration during the optimization process (prior to reaching the specified material volume $V_{pm}$); $n$ represents the iteration number; $V^{(n)}$ is the target material volume for the current iteration step; and $V_{igd}$ is the initial design material volume (e.g. 1 represents a full material design). Once the specified material volume $V_{pm}$ is reached, the algorithm maintains this volume constant, and only the topological layout is altered in subsequent iterations.

In this article, the LSF update is governed by an ordinary differential equation (ODE; Liu et al. 2021). Combining the Optimality Criteria (OC) derived from the Karush-Kuhn-Tucker (KKT) conditions with the material volume constraint in Eq. (1), the nodal compliance ratio $C_k^{NP(n)}$ in Eq. (15) is used to control the update of the LSF:

$$\frac{d\Phi}{dt} - C_k^{NP(n)} = 0 \tag{23}$$

In the OC method, the optimality criterion for elements is equivalent to that for nodes; therefore, the element optimality criterion is replaced by the nodal optimality criterion. The nodal optimality criterion used here is constructed as follows:

$$C_k^{NP(n)} - \lambda = 0 \tag{24}$$

where $\lambda$ represents a Lagrange multiplier calculated via the bisection method so that the prescribed volume is met. Since $C_k^{NP(n)} \geq 0$, Eq. (23) alone would increase $\Phi$ at every node; the LSF is therefore driven by the residual of the optimality condition (24) instead of by $C_k^{NP(n)}$ itself, i.e. the driving term of Eq. (23) is replaced by the left-hand side of Eq. (24):

$$\frac{d\Phi}{dt} - \left(C_k^{NP(n)} - \lambda\right) = 0 \tag{25}$$

The evolution then stops exactly where the optimality condition holds, and $\lambda$ shifts the zero level of the velocity so that the volume constraint is enforced. When the LSF update concludes, the compliance ratio values of all nodes equal the value of the Lagrange multiplier. Mathematically, this is described as OC( $C_k^{NP(n)} - \lambda$ ) being equal to 0, which corresponds to a KKT point. Consequently, the update of the LSF is defined as:

$$\Phi^{(n+1)} = \Phi^{(n)} + \Delta t\left(C_k^{NP(n)} - \lambda\right) \tag{26}$$

where $n$ is the iteration number, and $\Delta t$ is the time step size defined by the designer.

Note that the nodal compliance proportion $C_k^{NP(n)}$ is, up to a positive factor, the element strain energy, i.e. the magnitude of the sensitivity of the compliance with respect to the density； Eq. (26) therefore adds material where this proportion exceeds the threshold $\lambda$; the same form holds for objectives that are maximized (e.g. a homogenized modulus), for which the proportion is the sensitivity itself.

The convergence criterion for the present algorithm is defined as minimizing the variation in compliance values during the iteration process, or reaching the maximum number of iterations specified by the designer ( $NoI_{\max}$ ):

$$\frac{\left|C^{(n)} - C^{(n-l)}\right|}{C^{(n)}} \leq \delta, \text{ or } n > NoI_{\max} \tag{27}$$

where $\delta$ is a small positive number, typically set to 0.001, and $l$ is an integer (set to 5 in this article), implying that the variation in compliance values over the last 5 iterations is negligible. Here "iteration" refers to the outer iteration of the AAPA (Fig. 3), i.e. one sweep over all $M(M-1)/2$ two-phase sub-problems. This criterion is used to monitor convergence, while all runs are continued up to $NoI_{\max}$ so that the different cases can be compared at the same number of iterations.

In summary, the flowchart of the non-overlapping multi-material topology optimization using the level set description and negative-mapping interpolation is presented in Fig 3.

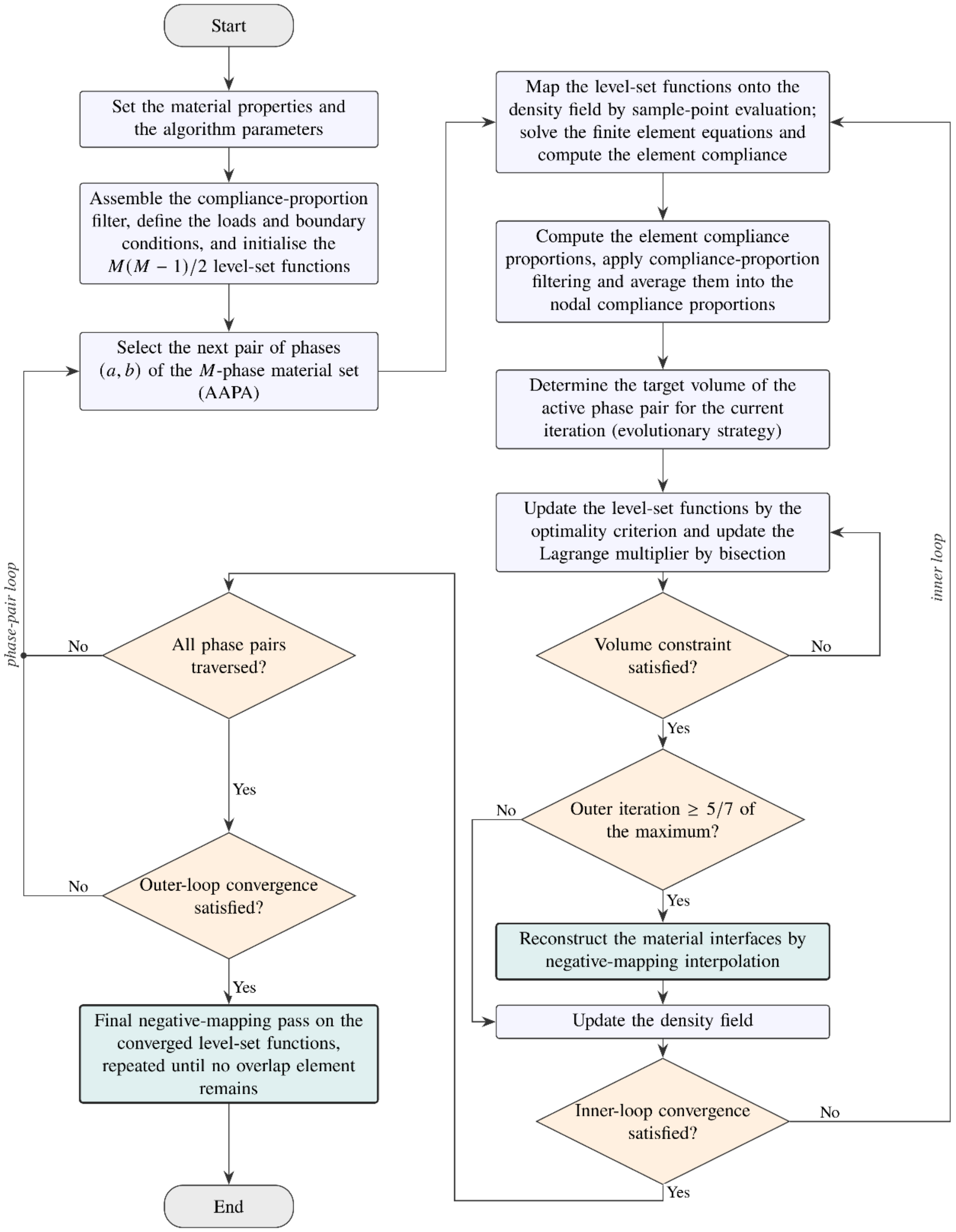

Fig 3. Flowchart of the present method.

## 5. Numerical examples

In this section, the 2D cantilever beam and the half-MBB beam of Fig 4 are employed as benchmark examples to test structures composed of three-phase materials (i.e. two solid materials and one void phase) and four-phase materials (i.e. three solid materials and one void phase). Both beams are used for both material systems, and the results are compared with those of the traditional SIMP method. The data and code for the 2D SIMP method are those of Tavakoli and Mohseni (2014). Sections 5.6–5.8 further discuss the convergence behaviour, the influence of the material properties and the extension of the method to 3D problems, the last with an implementation of the same algorithm for hexahedral meshes. In all figures the solid phases are distinguished by shading, the stiffest phase being the darkest and the weakest the lightest, so that the designs remain legible in black and white.

Fig 4 illustrates the 2D models of the half-MBB beam and the cantilever beam, including their dimensions, boundary conditions, and external loads. Both structures are discretized into a mesh of 100×40 elements, using square quadrilateral elements. The basic parameters for these algorithms are set as follows: to avoid singularity, the Young's modulus of the void phase is set to $10^{-9}$; the Poisson's ratio $v$ is set to 0.3; the load $F$ is set to 1; the material volume evolutionary rate ($ER$) is set to 2%; and the time step size $\Delta t$ is set to 0.01. In the present method, the number of inner iterations is set to 4, and one outer iteration consists of $4M(M-1)/2$ inner iterations. Because the initial design does not satisfy the volume constraints, the number of inner iterations performed in the first outer iteration is set to 150. The maximum number of outer iterations is 200 and all runs are continued up to this number, so that the different cases are compared at the same iteration count; the iteration at which criterion (27) is first met is reported in Section 5.6. The negative-mapping interpolation strategy is activated at iteration 142 (≈ 5/7 of the maximum number of outer iterations) and applied in every subsequent inner iteration of the ($m$, void) sub-problems.

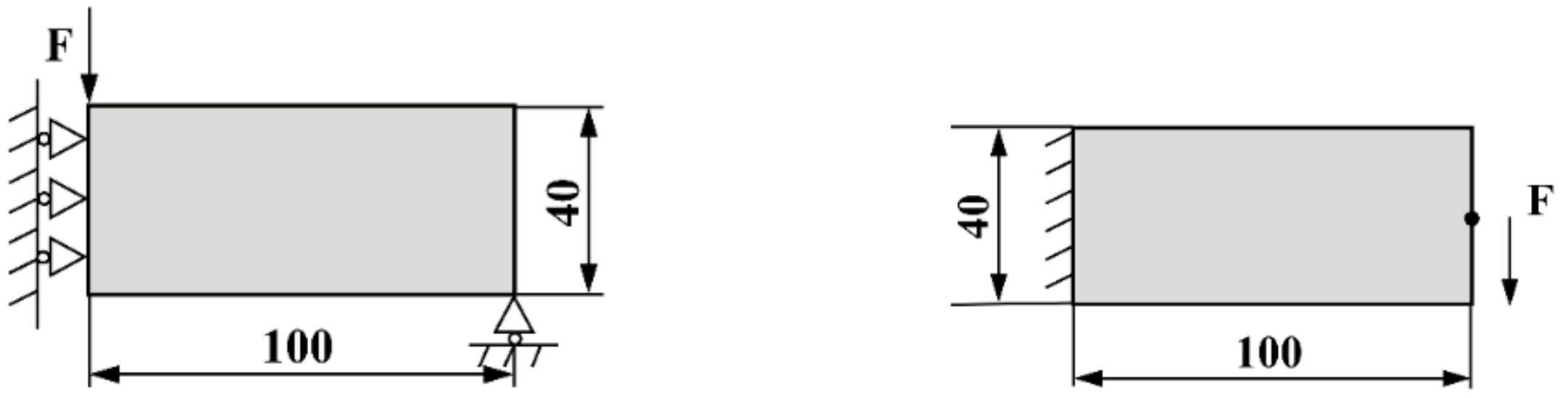

(a) 2D half-MBB beam (b) 2D cantilever beam

Fig 4. 2D structural models

## *5.1. Comparison of Three-Phase Material Optimization Results*

Two solid materials plus a void phase (three material phases) are considered here. Their Young's moduli are $E$=2 and $E$=1, the stiffer phase being drawn in the darker shade throughout. Two volume fraction combinations are tested: (1) $v_{E=2}=0.1, v_{E=1}=0.3$; (2) $v_{E=2}=0.2$, $v_{E=1}=0.2$. Here, the filter radius $r_f$ is set to 5 for both the SIMP method and the present method.

Table 1 presents the optimization results for the 2D cantilever beams and 2D half-MBB beams obtained by the SIMP method, by the AAPA only and by the present method; the comparison of the last two is discussed in Section 5.3. After 200 iterations, the compliance values for the optimized 2D half-MBB beams converge to 132.32 and 115.13 under the two different volume fraction combinations, respectively. For the 2D cantilever beams, the corresponding compliance values converge to 121.83 and 114.06, respectively.

The boundaries are smooth and the interface between the two materials is sharp. With the volume fraction of the void phase held constant, the compliance decreases as the volume fraction of the stiffer material grows. The material distribution is also sensible: the stiffer material (E=2) is placed at the supports and at the loading point, where the deformation is largest, and the weaker one ($E$=1) reinforces the rest.

Table 1. Three-phase 2D beams: optimized structures and compliance obtained with SIMP, with the AAPA only and with the negative-mapping interpolation.

| Volume fractions | Method | Half-MBB beam | Cantilever beam |
|---|---|---|---|
| $v_{E=2}=0.1$, $v_{E=1}=0.3$ | SIMP | 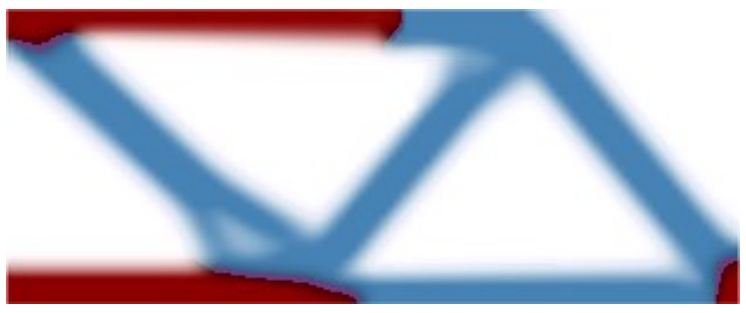 $C=137.76$ | 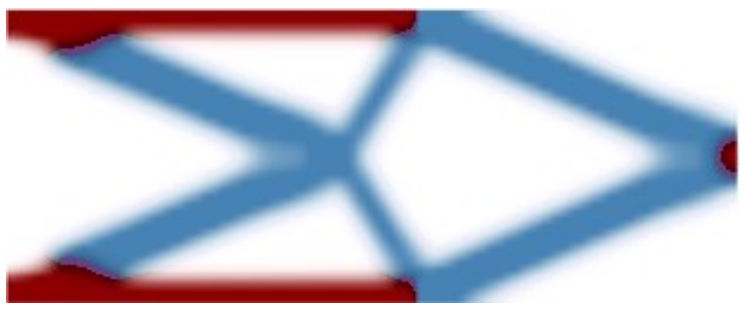 $C=121.32$ |
| | AAPA only | 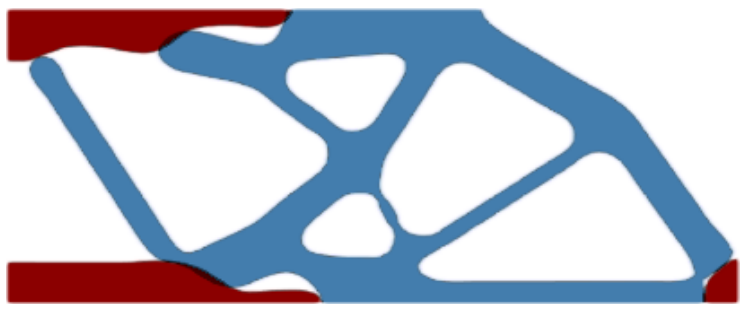 $C=132.54$ | 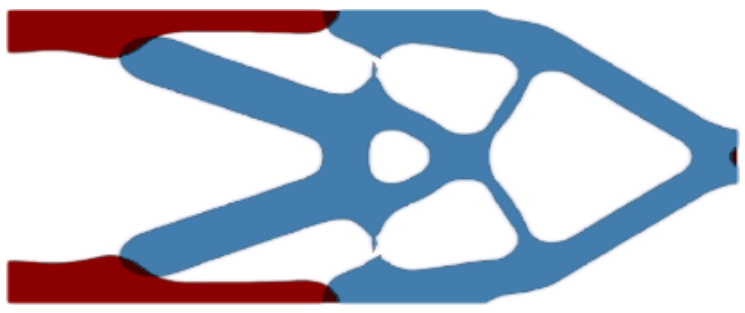 $C=122.96$ |
| | Negative-mapping interpolation | 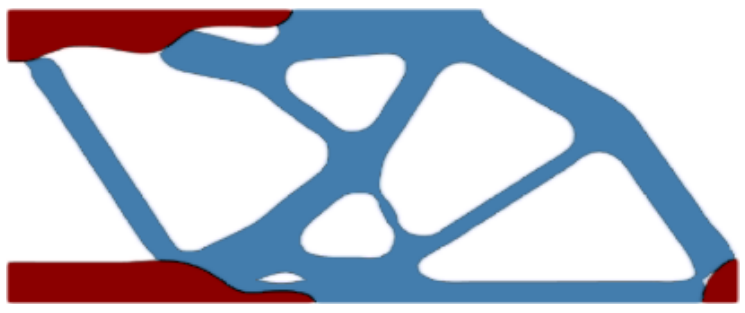 $C=132.32$ | 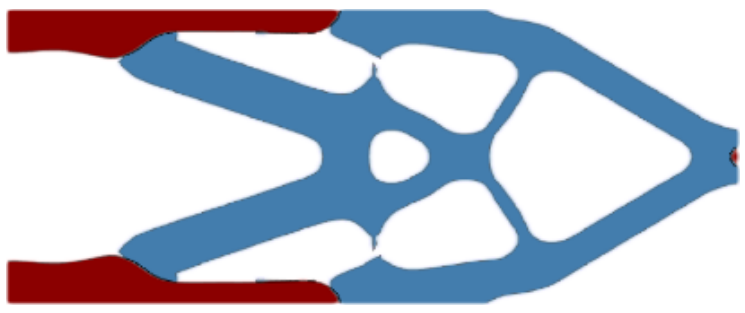 $C=121.83$ |
| $v_{E=2}=0.2$, $v_{E=1}=0.2$ | SIMP | 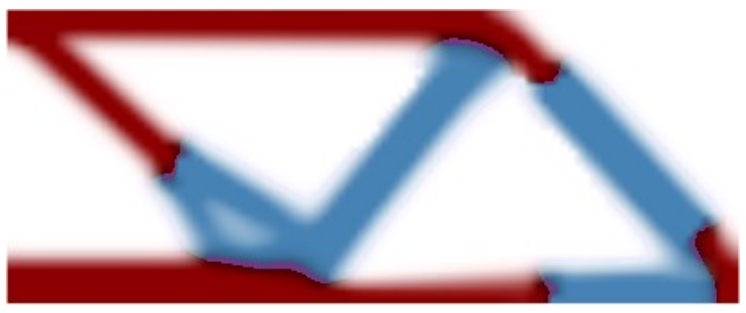 $C=114.83$ | 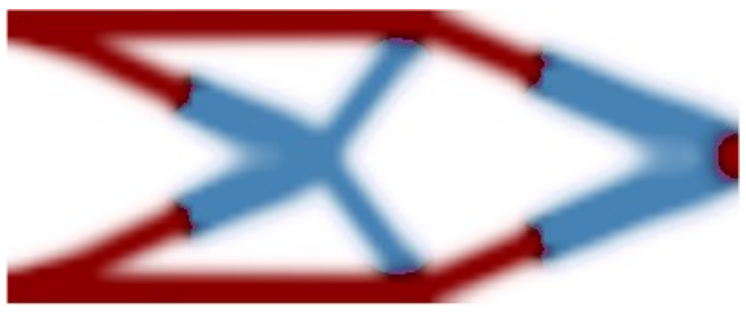 $C=101.19$ |
| | AAPA only | 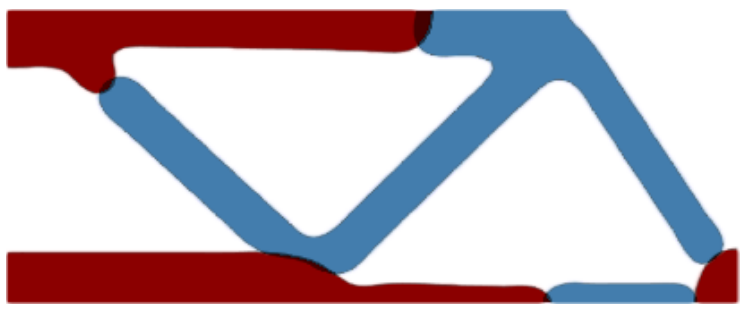 $C=113.33$ | 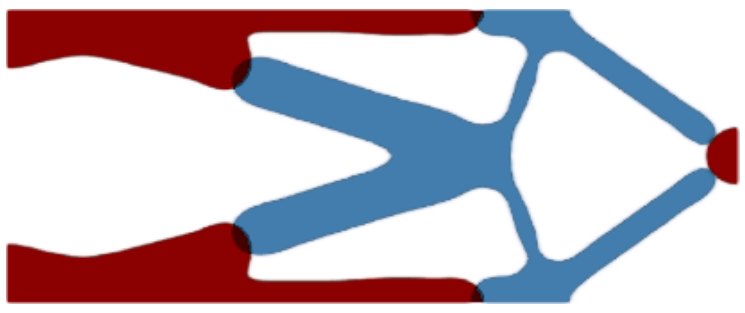 $C=114.31$ |
| | Negative-mapping interpolation | 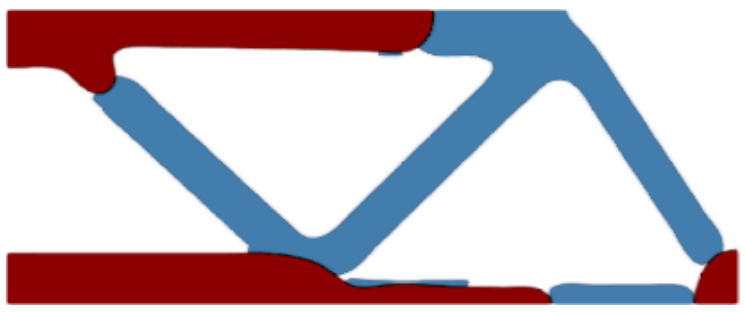 $C=115.13$ | 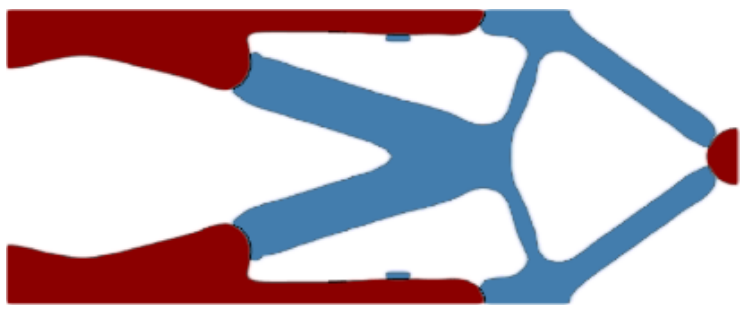 $C=114.06$ |

Fig 5 shows the evolution of the topology and of the compliance of the three-phase 2D half-MBB beam, and Fig 6 the iteration history of the volume fractions of each material phase.

At the start of the iteration, the material with Young's modulus $E$=2 occupies the entire design domain, and the material with Young's modulus $E$=1 has a volume fraction of 0. After the first iteration, the volume fractions approach the prescribed values and remain stable, while the compliance decreases rapidly with oscillation; after the 40th iteration it reduces further and then converges slowly. The main topological structure is formed within the first 3 iteration steps, with only minor local changes thereafter.

Fig 7 presents the optimization process for the three-phase 2D cantilever beam. Fig 8 shows the iteration history of the volume fractions for each material phase of the same beam. The beam starts, like the half-MBB beam, with *E*=2 filling the design domain and *E*=1 absent; the volume fractions reach their prescribed values after the first iteration and the compliance falls with oscillation. After the 80th iteration the compliance oscillates less and converges slowly, and after about 150 iteration steps the structural topology remains essentially unchanged.

Because the evolutionary rate ER = 2% acts on the inner iterations, the prescribed volume fractions are already reached at the end of the first outer iteration (Figs. 6 and 8); they stay slightly below the prescribed values because the volume constraint is enforced on the fraction of nodes with a non-negative level set function rather than on the element densities. The volume being essentially fixed from then on, the algorithm only redistributes the phases: the compliance falls steeply during the first 40 outer iterations and then decreases slowly, converging after about 90 iterations up to a small periodic oscillation caused by the alternation of the AAPA sub-problems, which is analysed quantitatively in Section 5.6. The interfaces are smooth and free of overlap.

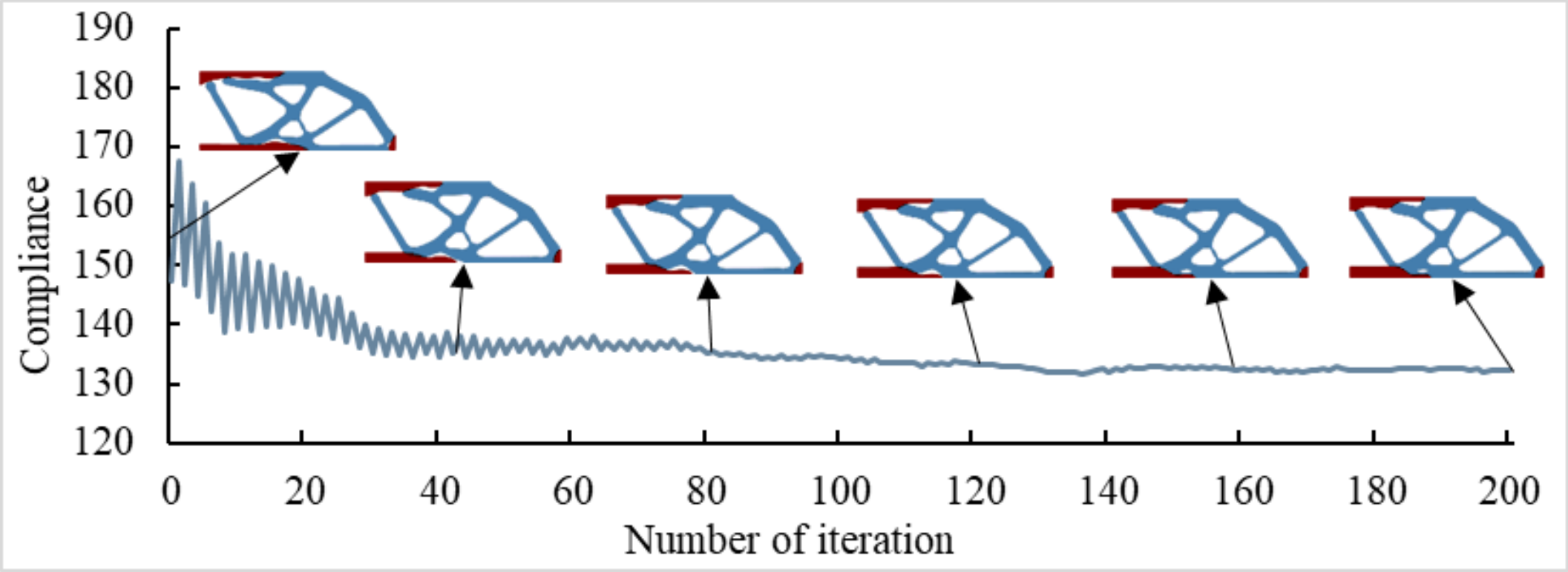

(a) $v_{E=2} = 0.1$ , $v_{E=1} = 0.3$

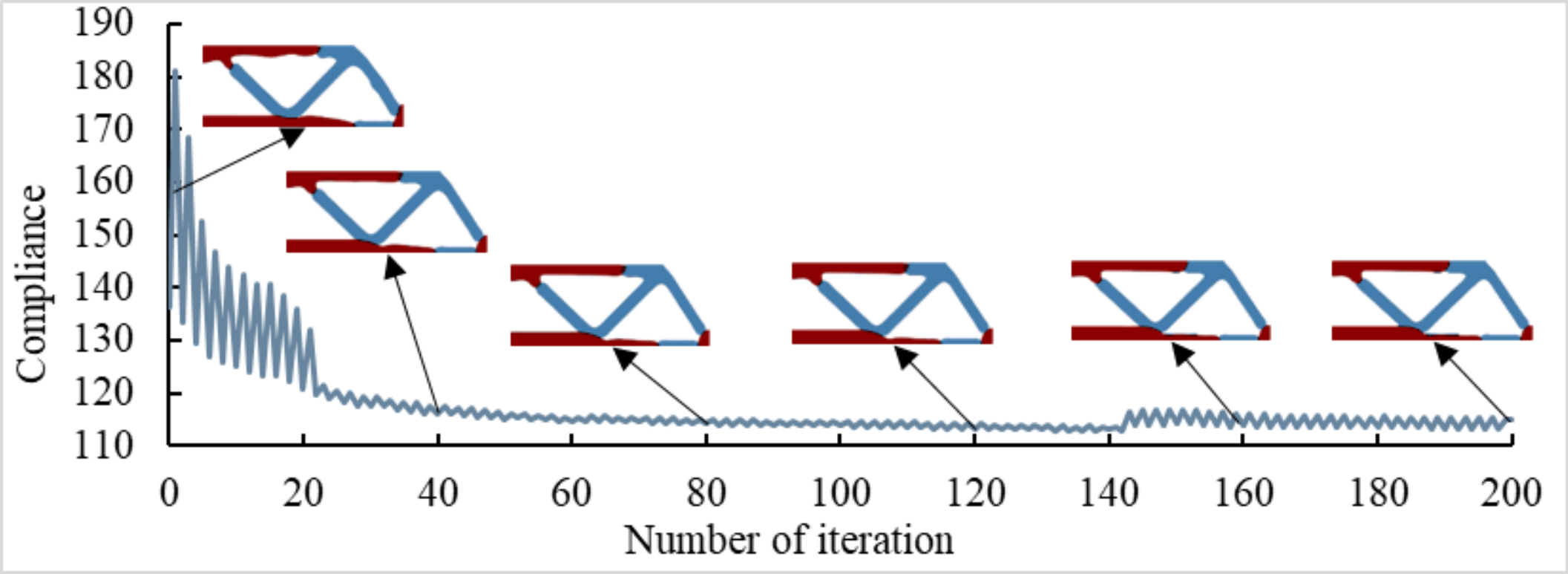

(b) $v_{E=2} = 0.2$ , $v_{E=1} = 0.2$

Fig 5. Optimization Process of the Three-Phase 2D Half-MBB Beam

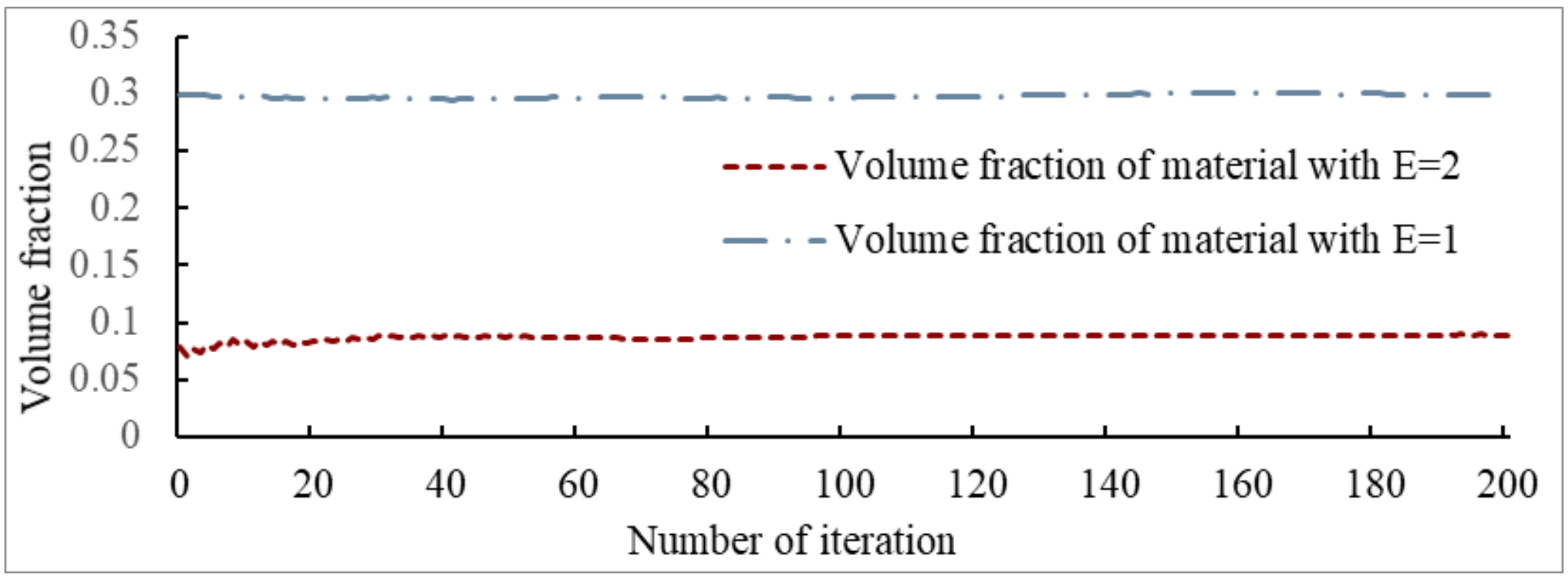

(a) $v_{E=2}=0.1$, $v_{E=1}=0.3$

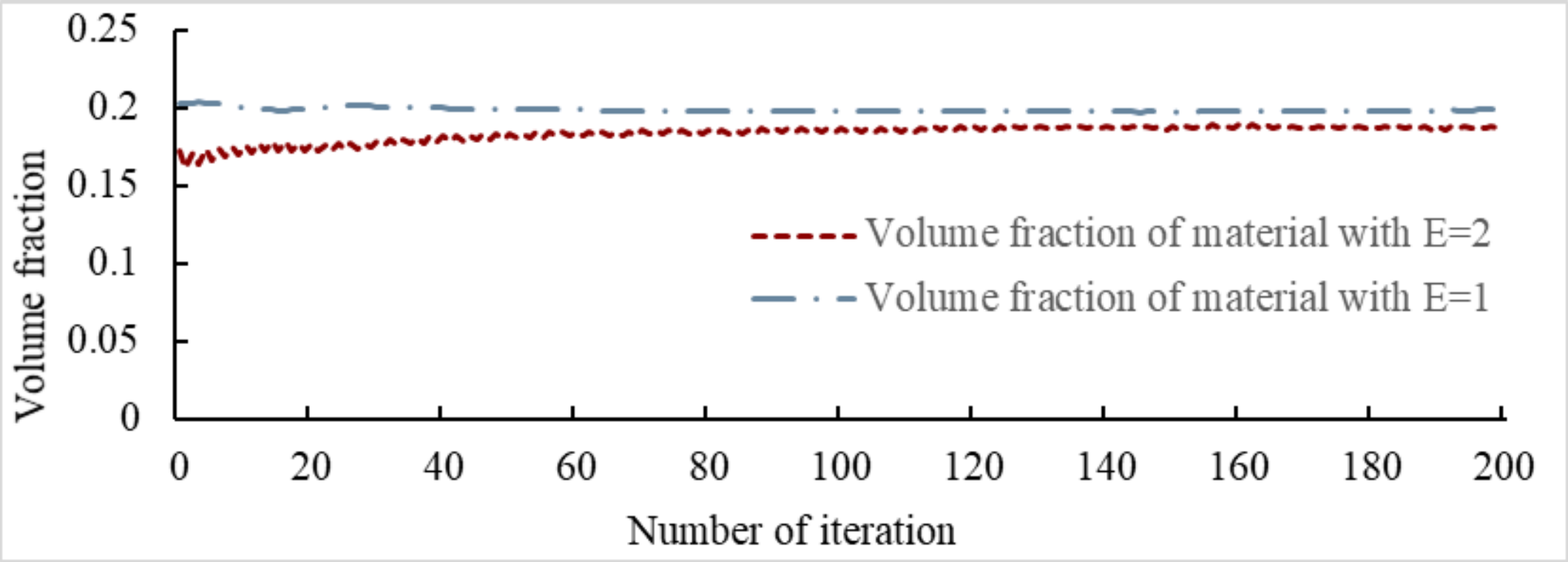

(b) $v_{E=2}=0.2$, $v_{E=1}=0.2$

Fig 6. Iteration History of Volume Fractions for Each Material Phase of the Three-Phase 2D Half-MBB Beam

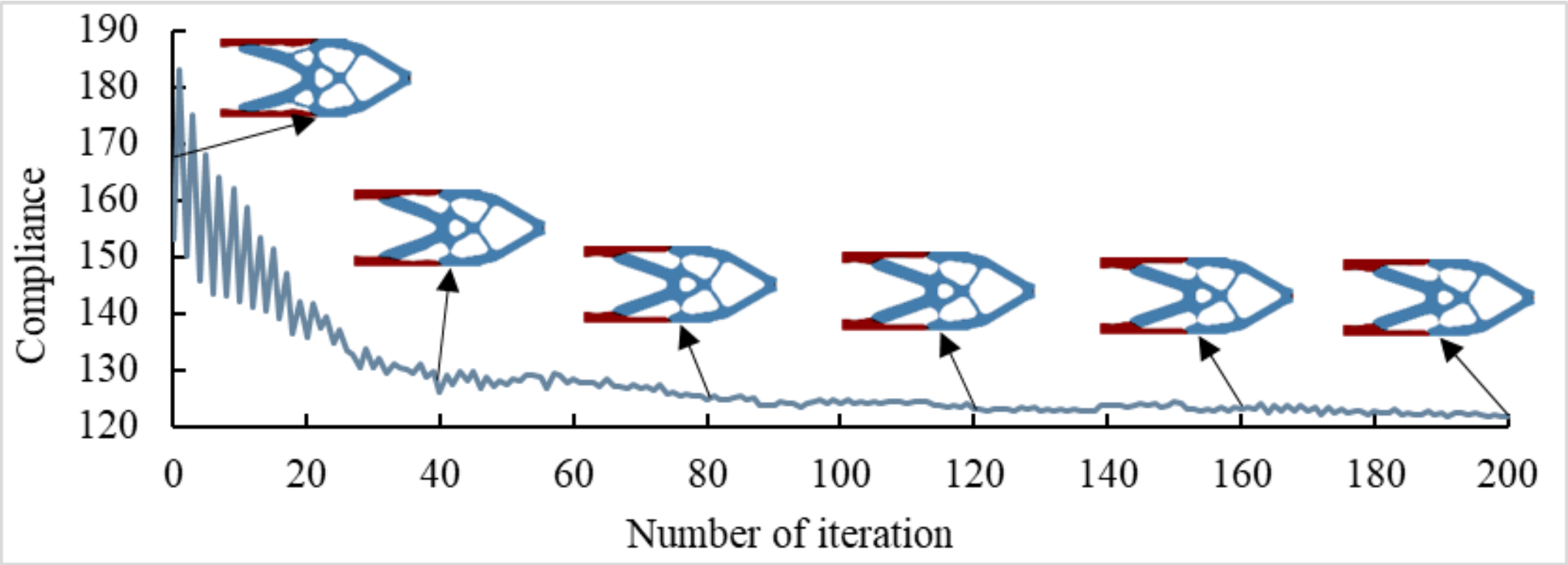

(a) $v_{E=2}=0.1$, $v_{E=1}=0.3$

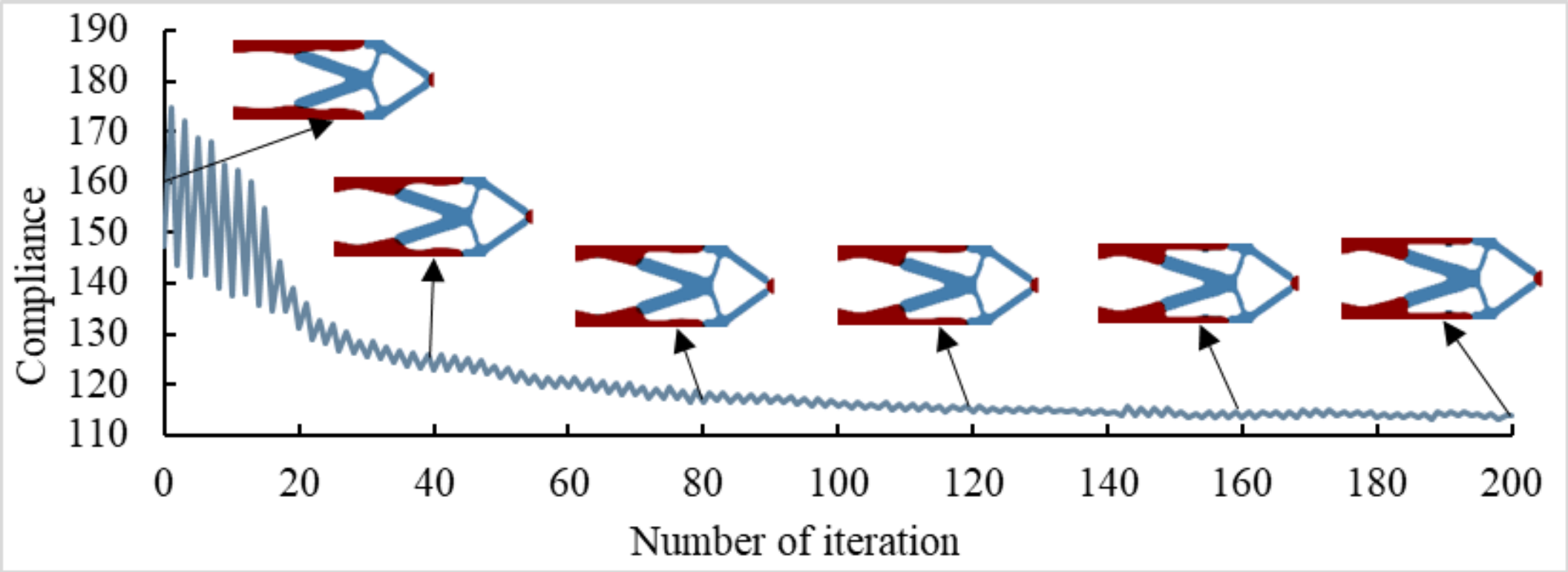

(b) $v_{E=2} = 0.2$ , $v_{E=1} = 0.2$

Fig 7. Optimization Process of the Three-Phase 2D Cantilever Beam

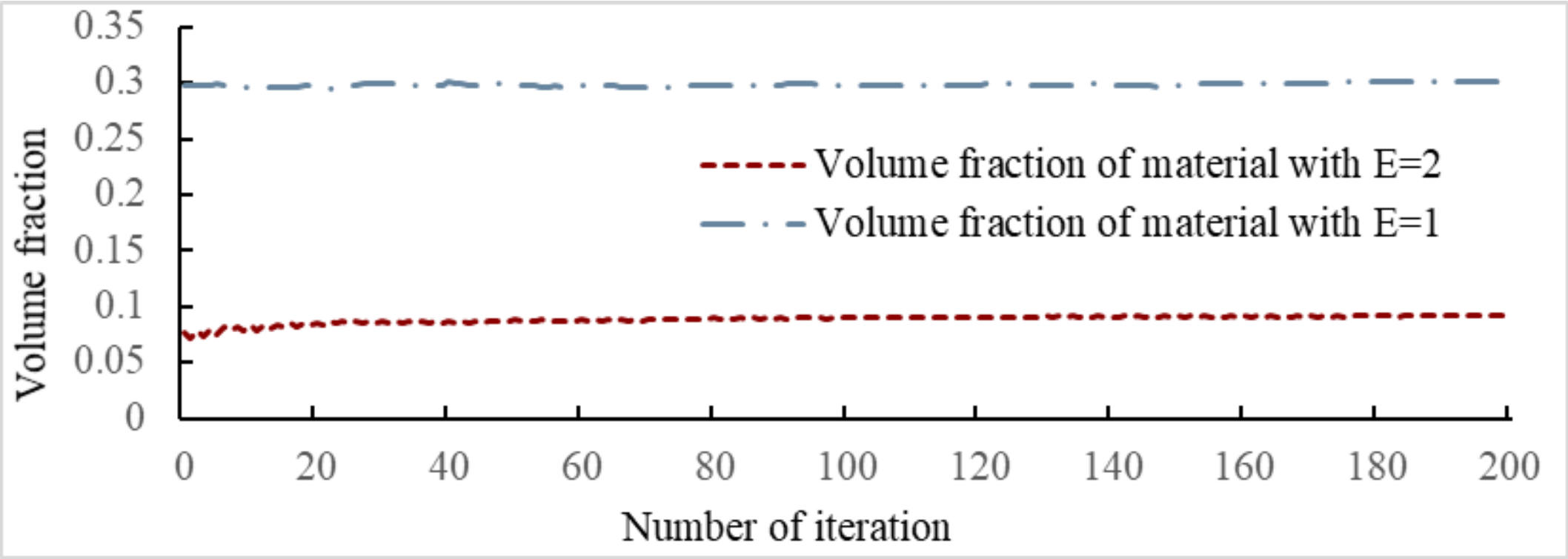

(a) $v_{E=2} = 0.1$ , $v_{E=1} = 0.3$

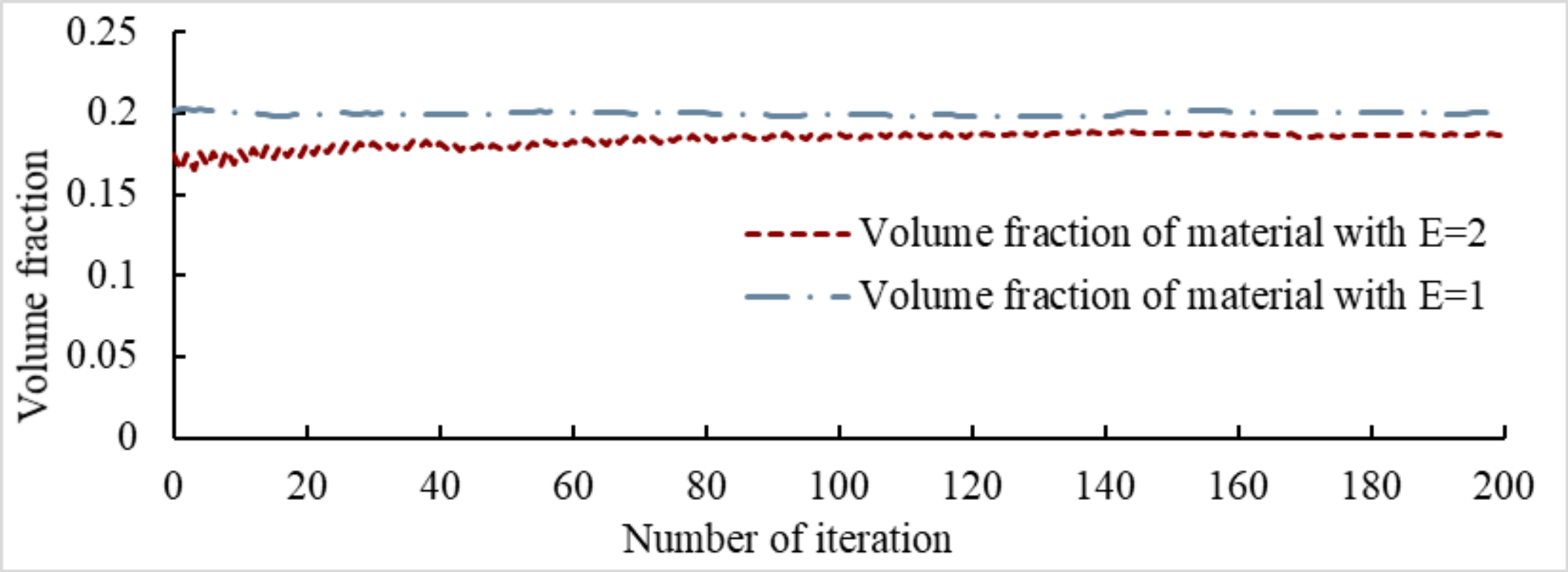

(b) $v_{E=2} = 0.2$ , $v_{E=1} = 0.2$

Fig 8. Iteration History of Volume Fractions for Each Material Phase of the Three-Phase 2D Cantilever Beam

### *5.2. Comparison of Four-Phase Material Optimization Results*

Three solid materials plus a void phase (four material phases) are now considered. Their Young's moduli are E=4, E=2 and E=1, drawn from the darkest to the lightest shade in order of decreasing stiffness.

The same two beams are optimized for two volume fraction combinations: (1) $v_{E=4} = 0.05$, $v_{E=2} = 0.1$, $v_{E=1} = 0.3$; (2) $v_{E=4} = 0.05$, $v_{E=2} = 0.1$, $v_{E=1} = 0.4$. As in Section 5.1, the filter radius $r_f$ is set to 5.

Table 2 displays the optimization results for the four-phase 2D cantilever and 2D half-MBB beams obtained by the SIMP method, by the AAPA only and by the present method. After 200 iterations, the compliance values for the optimized 2D half-MBB beams converge to 104.76 and 77.98 under the two different volume fraction combinations, respectively. For the 2D cantilever beams, the corresponding compliance values converge to 80.36 and 68.94, respectively.

As in the three-phase case, the boundaries are smooth and the interfaces between the solid materials are clear. With the other solid volume fractions held constant, the compliance decreases as the volume fraction of any one solid phase grows. Adding a third solid material (E=4) changes none of this: it occupies the supports and the loading point, and the two weaker materials reinforce the rest.

Table 2. Four-phase 2D beams: optimized structures and compliance obtained with SIMP, with the AAPA only and with the negative-mapping interpolation.

| **Volume fractions** | **Method** | **Half-MBB beam** | **Cantilever beam** |
|---|---|---|---|
| $v_{E=4} = 0.05$, $v_{E=2} = 0.1$, $v_{E=1} = 0.3$ | SIMP | 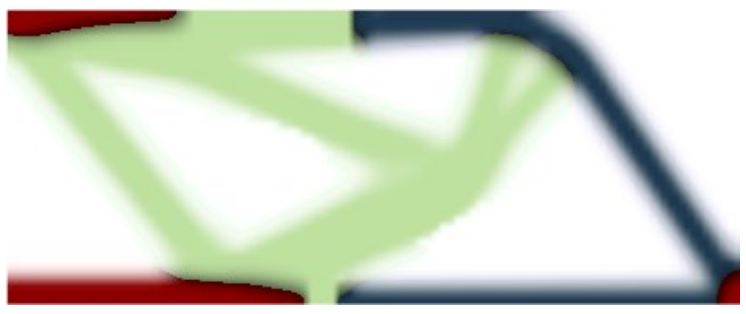 $C = 99.99$ | 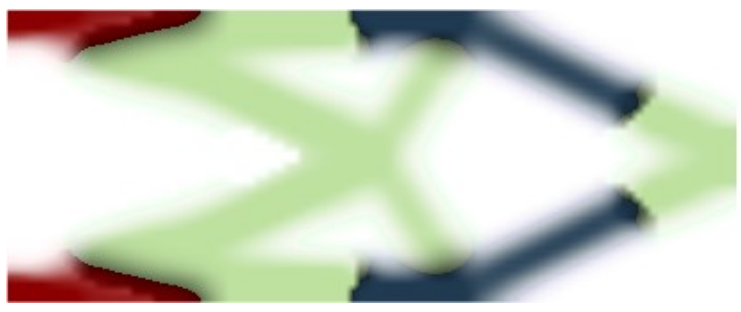 $C = 105.41$ |
| | AAPA only | 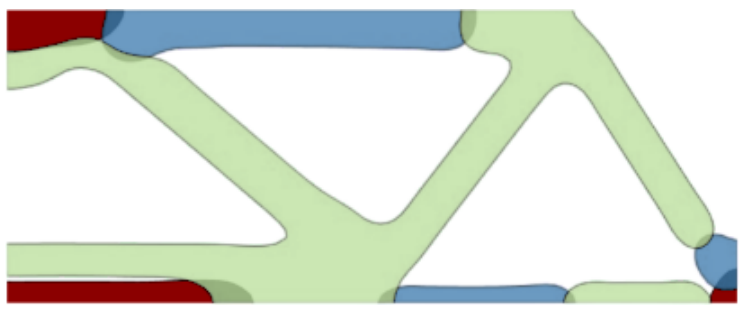 $C = 105.86$ | 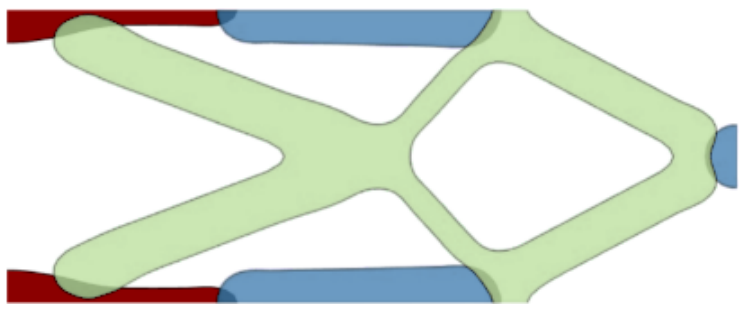 $C = 82.61$ |
| | Negative-mapping interpolation | 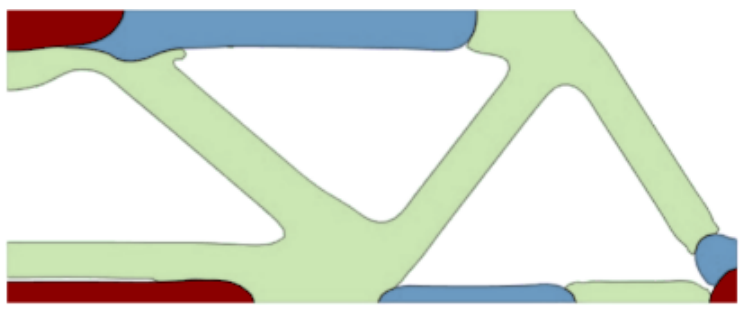 $C = 104.76$ | 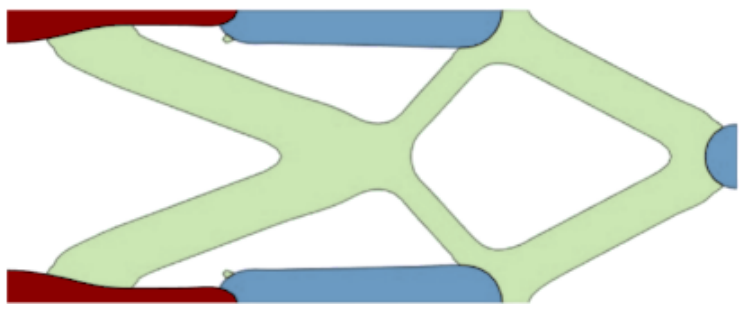 $C = 80.36$ |
| $v_{E=4} = 0.05$, $v_{E=2} = 0.1$, $v_{E=1} = 0.4$ | SIMP | 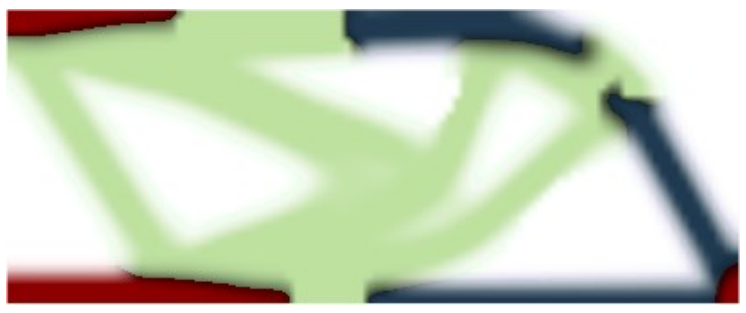 $C = 85.59$ | 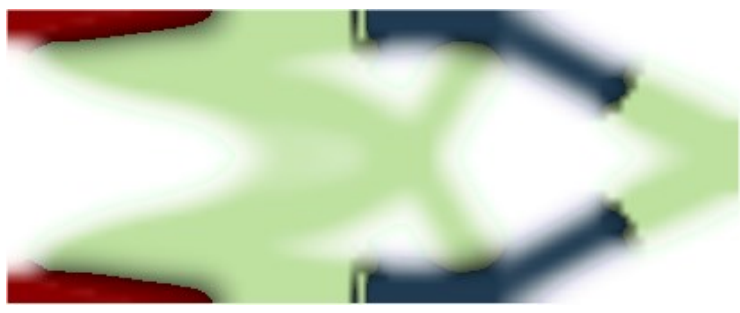 $C = 87.06$ |
| | AAPA only | 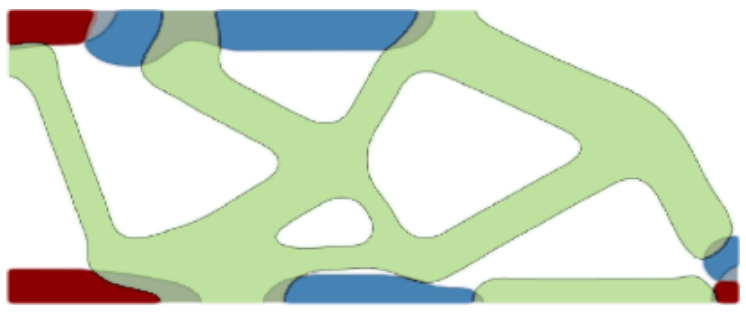 $C = 79.72$ | 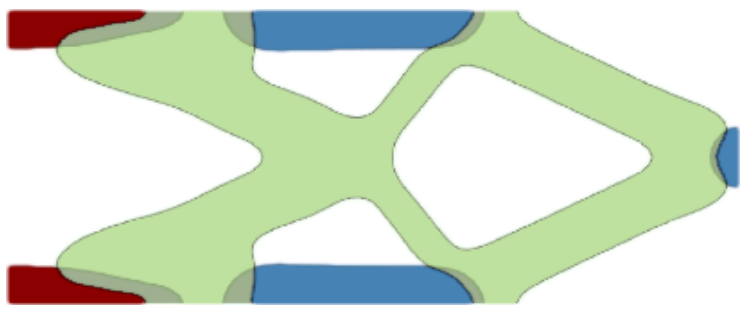 $C = 71.39$ |

| Volume fractions | Method | Half-MBB beam | Cantilever beam |
|---|---|---|---|
| | Negative-mapping interpolation | 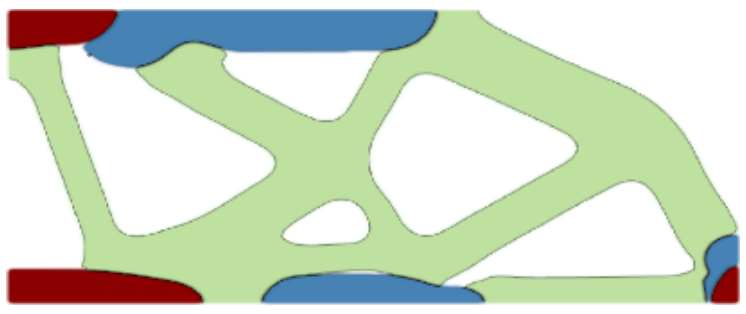 $C = 77.98$ | 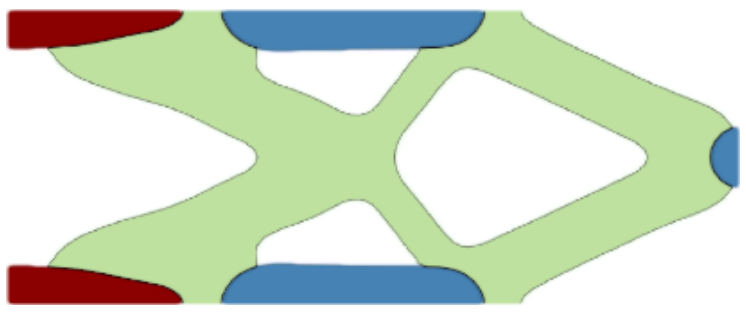 $C = 68.94$ |

Fig 9 shows the evolution of the topology and of the compliance of the four-phase half-MBB beam, and Fig 10 the iteration history of the volume fractions of each material phase.

At the start of the iteration, the material with Young's modulus $E$=4 occupies the entire design domain, while the materials with $E$=2 and $E$=1 have volume fractions of 0. This initial state gives the stiffest possible structure. After the first iteration the volume fractions approach their prescribed values, that of the $E$=4 material decreasing slightly and those of the $E$=2 and $E$=1 materials increasing, while the compliance decreases with oscillation. From about the 10th iteration the volume fraction of the $E$=4 material increases again slightly and the compliance settles. After about 150 iteration steps the topology of the half-MBB beam remains essentially unchanged.

Fig 11 shows the same two histories for the four-phase cantilever beam, and Fig 12 the corresponding volume fractions. The first iterations are similar to those of the half-MBB beam. From about the 15th iteration the volume fraction of the E=4 material rises slightly again and the compliance keeps falling; after about 180 iterations the topology no longer changes and the compliance settles.

The figures show that the present method solves the 2D four-phase problem for both beams and yields convergent results, the main topological structure being formed within about 40 outer iterations, with only minor local changes thereafter. Here too the interfaces are smooth and free of overlap.

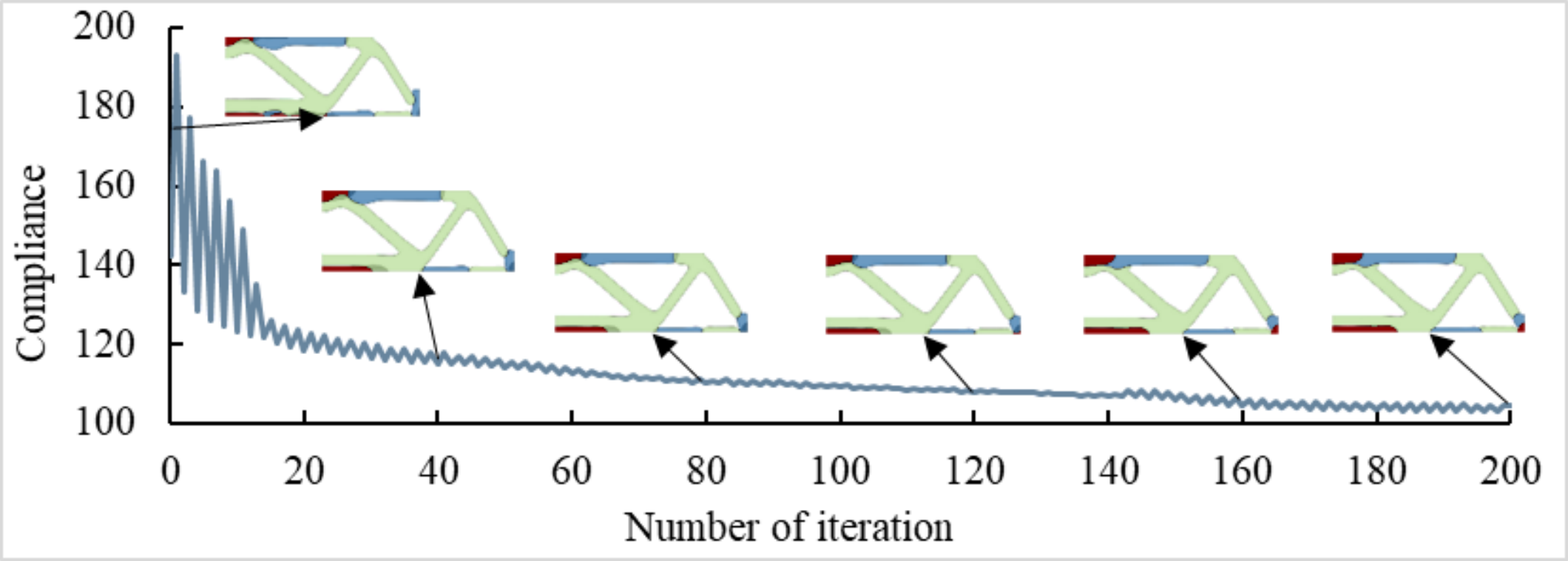

(a) $v_{E=4} = 0.05$, $v_{E=2} = 0.1$, $v_{E=1} = 0.3$

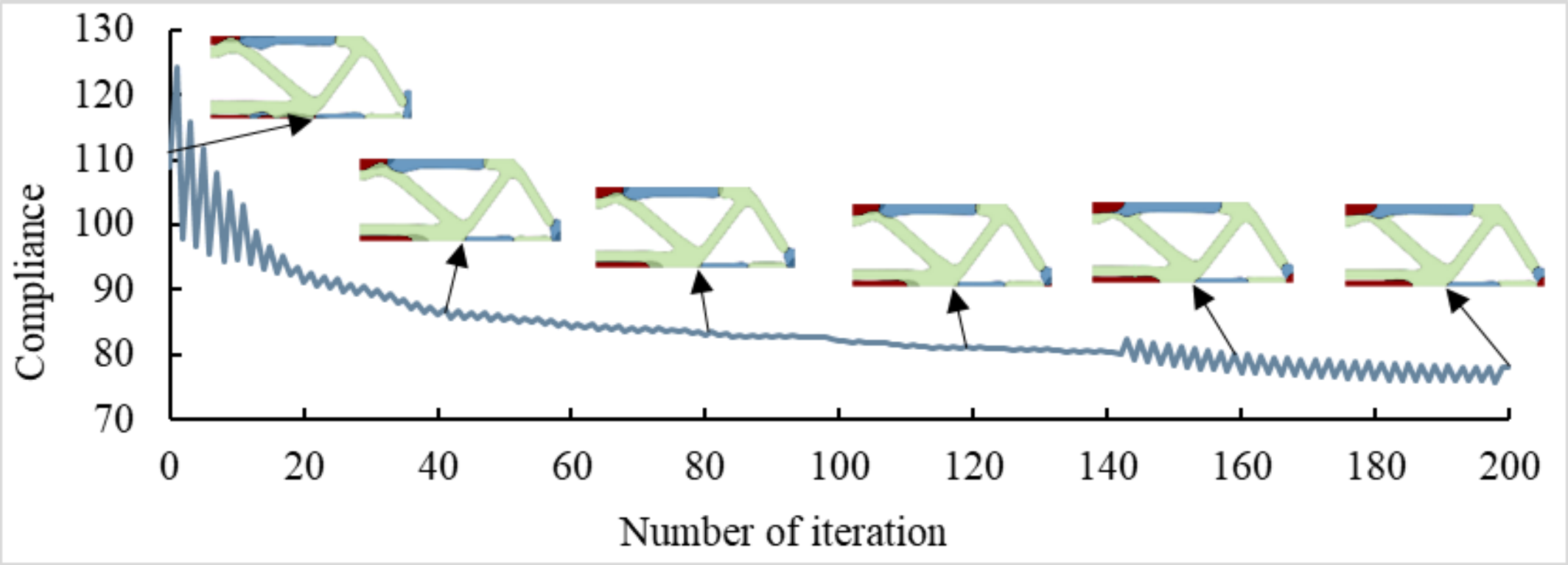

(b) $v_{E=4} = 0.05$, $v_{E=2} = 0.1$, $v_{E=1} = 0.4$

Fig 9. Optimization Process of the Four-Phase 2D Half-MBB Beam

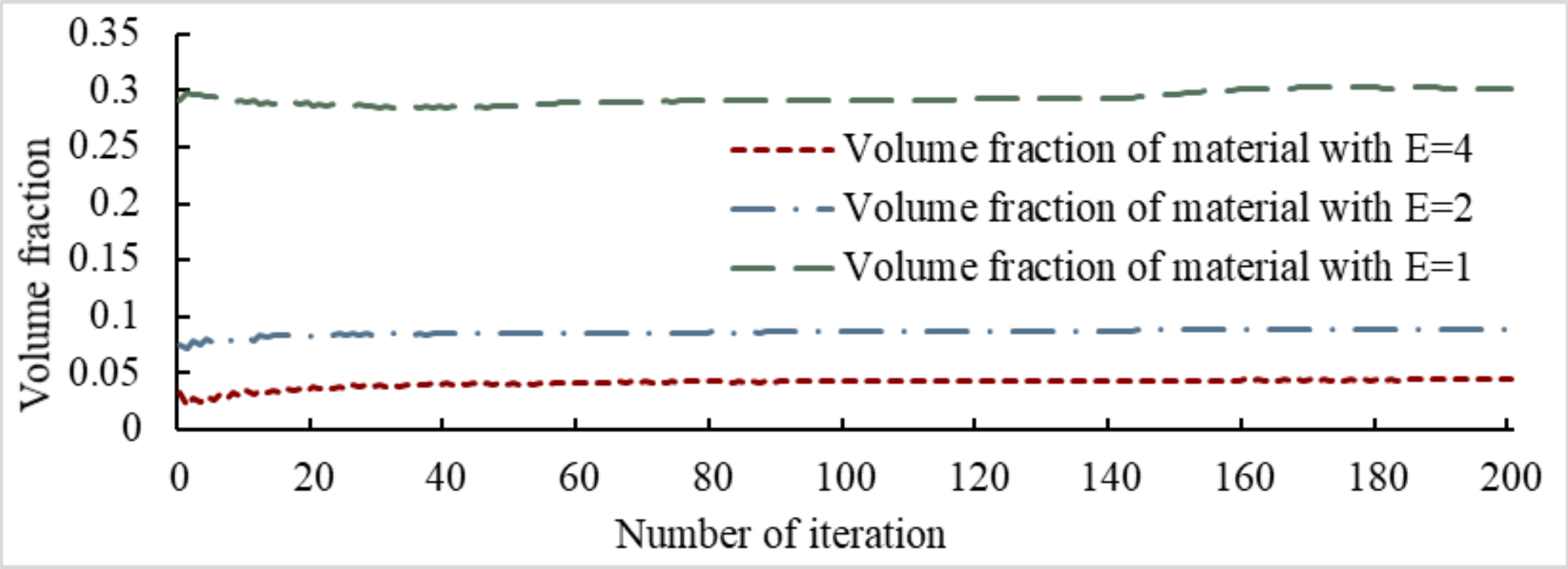

(a) $v_{E=4} = 0.05$, $v_{E=2} = 0.1$, $v_{E=1} = 0.3$

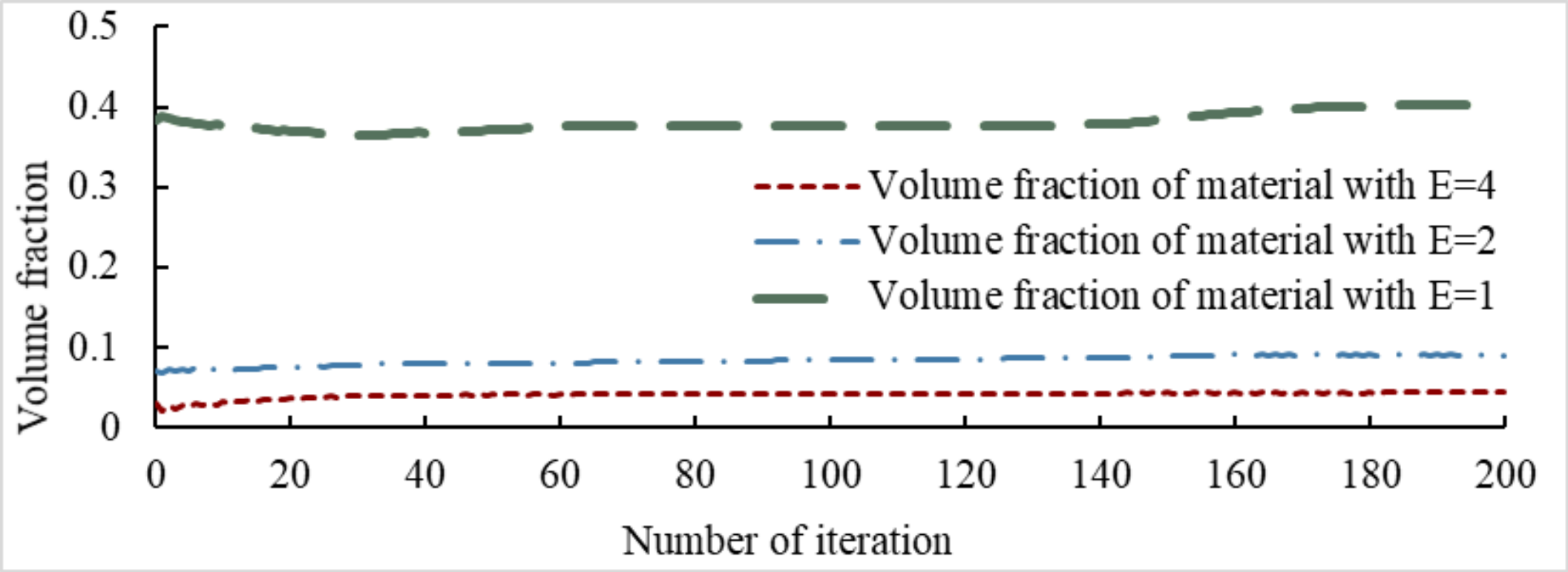

(b) $v_{E=4} = 0.05$, $v_{E=2} = 0.1$, $v_{E=1} = 0.4$

Fig 10. Iteration History of Volume Fractions for Each Material Phase of the Four-Phase 2D Half-MBB Beam

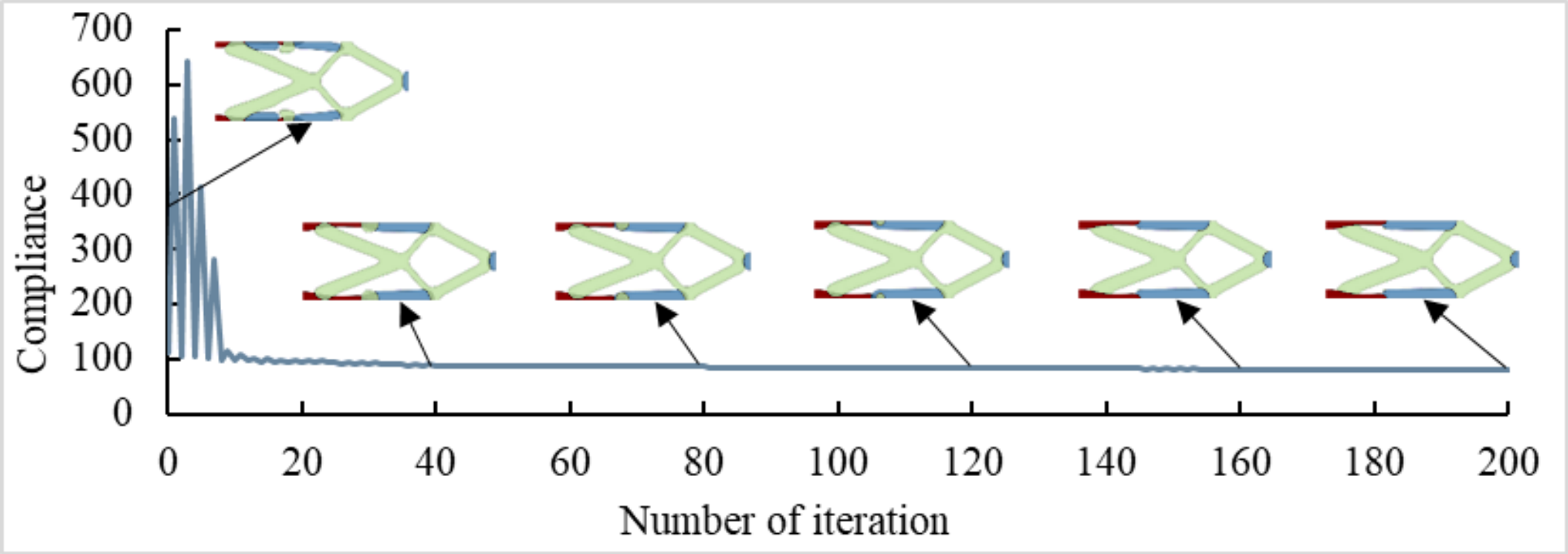

(a) $v_{E=4}=0.05$, $v_{E=2}=0.1$, $v_{E=1}=0.3$

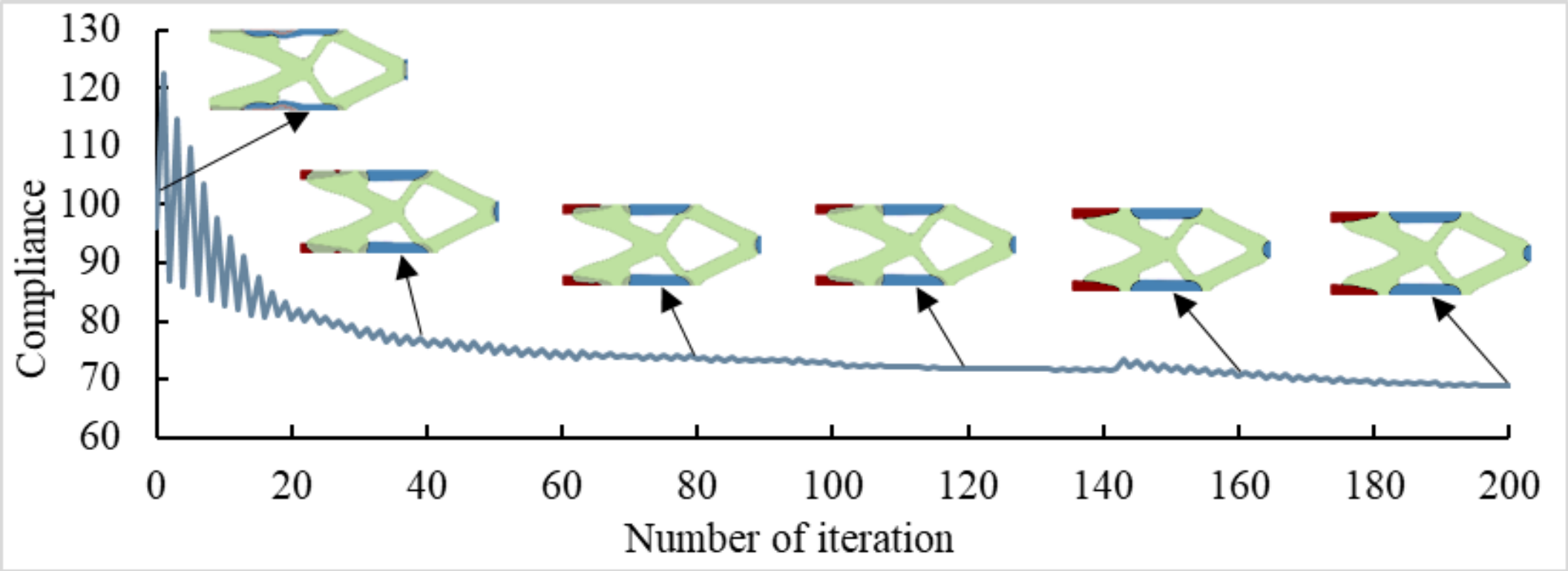

(b) $v_{E=4}=0.05$, $v_{E=2}=0.1$, $v_{E=1}=0.4$

Fig 11. Optimization Process of the Four-Phase 2D Cantilever Beam

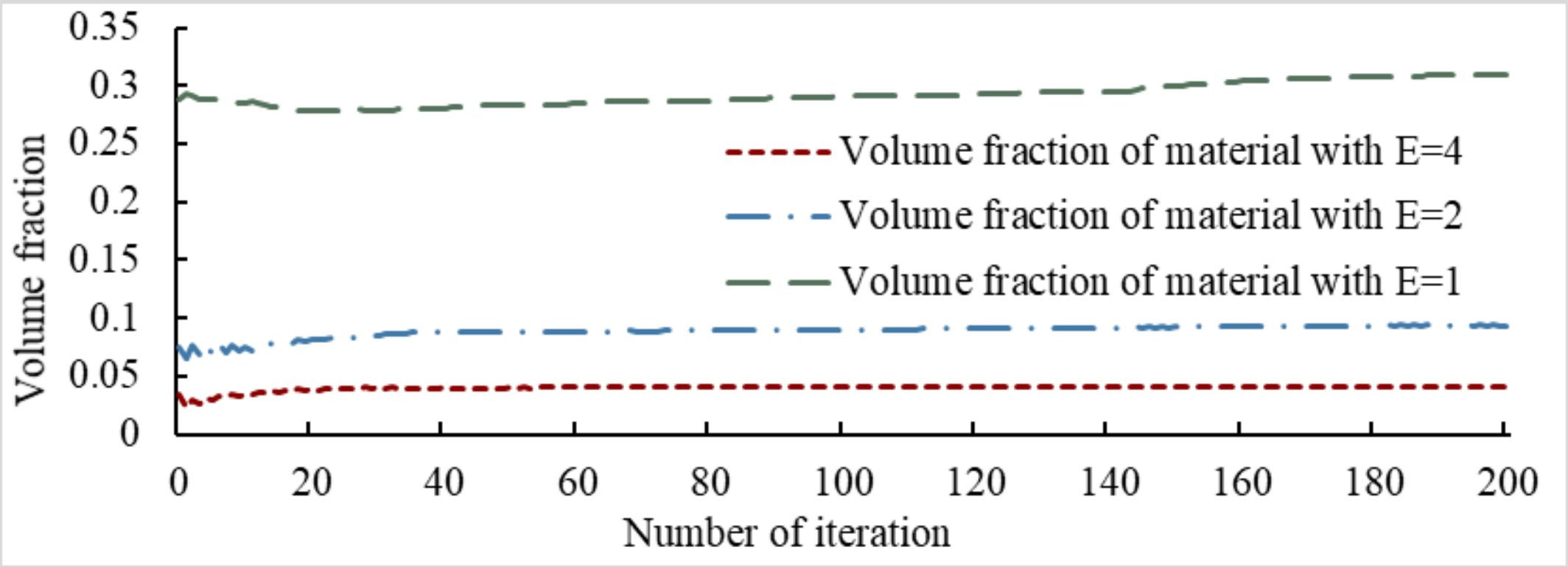

(a) $v_{E=4}=0.05$, $v_{E=2}=0.1$, $v_{E=1}=0.3$

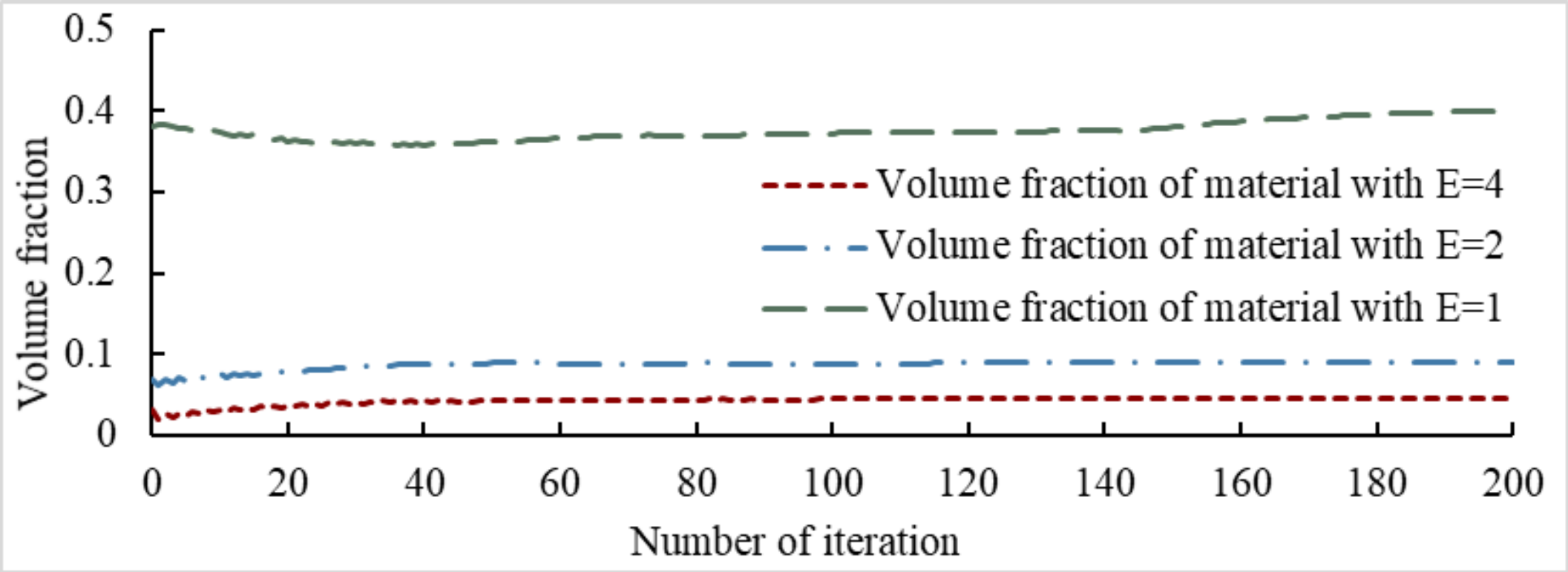

(b) $v_{E=4}=0.05$ , $v_{E=2}=0.1$ , $v_{E=1}=0.4$

Fig 12. Iteration History of Volume Fractions for Each Material Phase of the Four-Phase 2D Cantilever Beam

### *5.3. Influence of Interface Reconstruction on the Optimized Structure*

This section examines how the interface reconstruction affects the boundaries, on the two beams with different volume fractions and numbers of phases. The volume fractions for the three-phase materials are set as: (1) $v_{E=2}=0.1$ , $v_{E=1}=0.3$ ; (2) $v_{E=2}=0.2$ , $v_{E=1}=0.2$ . The volume fractions for the four-phase materials are set as: (1) $v_{E=4}=0.05$ , $v_{E=2}=0.1$ , $v_{E=1}=0.3$ ; (2) $v_{E=4}=0.05$ , $v_{E=2}=0.1$ , $v_{E=1}=0.4$ . All other parameters are those of Section 5.

Two interface reconstruction schemes are compared. The first constrains the volume of the materials inside an element with the AAPA constraint alone, as in Eq. (5). The second adds the negative-mapping interpolation at the interfaces between different materials, as in Eq. (20).

Table 1 compares, in the last two rows of each block, the two interface reconstruction schemes (AAPA only and negative-mapping interpolation) for the three-phase materials. As shown there, when the material volume fractions are (1) $v_{E=2}=0.1$ , $v_{E=1}=0.3$ ; (2) $v_{E=2}=0.2$ , $v_{E=1}=0.2$ , the two schemes give similar objective values: the flip acts on the overlapping parts only and leaves the topology as it was.

Table 2 gives the same comparison for the four-phase materials.

The optimization results for the four-phase materials are similar to those of the three-phase materials. When the material volume fractions are (1) $v_{E=4}=0.05$ , $v_{E=2}=0.1$ , $v_{E=1}=0.3$ ; (2) $v_{E=4}=0.05$ , $v_{E=2}=0.1$ , $v_{E=1}=0.4$ , the objective function values obtained with the AAPA only are smaller than those obtained with the negative-mapping interpolation, and the difference grows with the total volume fraction of the solid materials, because more material overlap occurs in the four-phase problems.

The tables show that with the AAPA volume constraint alone an obvious overlap appears at the interfaces between different materials, because the AAPA controls only the volume fractions within an element and not the regions that the phases occupy inside it: an interface element becomes a mixture of the phases, which in the level set representation appears as material overlap. The negative-mapping interpolation of Eq. (20) adjusts the regions occupied by the phases within the element and gives a clear separation of the materials in the interface element.

### *5.4. Discussion on the nodal compliance ratio filtering technique*

This subsection examines how the nodal compliance proportional filter affects the results, on the same two beams. The volume fractions (constraints) for the three-phase material are set as (1) $v_{E=2}=0.1$, $v_{E=1}=0.3$; (2) $v_{E=2}=0.2$, $v_{E=1}=0.2$. The volume fractions (constraints) for the four-phase material are set as (1) $v_{E=4}=0.05$, $v_{E=2}=0.1$, $v_{E=1}=0.3$; (2) $v_{E=4}=0.05$, $v_{E=2}=0.1$, $v_{E=1}=0.4$. All other parameters are those of Section 5.

Table 3 presents the structural topology and objective function values obtained for the three-phase material structure using different filtering schemes. The sensitivity filtering scheme is not applicable, since the method does not compute sensitivities; the discussion therefore compares density filtering, compliance proportional filtering and no filtering.

For the 2D half-MBB beam, the compliance obtained with the compliance proportional filtering scheme ($C$ = 132.32) is larger than those obtained with density filtering ($C$ = 126.60) and without filtering ($C$ = 123.79); the same holds for the 2D cantilever beam ($C$ = 121.83 against 108.98 and 107.10).

Table 3. Optimization Results of 2D Structures Obtained Using Different Filtering Schemes (Three-Phase Material)

| **Filtering Scheme** | **No Filtering Scheme** | **Density Filtering** | **Compliance Proportional Filtering** |
|---|---|---|---|
| Half-MBB Beam | 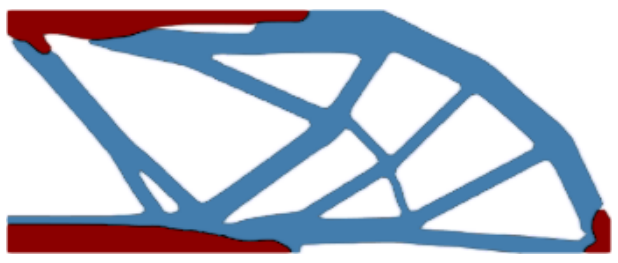<br>$C$ = 123.79 | 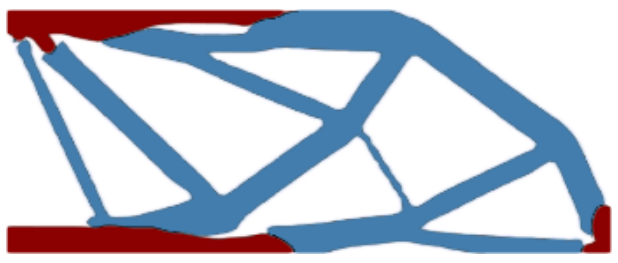<br>$C$ = 126.60 | 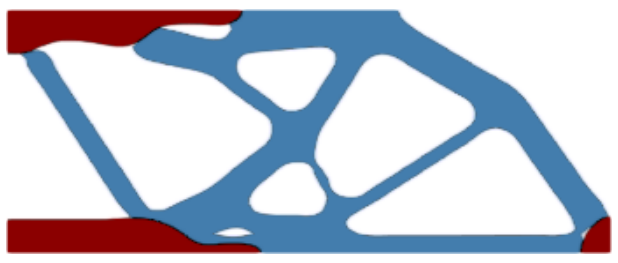<br>$C$ = 132.32 |
| Cantilever Beam | 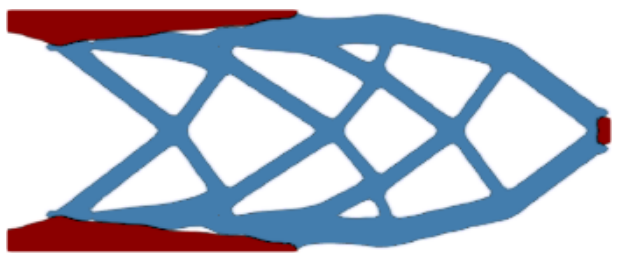<br>$C$ = 107.10 | 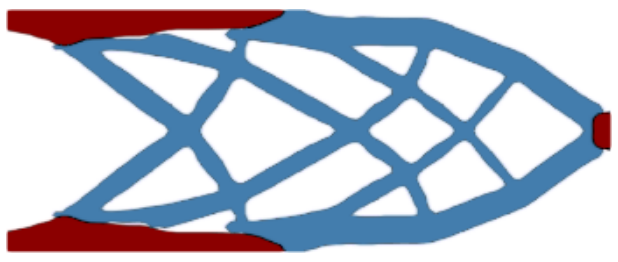<br>$C$ = 108.98 | 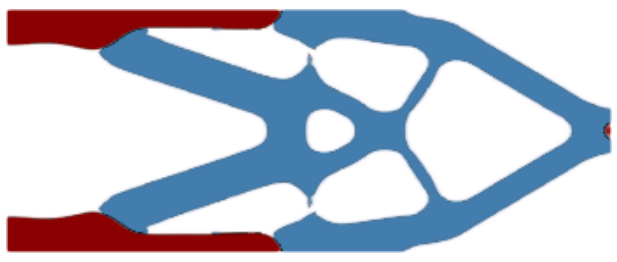<br>$C$ = 121.83 |

Table 4 presents the structural topology and objective function values obtained for the four-phase material structure using different filtering schemes (No Filtering, Density Filtering, and Nodal Compliance Proportional Filtering).

For the 2D half-MBB beam, the compliance obtained with the compliance proportional filtering scheme (C = 104.76) is larger than those obtained with density filtering ($C$ = 93.10) and without filtering ($C$ = 89.57). For the 2D cantilever beam it ($C$ = 80.36) is slightly smaller than with density filtering ($C$ = 80.54) and larger than without filtering ($C$ = 78.67).

The two tables show that, although the compliance obtained with the nodal compliance proportional filtering scheme is larger than those obtained with density filtering or without filtering, those two strategies produce topologies with numerous small holes and chaotic branches, which are unsuitable for practical use. The nodal compliance proportional filter leaves no redundant branches and so reduces the post-processing work.

Table 4. Optimization Results of 2D Structures Obtained Using Different Filtering Schemes (Four-Phase Material)

| **Filtering Scheme** | **No Filtering Scheme** | **Density Filtering** | **Compliance Proportional Filtering** |
|---|---|---|---|
| Half-MBB Beam | 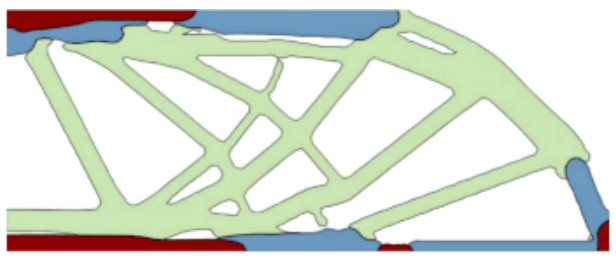 $C = 89.57$ | 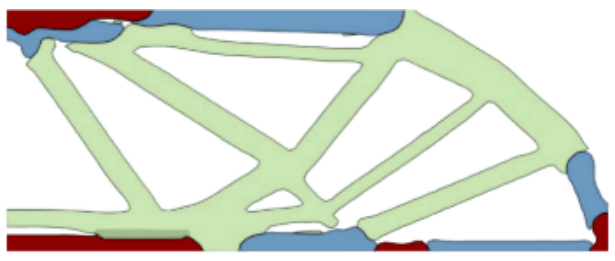 $C = 93.10$ | 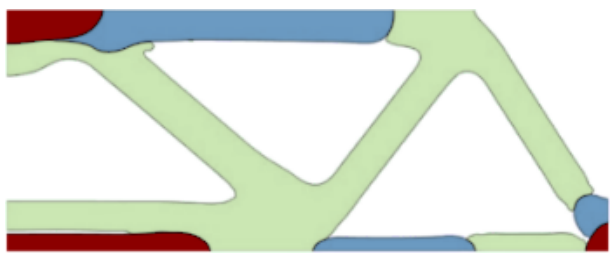 $C = 104.76$ |
| Cantilever Beam | 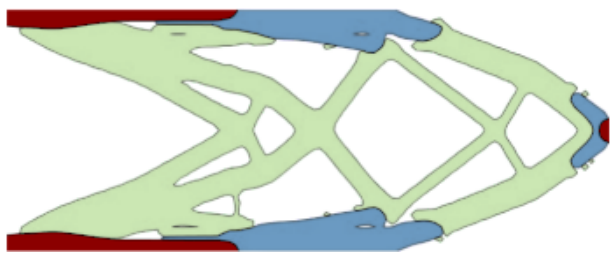 $C = 78.67$ | 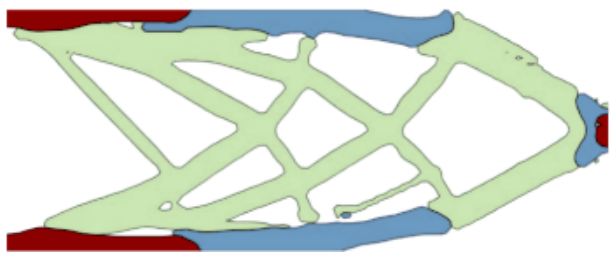 $C = 80.54$ | 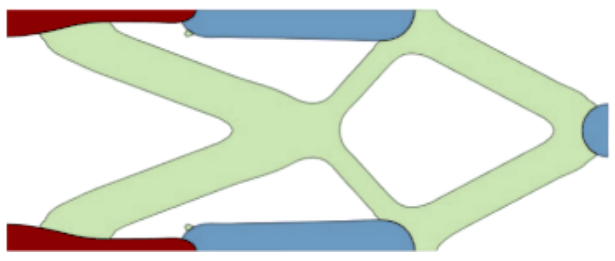 $C = 80.36$ |

## *5.5. Effect of the number of initial inner iterations*

This section examines, on the half-MBB beam, how the number of inner iterations performed in the first outer iteration (iter_max_in1, called the initial inner iterations below and set to 150 elsewhere in this article) affects the results under different volume constraints. The volume constraints for the three-phase material are set as: (1) $v_{E=2} = 0.1$, $v_{E=1} = 0.3$; (2) $v_{E=2} = 0.2$, $v_{E=1} = 0.2$. The volume constraints for the four-phase material are set as: (1) $v_{E=4} = 0.05$, $v_{E=2} = 0.1$, $v_{E=1} = 0.3$; (2) $v_{E=4} = 0.05$, $v_{E=2} = 0.1$, $v_{E=1} = 0.4$. Except for the different initial inner iterations $iter_max_in_1$, all other parameters are those of Section 5.

Table 5 and Table 6 respectively present the optimization results for the three-phase 2D half-MBB beam obtained under different volume fractions using various initial inner iterations $iter_max_in_1$, namely 4, 20, 50, 100, 150 and 200. With few initial inner iterations (e.g. 4 or 20) the compliance is larger and the topology is unstable, showing redundant, branched and void members; as the number increases, both the compliance and the configuration improve. When the initial inner iterations $iter_max_in_1$ are greater than 100, the topology optimization results tend to stabilize, with the compliance values of all results being similar and the topological configurations being nearly identical.

Table 5. Optimization Results of Three-Phase Material under Different Initial Inner Iterations ($v_{E=2} = 0.1$, $v_{E=1} = 0.3$)

| $iter_max_in_1 = 4$ | $iter_max_in_1 = 20$ | $iter_max_in_1 = 50$ |
|---|---|---|
| 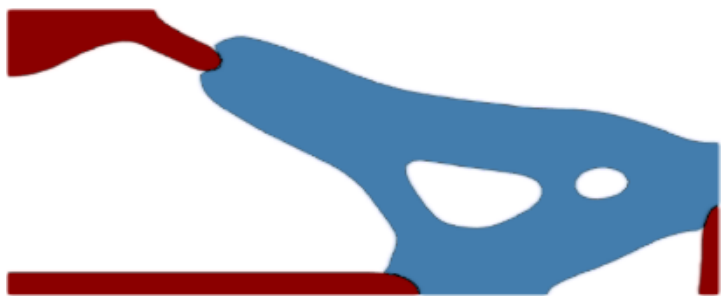 | 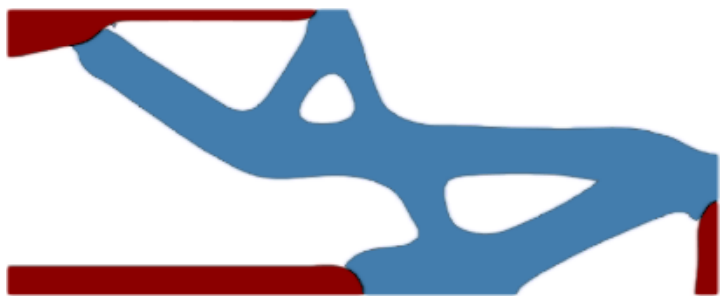 | 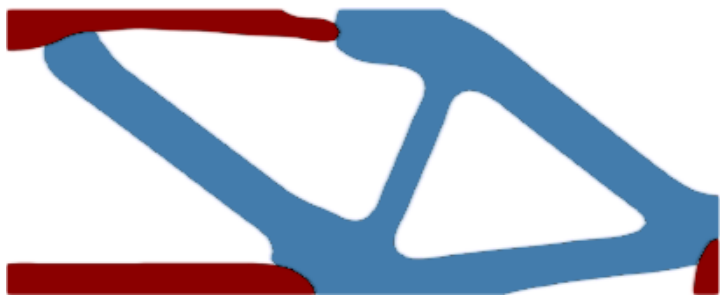 |
| $C = 179.77$ | $C = 199.01$ | $C = 137.54$ |

| $iter_max_in_1 = 100$ | $iter_max_in_1 = 150$ | $iter_max_in_1 = 200$ |
| --- | --- | --- |
| 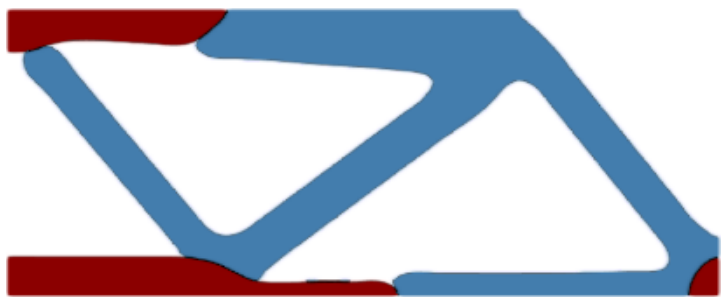 | 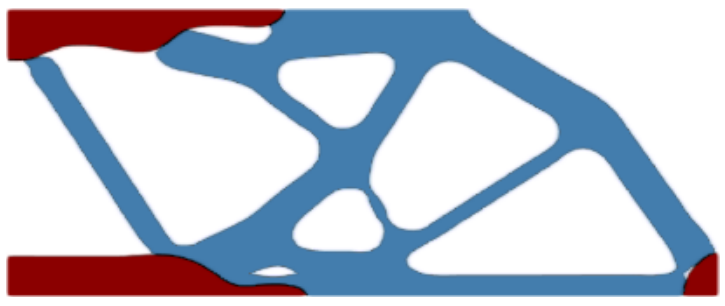 | 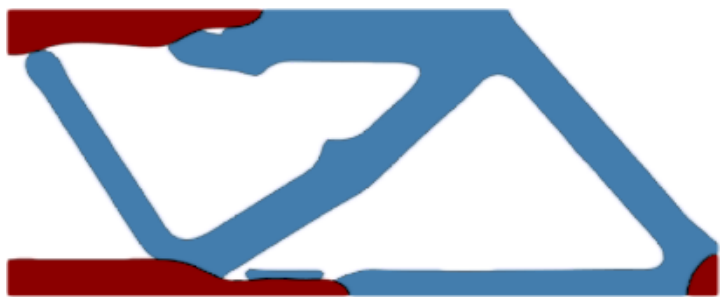 |
| $C = 126.06$ | $C = 132.32$ | $C = 135.14$ |

Table 6. Optimization Results of Three-Phase Material under Different Initial Inner Iterations

( $v_{E=2} = 0.2$ , $v_{E=1} = 0.2$ )

| $iter_max_in_1 = 4$ | $iter_max_in_1 = 20$ | $iter_max_in_1 = 50$ |
| --- | --- | --- |
| 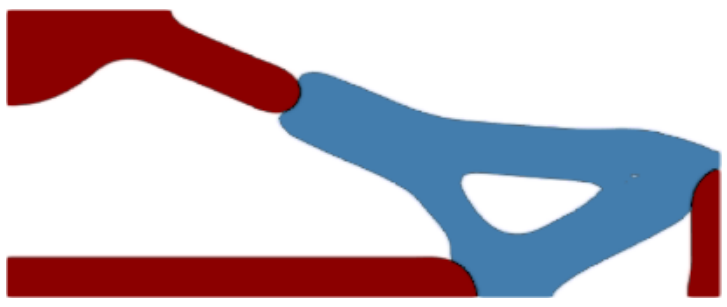 | 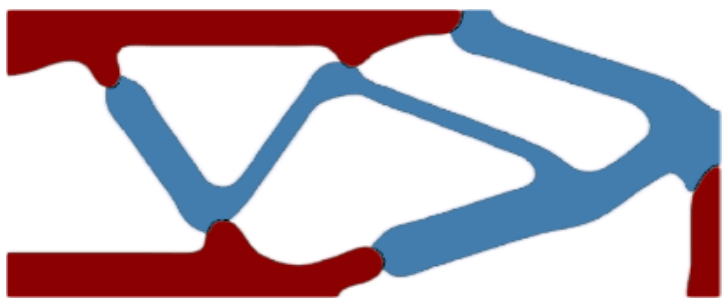 | 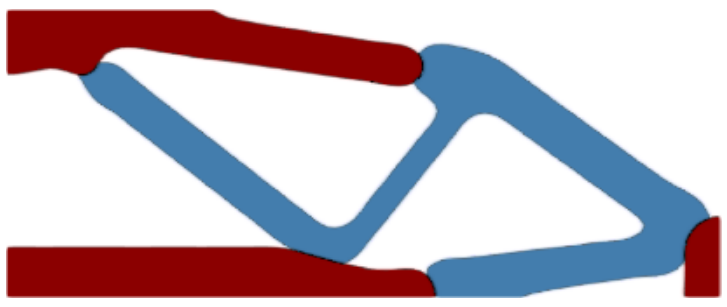 |
| $C = 136.43$ | $C = 134.95$ | $C = 111.44$ |
| $iter_max_in_1 = 100$ | $iter_max_in_1 = 150$ | $iter_max_in_1 = 200$ |
| 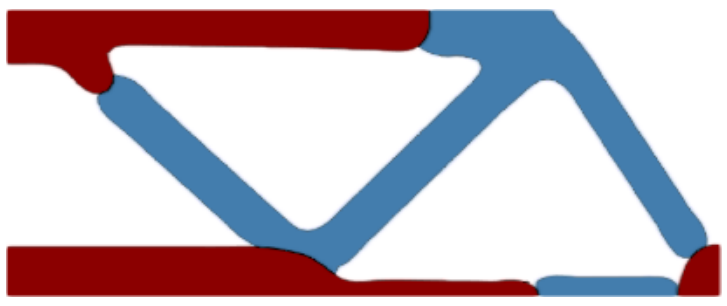 | 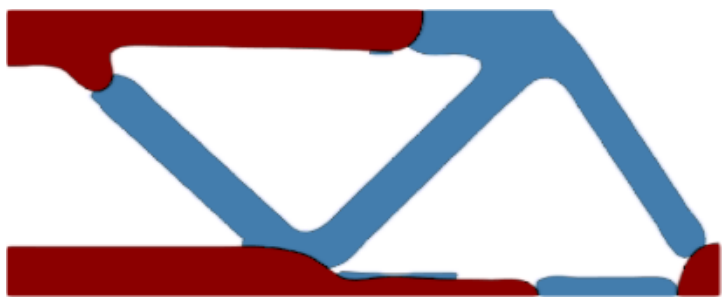 | 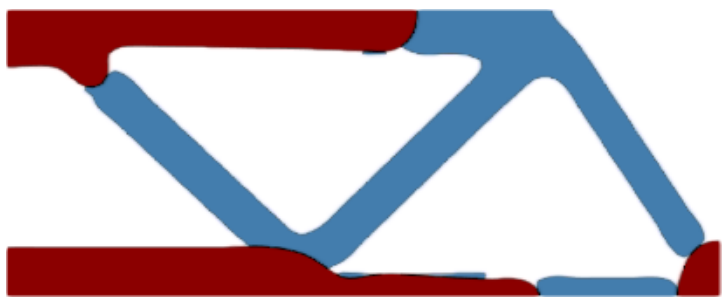 |
| $C = 112.35$ | $C = 115.13$ | $C = 118.09$ |

Table 7 and Table 8 respectively present the optimization results for the four-phase 2D half-MBB beam obtained under different volume fractions using various initial inner iterations $iter_max_in_1$. The behaviour is the same as for the three-phase material: with few initial inner iterations (e.g. 4 or 20) the compliance is larger and the structure exhibits redundant, branched, void and fractured components, which seriously impair its engineering application. When the initial inner iterations $iter_max_in_1$ reach 100, the compliance drops markedly and the design stabilizes: 100 and 150 initial inner iterations give practically the same layout and compliance values that differ by less than 1%. With 200 the prescribed volumes are met just as well, but the optimization settles in a different local solution and the compliance moves by 4–7% (downwards for the first volume set and upwards for the second); the value 150 used throughout this article lies in the stable range.

The optimization starts from a full-material layout and reaches the prescribed volume through the evolutionary strategy, which usually favours a good solution. With few initial inner iterations, however, the volume fractions have not yet reached their prescribed values when the outer iterations begin, and the excess material changes the material distribution. Since this parameter acts only within the first outer iteration, about 140 iterations before the negative-mapping is activated, the sensitivity belongs to the evolutionary volume strategy and not to the interface treatment. The optimization may then settle in a different local solution, which is why the compliance is not monotone in that range.

Table 7. Optimization Results of Four-Phase Material under Different Initial Inner Iterations

( $v_{E=4} = 0.05$ , $v_{E=2} = 0.1$ , $v_{E=1} = 0.3$ )

| $iter_max_in_1 = 4$ | $iter_max_in_1 = 20$ | $iter_max_in_1 = 50$ |
| --- | --- | --- |
| 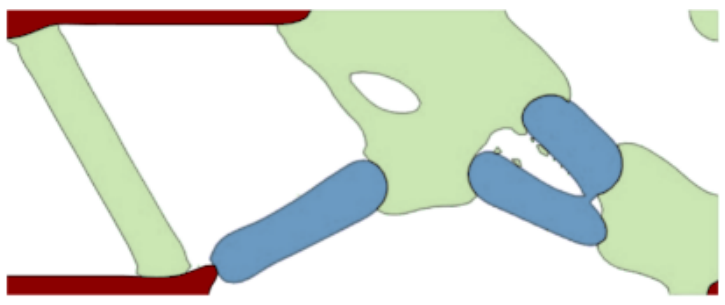 | 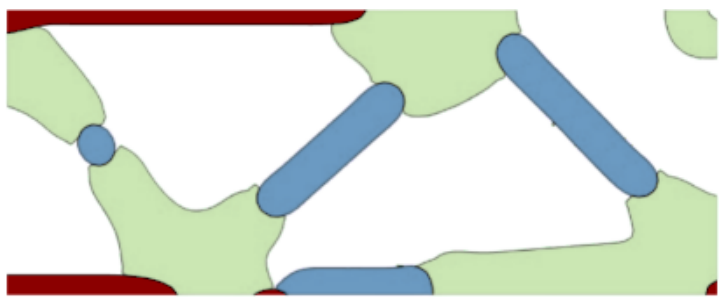 | 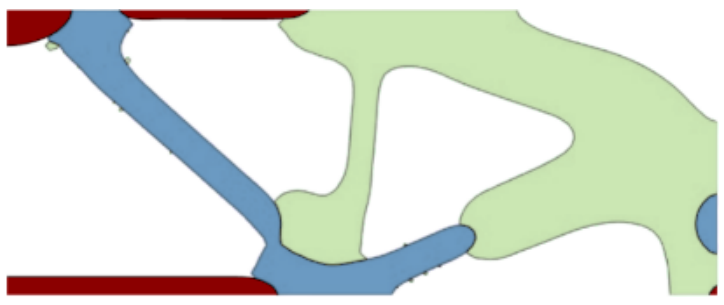 |
| $C$ = 133.32 | $C$ = 110.41 | $C$ = 119.09 |
| $iter_max_in_1$ = 100 | $iter_max_in_1$ = 150 | $iter_max_in_1$ = 200 |
| 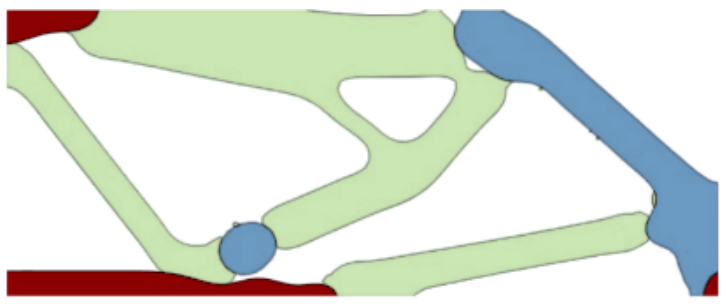 | 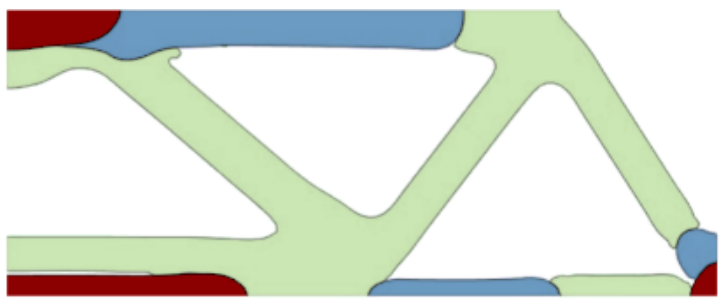 | 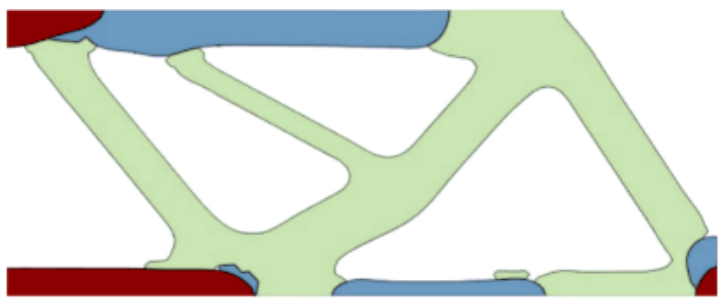 |
| $C$ = 103.51 | $C$ = 104.76 | $C$ = 96.53 |

Table 8. Optimization Results of Four-Phase Material under Different Initial Inner Iterations

( $v_{E=4} = 0.05$ , $v_{E=2} = 0.1$ , $v_{E=1} = 0.4$ )

| $iter_max_in_1$ = 4 | $iter_max_in_1$ = 20 | $iter_max_in_1$ = 50 |
|---|---|---|
| 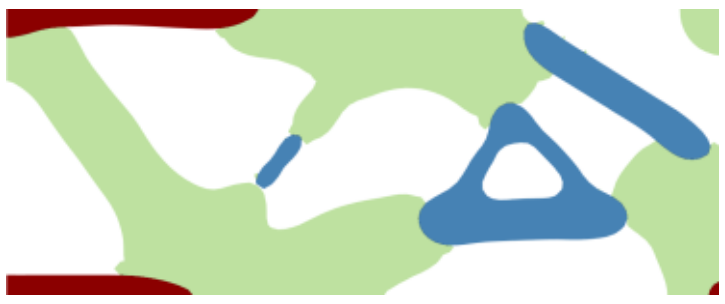 | 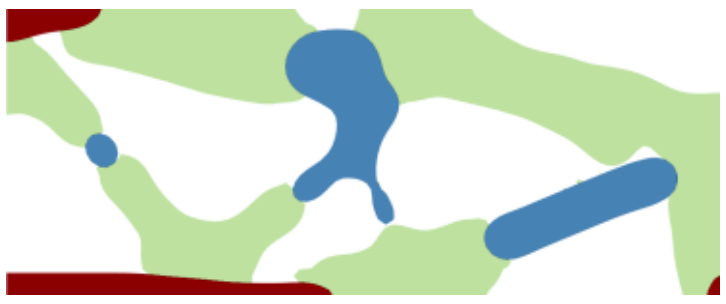 | 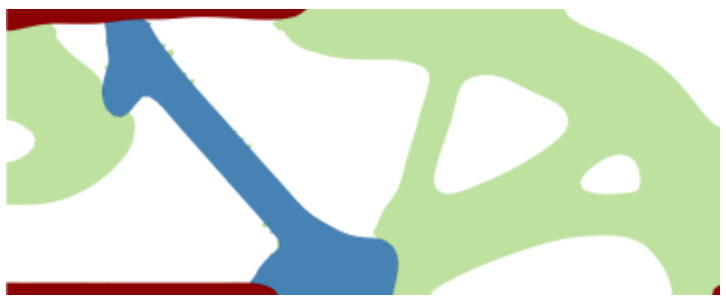 |
| $C$ = 98.58 | $C$ = 96.55 | $C$ = 109.59 |
| $iter_max_in_1$ = 100 | $iter_max_in_1$ = 150 | $iter_max_in_1$ = 200 |
| 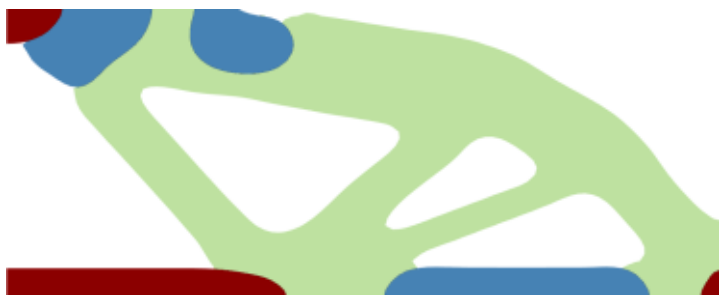 | 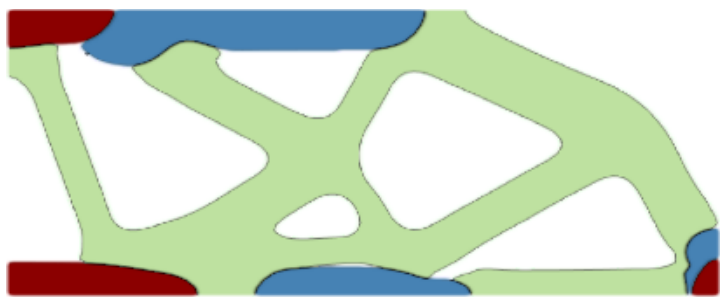 | 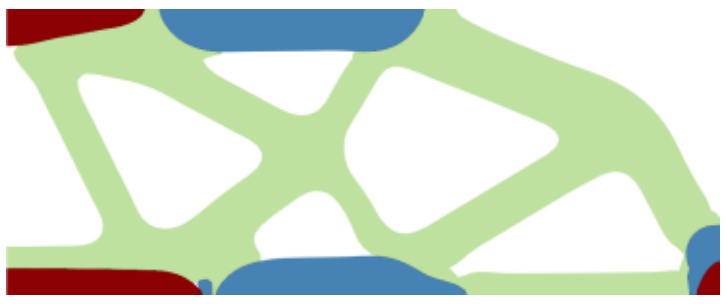 |
| $C$ = 77.66 | $C$ = 77.98 | $C$ = 81.09 |

### *5.6. Convergence behaviour and oscillation of the objective function*

The compliance histories in Figs. 5, 7, 9 and 11 show two features: the strong oscillation of the first iterations, and the small periodic oscillation that persists to the last iteration. Neither is a sign that the optimization has failed to converge; both follow from the alternation of the two-phase sub-problems.

(1) Early iterations. The design starts full of material and the prescribed volumes are reached at the evolutionary rate $ER$ = 2% per inner iteration, so during the first outer iteration the amount of material changes strongly, and with it the compliance. A second effect is superimposed on this one. Only two phases are active at a time, so the compliance recorded at the end of successive outer iterations alternates between two nearby states, with a period of one or two outer iterations (Fig. 13(c), where the dotted lines separate successive outer iterations). This second amplitude falls off quickly once the volume constraints are met.

(2) Activation of the negative-mapping. From iteration 142 on, the flip and the OC update (26) counteract each other: the update moves the flipped nodal values, and the flip restores them in the next inner iteration. The periodic oscillation therefore grows, while the topology stays as it was, compare the histories on either side of the dashed line in Fig. 13(a, b). Over the last 20 outer iterations the peak-to-peak amplitude is 0.4% of the compliance for the three-phase half-MBB beam and 2.9% for the four-phase beam, against 1.2% and 1.4% in the 20 iterations preceding the activation.

(3) Convergence. To check that the oscillation is not a slow drift, the two half-MBB examples of Sections 5.1–5.2 were continued to 400 outer iterations (Fig. 13(a, b)). For the three-phase beam the compliance reads 132.32, 130.96 and 130.16 at the 200th, 300th and 400th iteration: a change of 1.7% over 200 further iterations, with no change of topology. The four-phase values are 105.47, 103.12 and 101.57. Here too the topology is unchanged, but the members keep being reshaped and the compliance falls by another 2.8%, so the values quoted at 200 iterations are slightly conservative. This continuation run and Table 2 describe the same case; their 105.47 and 104.76 differ by 0.7%, less than the oscillation just quantified. Criterion (Eq. (27)) is first met at iteration 80 for the three-phase beam, whereas for the four-phase beam the oscillation keeps it from being met at all within 400 iterations, its smallest value being $1.3\times10^{-3}$. All results are therefore quoted at a fixed count of 200 outer iterations, the compliance being the value at the end of an outer iteration.

(4) Overlap elements. Fig. 13(d) follows the number of overlap elements and the overlap volume fraction of the four-phase half-MBB beam. Without the negative-mapping the AAPA leaves about 129 of them, 3.2% of the elements, and an overlap volume of 1.5% of the design domain. Once the negative-mapping is switched on the overlap volume drops by three orders of magnitude: at the 200th iteration three overlap elements are left, with an overlap volume of $6\times10^{-4}$%. The final pass of Section 3.1 removes the rest, the compliance moving from 105.47 to 104.88. The three-phase beam and the two cantilever beams behave in the same way: 41 overlap elements before the activation, none at the 200th iteration.

(5) How often the flip should be applied. Table 9 compares four schemes for the four-phase half-MBB beam: the flip in every inner iteration after its activation, as used throughout this article; only in the last inner iteration of each sub-problem; once at iteration 142; and the AAPA on its own. Applying the flip less often reduces the late oscillation but leaves more overlap for the final pass, and the four compliance values still lie within 1.8% of each other. The comparison also shows why the flip is not applied only at the end: on a design produced by the AAPA only the final pass has to correct 116 elements instead of three, and the compliance ends 2.7% higher. A relaxed flip, with a weight below one, is left for future work.

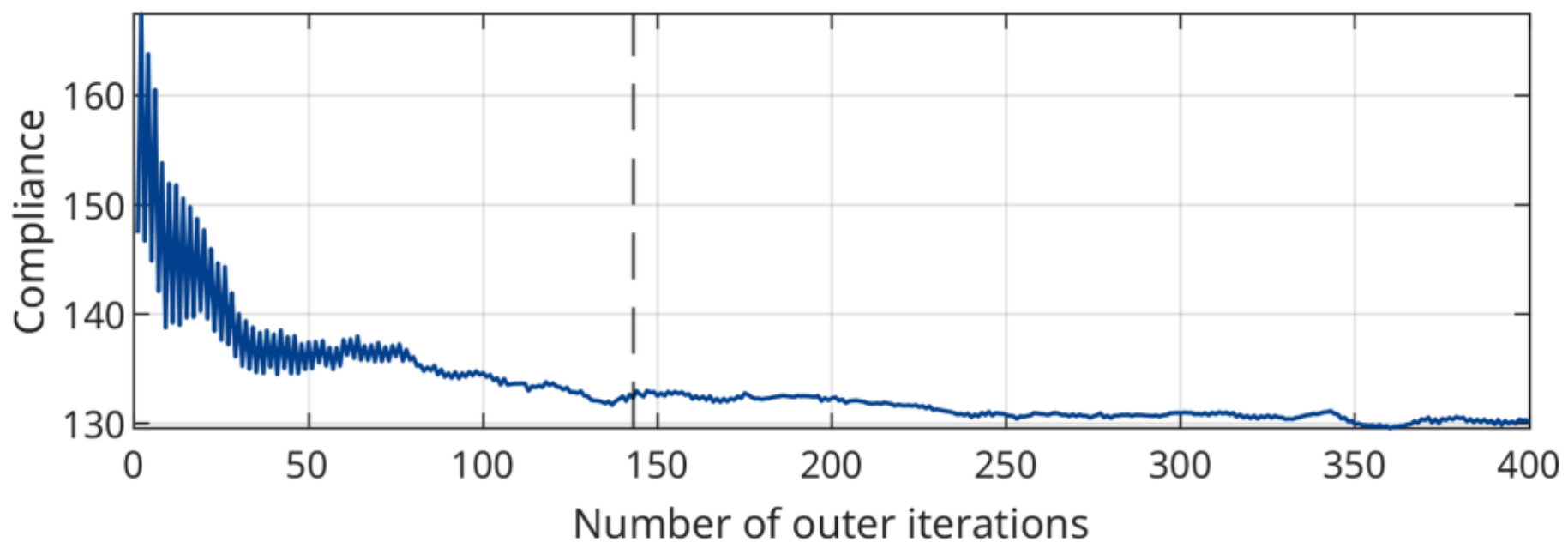

(a)

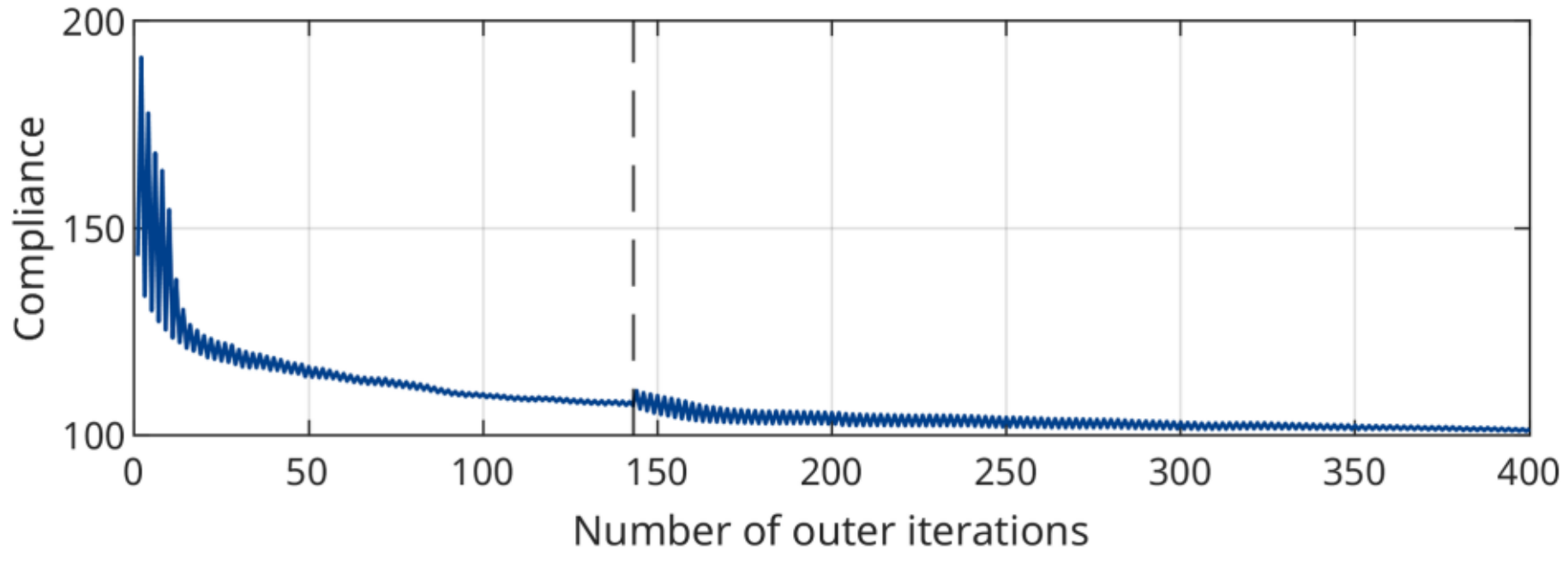

(b)

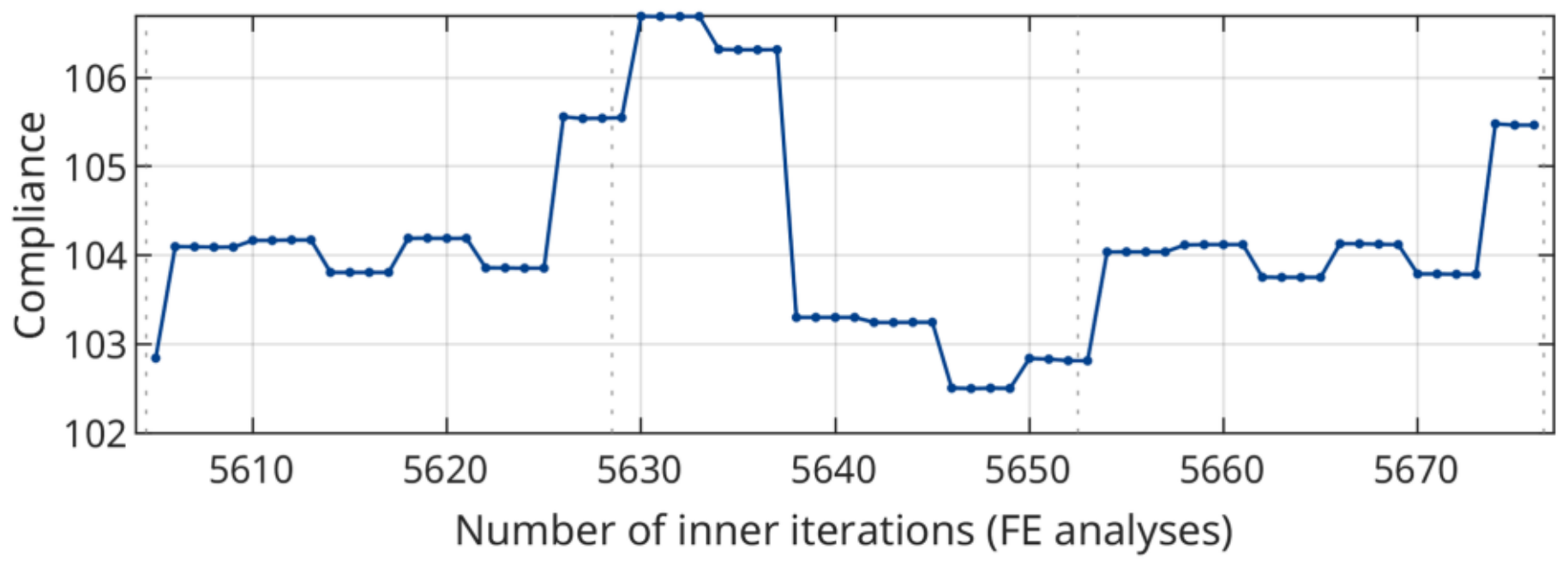

(c)

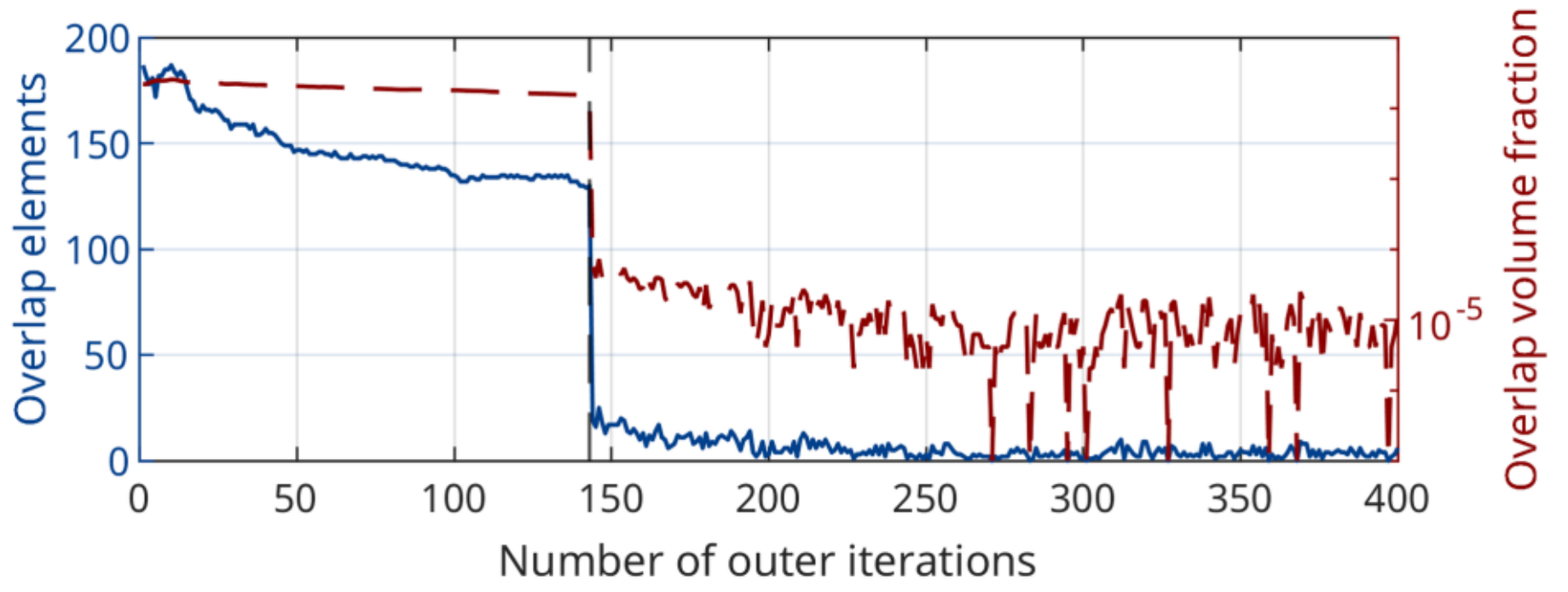

(d)

Fig 13. Convergence of the half-MBB beams: compliance over 400 outer iterations for (a) the three-phase and (b) the four-phase beam; (c) compliance after every inner iteration and (d) number of overlap elements and overlap volume fraction of the four-phase beam (vertical dashed line: activation of the negative-mapping).

Table 9. Influence of the frequency of application of the negative-mapping (four-phase half-MBB beam, $f$ = {0.05, 0.1, 0.3})

| **Variant** | **C at iteration 200** | **Mean C, iterations 181–200** | **Oscillation amplitude, last 20 iterations (%)** | **Overlap elements at iteration 200 / after final pass** | **Final C** |
|---|---|---|---|---|---|
| Negative-mapping in every inner iteration (adopted) | 105.47 | 104.29 | 2.9 | 3 / 0 | 104.88 |
| Negative-mapping once per sub-problem | 104.50 | 103.93 | 2.6 | 28 / 0 | 103.89 |
| Negative-mapping once (iteration 142) | 104.33 | 104.13 | 0.8 | 91 / 0 | 105.51 |
| AAPA only (no negative-mapping) | 106.13 | 106.00 | 1.1 | 116 / 0 | 107.72 |

## *5.7. Influence of the material properties*

Sections 5.1–5.5 keep the moduli of the solid phases at 2 : 1 and 4 : 2 : 1. To see how much the material properties matter, the three-phase problem ($f$ = {0.1, 0.3}) was solved for E1 : E2 = 1.5 : 1, 2 : 1, 5 : 1 and 10 : 1, and the four-phase problem ($f$ = {0.05, 0.1, 0.3}) for E1 : E2 : E3 = 3 : 2 : 1, 4 : 2 : 1 and 10 : 4 : 1, the weakest solid phase always having modulus 1. All other parameters are those of Section

5. Table 10 collects the designs and their compliance.

Table 10. Optimized structures and compliance for different ratios of the Young's moduli of the solid phases (dark to light: stiffest to weakest phase); the 2 : 1 and 4 : 2 : 1 rows are the reference cases of Tables 1 and 2.

| **Case** | **Modulus ratio** | **Half-MBB beam** | **Cantilever beam** |
|---|---|---|---|
| Three-phase | 1.5 : 1 | 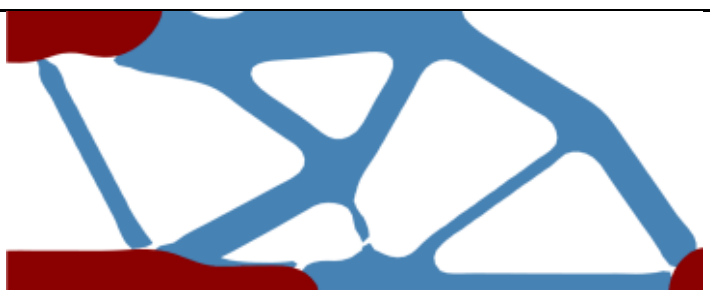 C = 148.16 | 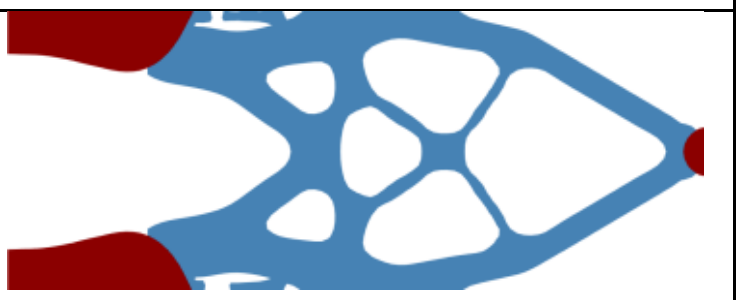 C = 145.33 |
| | 2 : 1 | 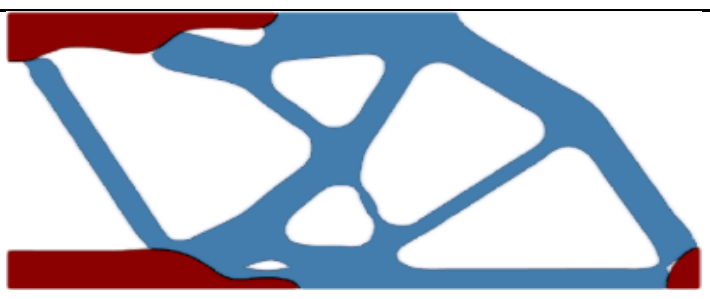 C = 132.32 | 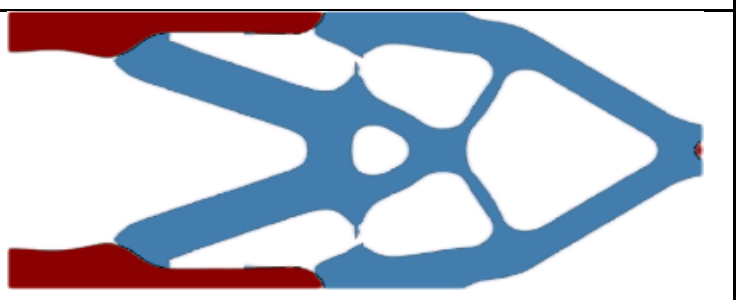 C = 121.83 |
| | 5 : 1 | 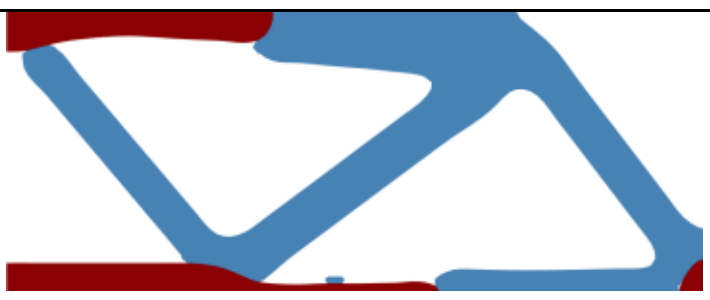 C = 90.04 | 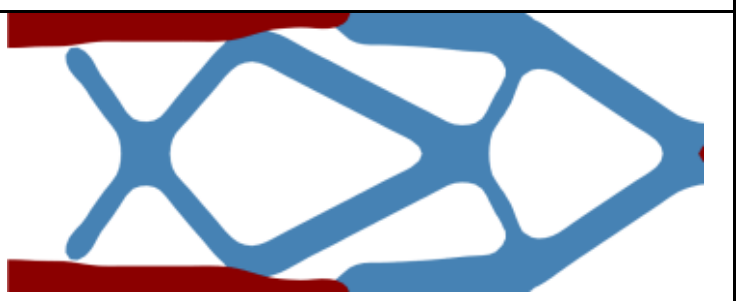 C = 81.35 |
| | 10 : 1 | 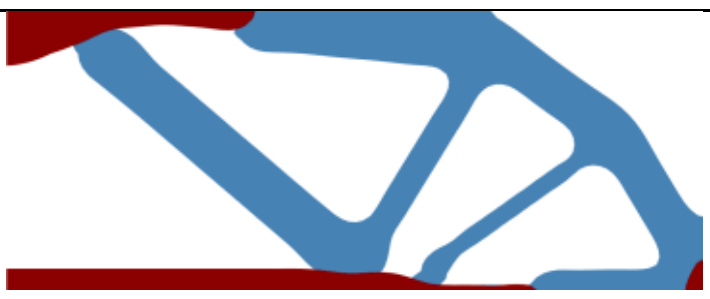 C = 68.72 | 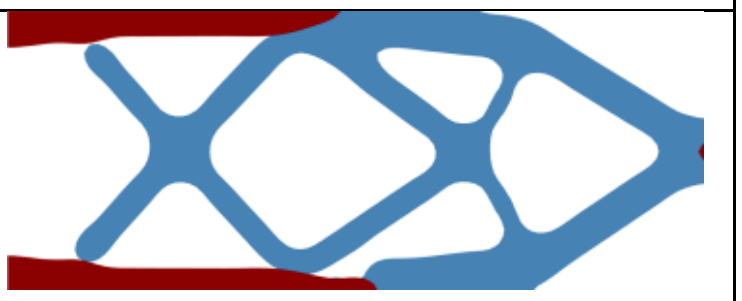 C = 62.31 |
| Four-phase | 3 : 2 : 1 | 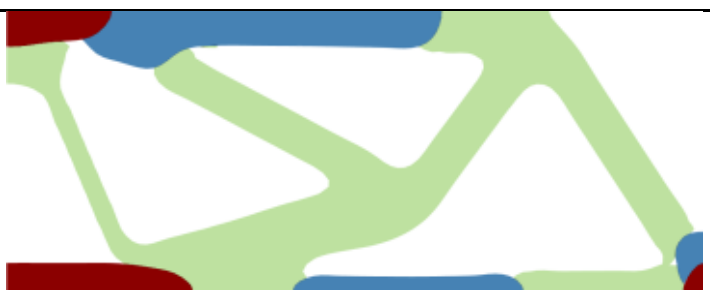 C = 102.01 | 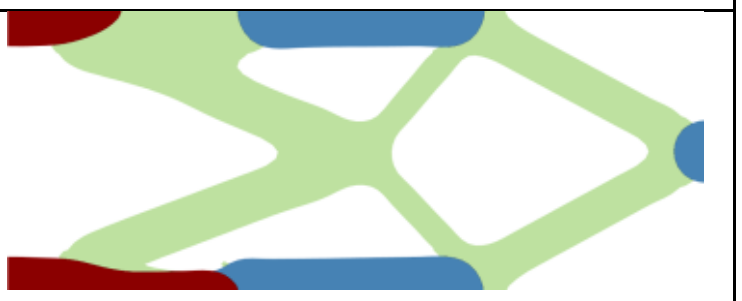 C = 87.28 |
| | 4 : 2 : 1 | 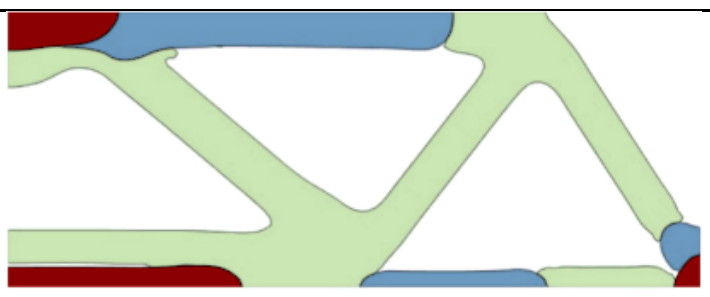 C = 104.76 | 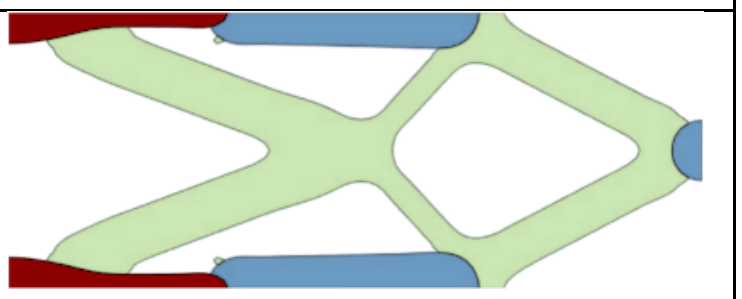 C = 80.36 |
| | 10 : 4 : 1 | 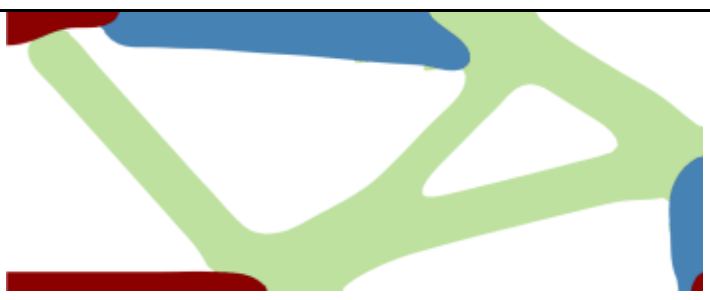 C = 69.76 | 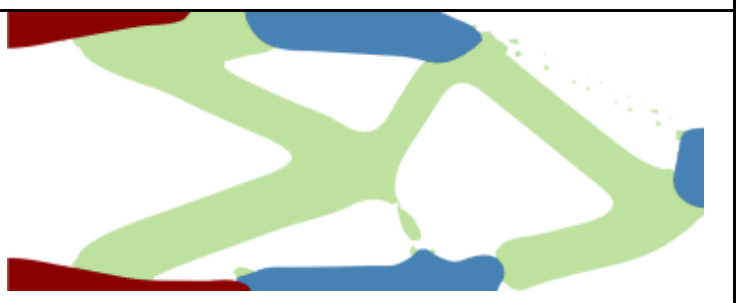 C = 70.37 |

For the three-phase problems the compliance falls monotonically as the contrast grows, the phase volumes being fixed. The stiffest phase always occupies the supports and the loading point, where the strain energy density is highest; as the contrast increases it spreads along the adjacent chord and the layout of the weaker phase becomes coarser. The four-phase problems behave in the same way, the intermediate phase forming the chord next to the load and the weakest phase the diagonals. The exception is the 4 : 2 : 1 half-MBB case, whose compliance lies slightly above that of 3 : 2 : 1 because it converges to the alternative local solution. More important for the present method, the modulus ratio leaves the interfaces untouched: the boundaries are smooth in every case, no gray elements appear, and the final negative-mapping pass leaves no overlap element at all, at a cost of at most 0.6% in compliance. Before that pass the three-phase designs carry at most three overlap elements and the four-phase designs at most fifteen, all of them junctions of three solid phases, which the final pass of Section 3.1 handles through the generalization of Eq. (20).

### *5.8. Extension to three-dimensional problems*

No part of the method depends on the number of spatial dimensions. The AAPA decomposition, the level set description, the compliance proportion and its filtering, the ODE/OC update (23)–(26), the evolutionary strategy and the negative-mapping are all written in terms of nodes and elements. Going to three dimensions therefore requires only four changes: the four-node quadrilateral becomes an eight-node hexahedron, with the corresponding connectivity; the compliance-proportion filter (Eq. (12)) uses a spherical neighbourhood; the element density (18)–(19) is evaluated at 11×11×11 sample points, with trilinear interpolation of the eight nodal values; and the negative-mapping acts on the eight nodes of an overlap element, the enclosed-node clean-up on the 26 neighbours of a node. The number of sub-problems, the number of inner iterations and every algorithmic parameter are unchanged.

The three-dimensional cantilever beam of Fig. 14 has a design domain of 40 × 20 × 8, discretized into 6400 cubic elements with 23 247 degrees of freedom. The face x = 0 is clamped and a unit load is applied in the negative y-direction at the centre of the free end face. The material data are those of Sections 5.1 and 5.2 (three-phase: $E$ = {2, 1}, $f$ = {0.1, 0.3}; four-phase: E = {4, 2, 1}, $f$ = {0.05, 0.1, 0.3}) and the filter radius is rf = 3 element sizes; every other parameter — $ER$ = 2%, $\Delta t$ = 0.01, 150 inner iterations in the first outer iteration and 4 afterwards, 200 outer iterations, negative-mapping from iteration 142, final pass — is the one used in two dimensions. Each problem was solved a second time with the AAPA only, for comparison.

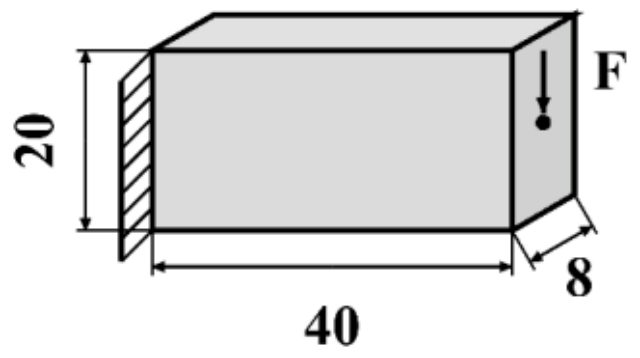

Fig 14. 3D cantilever beam: design domain, boundary conditions and load

Table 11. Optimized three-dimensional cantilever beams: AAPA only versus the negative-mapping interpolation.

| **Case** | **Method** | **Optimized structure (level sets of the solid phases)** | **Mid-thickness section (zero level contours of the solid phases at mid-thickness)** | **C** | **Overlap elements at iteration 200 / after final pass** | **Final volume fractions of the solid phases** |
|---|---|---|---|---|---|---|
| Three-phase, $E$ = {2, 1}, $f$ = {0.1, 0.3} | AAPA only | 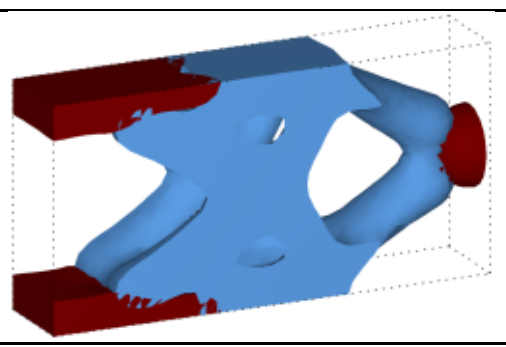 | 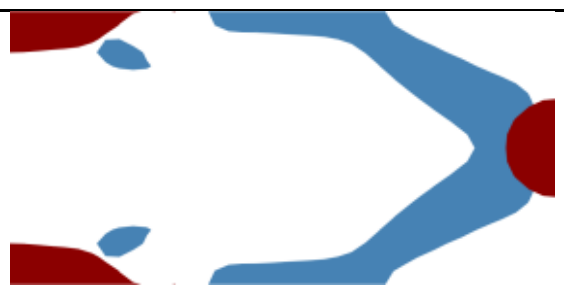 | 10.39 | 213 / – | {0.087, 0.288} |
| | AAPA + negative-mapping | 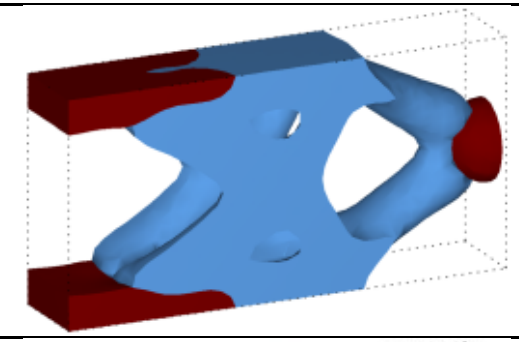 | 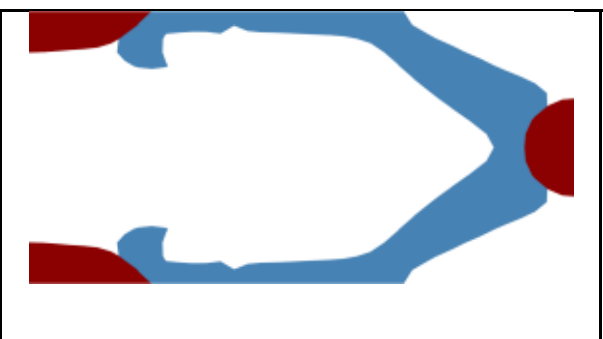 | 10.18 | 36 / 0 | {0.085, 0.300} |
| Four-phase, E = {4, 2, 1}, $f$ = {0.05, 0.1, 0.3} | AAPA only | 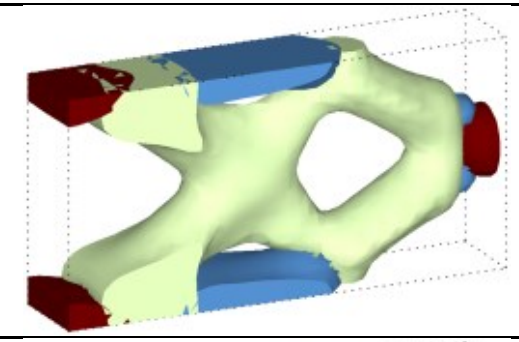 | 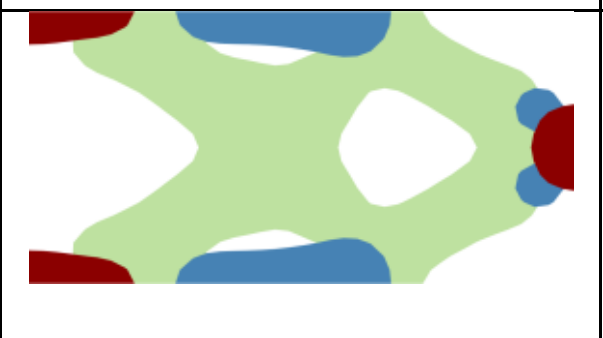 | 7.00 | 506 / – | {0.040, 0.086, 0.316} |
| | AAPA + negative-mapping | 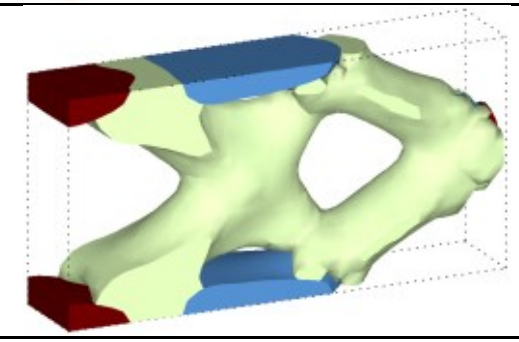 | 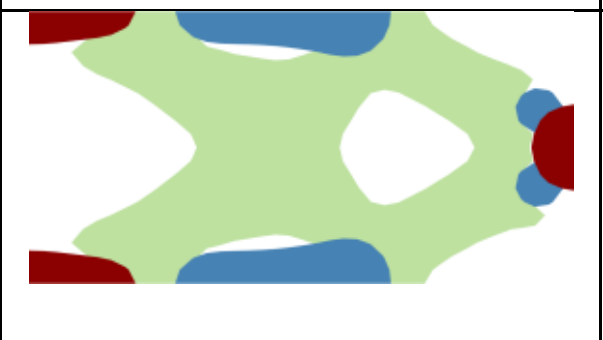 | 6.48 | 155 / 0 | {0.040, 0.089, 0.361} |

Table 11 shows the designs, the compliance and the number of overlap elements, the stiffest, intermediate and weakest phase being drawn from dark to light; in the mid-thickness sections a blended shade marks where two phases overlap. In the three-phase beam the stiffest phase takes the upper and lower edges of the clamped face and the region around the load, where the bending stresses are largest, and the weaker phase forms a perforated web between supports and load. The AAPA on its own leaves 213 overlap elements along the interfaces, 3.3% of the elements; with the negative-mapping 36 remain at the end of the iterations and none after the final pass, the two designs differing by 2.0% in compliance. The four-phase beam behaves the same way, with 506, 155 and finally 0 overlap elements. The interface treatment thus acts on the interface elements without changing the topology. On the coarse three-dimensional mesh the element-volume fractions depart somewhat from the prescribed values (Table 11), the volume constraint being enforced on the fraction of nodes with a non-negative LSF. Fig. 15 gives the compliance and overlap histories of the three-phase problem, with the same features as in 2D. The cost is dominated by the finite element analyses, 2838 of them for the three-phase and 5676 for the four-phase problem; the negative-mapping is a nodal operation and adds no measurable time.

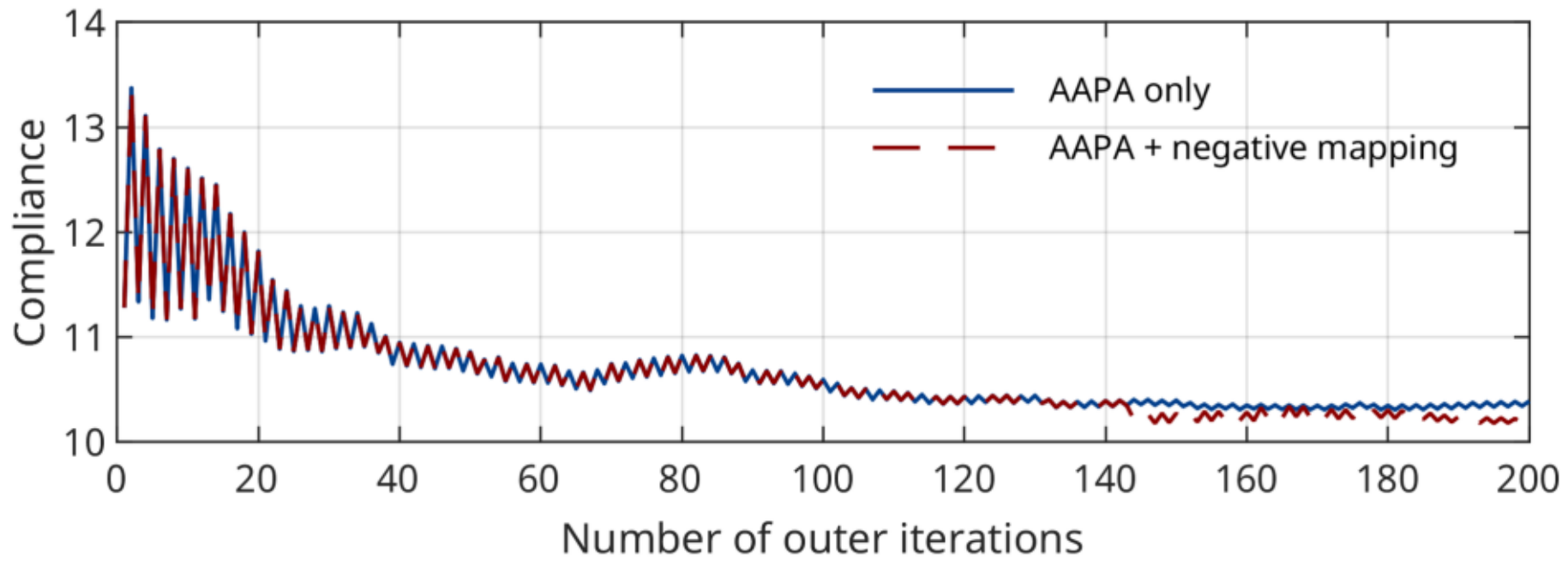

(a)

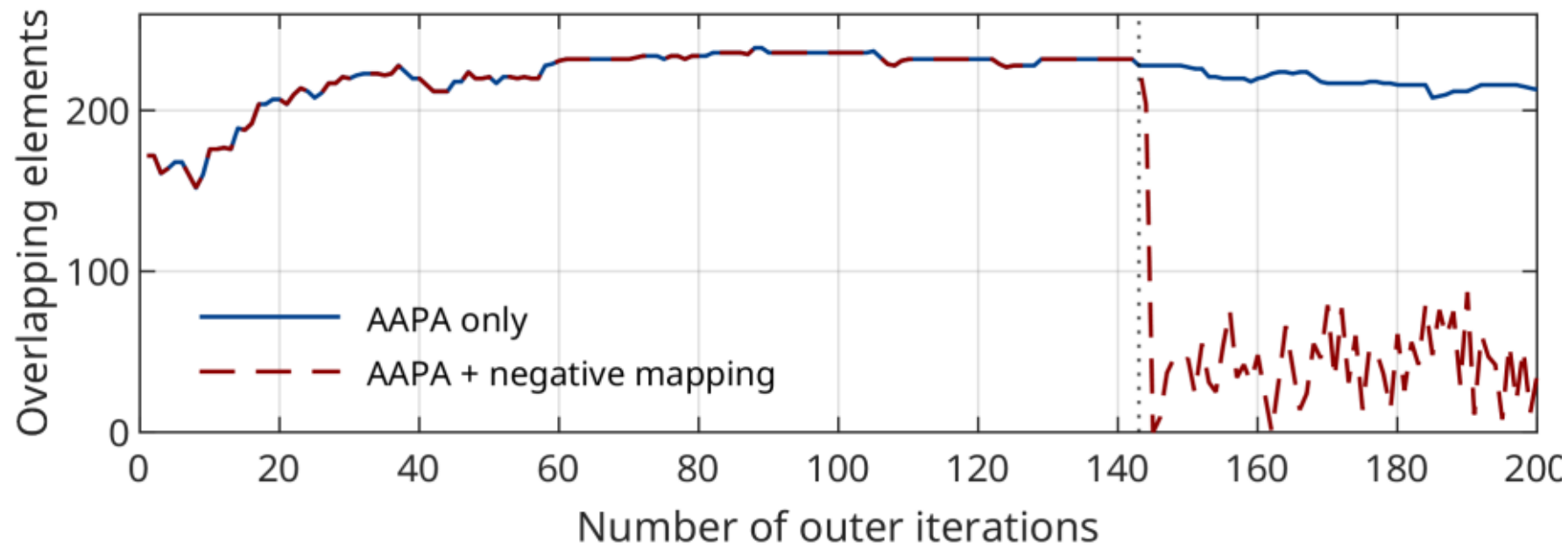

(b)

Fig 15. Three-phase 3D cantilever beam: (a) compliance and (b) number of overlap elements, with and without the negative-mapping interpolation (vertical dotted line: its activation).

Fig. 16 shows the overlap in three dimensions directly: the level sets of the solid phases are drawn semi-transparent, and the overlap volume, where two of them are both non-negative, opaque. With the AAPA only it runs along every interface between the stiff and the weaker phases (Fig. 16(a, c)). With the negative-mapping interpolation none is left anywhere in the design domain, for the four-phase beam as well as for the three-phase one (Fig. 16(b, d)).

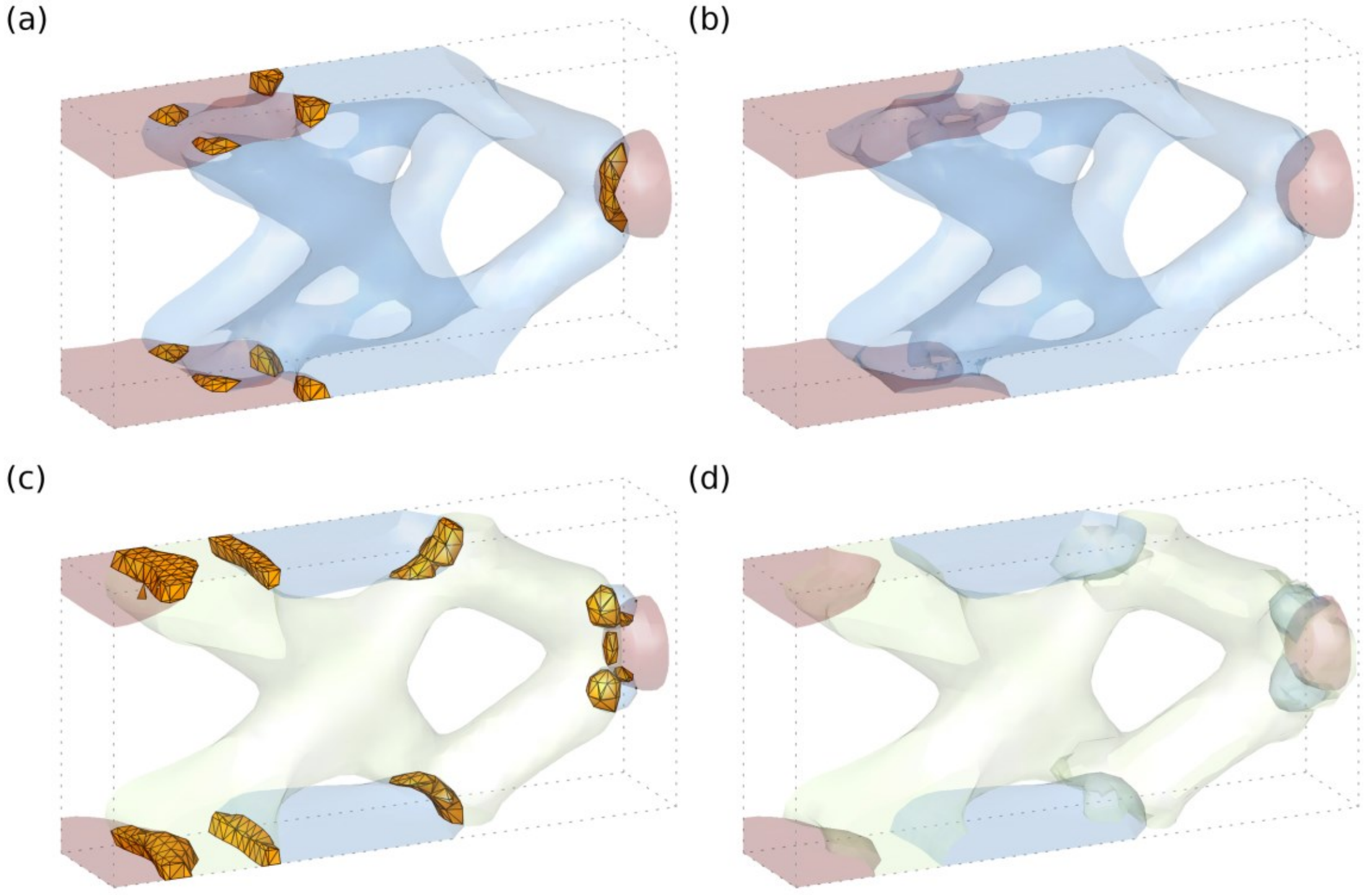

Fig 16. Overlap volume of the three-phase (a, b) and of the four-phase (c, d) beam, with the AAPA only (a, c) and with the negative-mapping interpolation (b, d).

## 6. Discussion

The main advantages of the method are: (1) no sensitivity analysis is required, the update being driven by the compliance proportion of each phase; (2) the level set description on the fixed mesh yields smooth boundaries without gray elements; (3) the negative-mapping interpolation removes, without changing the topology, the phase overlap that a purely density-based treatment of the AAPA sub-problems cannot avoid; (4) the M-phase problem is decomposed into two-phase sub-problems, each with few design variables and constraints; and (5) the algorithm is dimension-independent and was applied to 3D without changing its parameters.

Its main limitations are: (1) as a proportional (gradient-free) method it does not guarantee optimality: of the eight cases compared with SIMP in Tables 1 and 2 it gives a lower compliance in four, a slightly larger one in three (0.3%, 0.4% and 4.8%, the last for the four-phase half-MBB beam with $f = \{0.05, 0.1, 0.3\}$) and a 12.7% larger one in the remaining case (Table 1, cantilever beam with $f = \{0.2, 0.2\}$), the cost paid for overlap-free, gray-free layouts; (2) the objective oscillates periodically because of the alternation of the sub-problems, and the negative-mapping increases the amplitude of this oscillation (Section 5.6), so that results have to be read at the end of an outer iteration; (3) the results depend on empirically chosen parameters, namely the number of initial inner iterations (Section 5.5), the evolutionary rate and the iteration at which the negative-mapping is activated, and in some cases (four-phase half-MBB beam) different runs converge to different local solutions differing by a few percent; (4) the number of sub-problems grows as $M(M-1)/2$, so that the cost per outer iteration grows quadratically with the number of phases; and (5) only linear elastic compliance minimization with volume constraints has been considered. Extensions to other responses (stress, frequency, compliant mechanisms) and to a relaxed, weighted negative-mapping are the subject of ongoing work.

## 7. Conclusions

This article has extended the level set-based negative-mapping interpolation method (Du et al. 2026) to the multi-material proportional topology optimization of macro-scale structures in two and three dimensions. Combining the AAPA with the PTO framework breaks the M-phase problem into two-phase sub-problems and needs no sensitivity analysis. The material interfaces are reconstructed by the negative-mapping interpolation; with respect to Du et al. (2026), the present work adds the comparison with SIMP and with the AAPA only for load-bearing structures, the extension to three dimensions and the study of the convergence behaviour. The following conclusions are drawn:

(1) Elimination of Material Overlap and Gray Elements: The negative-mapping interpolation removes the overlap at the interface elements that density-based multi-material methods cannot avoid. Compared with SIMP and with the AAPA only, it also eliminates the gray elements and cuts the overlap volume by three to four orders of magnitude during the iterations; the final pass then leaves designs that are free of overlap, with smooth and continuous interfaces. The compliance is below the SIMP value in half of the two-dimensional test cases and at most 12.7% above it in the others.

(2) Effectiveness of the Optimization Framework: The method was tested on 2D cantilever and half-MBB beams and on a 3D cantilever beam, with three and four material phases; going to 3D asked for nothing beyond hexahedral elements and a three-dimensional filter. The stiffest material is consistently placed at the supports and at the loading point, the weaker materials reinforcing the secondary regions. The nodal compliance proportional filter gives a higher compliance than density filtering or no filtering, but suppresses the tiny holes and branched members that these leave and so reduces the post-processing work.

(3) Convergence Behaviour and Limitations: The compliance history shows a periodic oscillation with a period of one or two outer iterations, inherent to the alternation of the AAPA sub-problems and not reduced by further iterations; its amplitude is below 1% for the three-phase problems and a few percent for the four-phase ones, so the compliance should be read at the end of an outer iteration. The results also depend on empirically chosen parameters, in particular the number of initial inner iterations: fewer than 50 leave unstable topologies, 100 and 150 give practically identical designs, and a larger number may settle in a different local solution. Being gradient-free, the method does not guarantee optimality and the number of sub-problems grows as $M(M-1)/2$; the extension to other objectives and constraints is left for future work.

## Acknowledgments

The authors thank the National Natural Science Foundation of China, Sichuan Provincial Science and Technology Program and Southwest Jiaotong University.

## Disclosure statement

No conflict of interest to declare by the authors.

## Author contributions

**Dong Wang**: conceptualization, Methodology, Software, Writing - original draft, review & editing.
**Qianglin Ran**: conceptualization, Validation, Software, Writing - review & editing.
**Xuanliang Wang**: Methodology, Formal analysis, Writing - review & editing.
**Wei Xiang**: Methodology, Validation, Writing - review & editing.
**Wenming Cheng**: Methodology, Formal analysis, Writing - review & editing.
**Run Du**: conceptualization, Funding acquisition, Methodology, Project administration, supervision, Writing - original draft, review & editing.

## Data availability

Data available on request from the authors.

## Funding

The financial support of the National Natural Science Foundation of China (Grant No. 51405397), the Sichuan Provincial Science and Technology Program (Grant No. 2023YFG0182), and Southwest Jiaotong University Multidisciplinary Research Fund (No. 2682025ZD005), is gratefully acknowledged.

## About the authors

**Dong Wang**: School of Mechanical Engineering, Southwest Jiaotong University. Research interest: Topology optimization, computational methods.
**Qianglin Ran**: School of Mechanical Engineering, Southwest Jiaotong University. Research interest: Topology optimization, computational methods.
**Xuanliang Wang**: School of Mechanical Engineering, Southwest Jiaotong University. Research interest: Topology optimization, mechanics, computational methods.
**Wei Xiang**: School of Mechanical Engineering, Southwest Jiaotong University. Research interest: mechanics, computational dynamics, computational methods.
**Wenming Cheng**: School of Mechanical Engineering, Southwest Jiaotong University. Research interest: Topology optimization, mechanics, computational dynamics, computational methods.
**Run Du**: School of Mechanical Engineering, Southwest Jiaotong University. Research interest: Topology optimization, computational fluid dynamics, computational dynamics, computational methods.

## ORCID

Run Du https://orcid.org/0000-0002-6614-2752

## References

Bendsøe, M. P., and O. Sigmund. 1999. “Material interpolation schemes in topology optimization.” Archive of Applied Mechanics 69 (9–10): 635–654. https://doi.org/10.1007/s004190050248.

Biyikli, E., and A. C. To. 2015. “Proportional Topology Optimization: A New Non-Sensitivity Method for Solving Stress Constrained and Minimum Compliance Problems and Its Implementation in MATLAB.” PLoS ONE 10 (12): e0145041. https://doi.org/10.1371/journal.pone.0145041.

Chu, S., M. Xiao, L. Gao, H. Li, J. Zhang, and X. Zhang. 2019. “Topology optimization of multi-

material structures with graded interfaces." Computer Methods in Applied Mechanics and Engineering 346: 1096–1117. https://doi.org/10.1016/j.cma.2018.09.040.

Cui, M., H. Chen, and J. Zhou. 2016. "A level-set based multi-material topology optimization method using a reaction diffusion equation." Computer-Aided Design 73: 41–52. https://doi.org/10.1016/j.cad.2015.12.002.

Cui, M., W. Li, G. Li, and X. Wang. 2023. "The asymptotic concentration approach combined with isogeometric analysis for topology optimization of two-dimensional linear elasticity structures." Electronic Research Archive 31 (7): 3848–3878. https://doi.org/10.3934/era.2023196.

Cui, M., Y. Zhang, X. Yang, and C. Luo. 2018. "Multi-material proportional topology optimization based on the modified interpolation scheme." Engineering with Computers 34 (2): 287–305. https://doi.org/10.1007/s00366-017-0540-z.

Du, R., D. Wang, Q. Ran, L. Yang, X. Rao, W. Xiang, X. Wang, and W. Cheng. 2026. "Level-set-based negative-mapping interpolation for non-overlapping multi-material microstructure topology optimization." Engineering with Computers 42 (3): 94. https://doi.org/10.1007/s00366-026-02333-1.

Gao, J., X. Wu, M. Xiao, V. P. Nguyen, L. Gao, and T. Rabczuk. 2023. "Multi-patch isogeometric topology optimization for cellular structures with flexible designs using Nitsche's method." Computer Methods in Applied Mechanics and Engineering 410: 116036. https://doi.org/10.1016/j.cma.2023.116036.

Gao, M., Z. Peng, M. Cui, and X. Wang. 2026. "A phased strategy for structural optimization: Enhanced proportional topology optimization and post-processing of results by level set method." Mechanics Based Design of Structures and Machines 54 (1): 2661100. https://doi.org/10.1080/15397734.2026.2661100.

Gao, T., and W. Zhang. 2010. "Topology optimization involving thermo-elastic stress loads." Structural and Multidisciplinary Optimization 42 (5): 725–738. https://doi.org/10.1007/s00158-010-0527-5.

Gao, T., and W. Zhang. 2011. "A mass constraint formulation for structural topology optimization with multiphase materials." International Journal for Numerical Methods in Engineering 88 (8): 774–796. https://doi.org/10.1002/nme.3197.

Hu, C., S. Shi, C. Guo, W. Sun, Z. Sun, and B. Chen. 2026. "Dynamic topology optimization of three-dimensional multi-material structures under time-varying loads: an improved alternating active phase algorithm-optimality criteria method." Engineering Analysis with Boundary Elements 186: 106684. https://doi.org/10.1016/j.enganabound.2026.106684.

Hvejsel, C. F., and E. Lund. 2011. "Material interpolation schemes for unified topology and multi-material optimization." Structural and Multidisciplinary Optimization 43 (6): 811–825. https://doi.org/10.1007/s00158-011-0625-z.

Le, L. T., T. Dirgantara, L. R. Zuhal, and A. Jusuf. 2026. "A multi-material level set method using logical functions for phase controlling with the reaction–diffusion level set technique." Structural and Multidisciplinary Optimization 69 (8): 195. https://doi.org/10.1007/s00158-026-04381-z.

Li, Q., G. P. Steven, O. M. Querin, and Y. M. Xie. 1999. "Optimization of thin shell structures subjected to thermal loading." Structural Engineering and Mechanics 7 (4): 401–412. https://doi.org/10.12989/sem.1999.7.4.401.

Li, W., M. Cui, X. Wang, and M. Gao. 2025. "An enhanced proportional topology optimization method with new density filtering weight function for the minimum compliance problem." Mechanics Based Design of Structures and Machines 53 (4): 2666–2695.

https://doi.org/10.1080/15397734.2024.2412753.

Lin, D., L. Gao, and J. Gao. 2025. "The Lagrangian-Eulerian described Particle Flow Topology Optimization (PFTO) approach with isogeometric material point method." Computer Methods in Applied Mechanics and Engineering 440: 117892. https://doi.org/10.1016/j.cma.2025.117892.

Liu, J., and Y. Ma. 2018. "A new multi-material level set topology optimization method with the length scale control capability." Computer Methods in Applied Mechanics and Engineering 329: 444–463. https://doi.org/10.1016/j.cma.2017.10.011.

Liu, Y., C. Yang, P. Wei, P. Zhou, and J. Du. 2021. "An ODE-driven level-set density method for topology optimization." Computer Methods in Applied Mechanics and Engineering 387: 114159. https://doi.org/10.1016/j.cma.2021.114159.

Luo, Z., L. Tong, J. Luo, P. Wei, and M. Y. Wang. 2009. "Design of piezoelectric actuators using a multiphase level set method of piecewise constants." Journal of Computational Physics 228 (7): 2643–2659. https://doi.org/10.1016/j.jcp.2008.12.019.

Nguyen, M. N., M. T. Tran, H. Q. Nguyen, and T. Q. Bui. 2023. "A multi-material Proportional Topology Optimization approach for compliant mechanism problems." European Journal of Mechanics – A/Solids 100: 104957. https://doi.org/10.1016/j.euromechsol.2023.104957.

Nguyen, M. N., N. Nguyen-Thanh, S. Chen, and T. Q. Bui. 2025. "Multi-material topology optimization of thermoelastic structures by an ordered SIMP-based phase field model." Computers & Mathematics with Applications 186: 84–100. https://doi.org/10.1016/j.camwa.2025.03.005.

Rao, X., W. Cheng, and R. Du. 2024. "A proportional topology optimization method with level-set description and evolutionary strategy." Engineering Analysis with Boundary Elements 166: 105853. https://doi.org/10.1016/j.enganabound.2024.105853.

Sha, W., M. Xiao, L. Gao, and Y. Zhang. 2021. "A new level set based multi-material topology optimization method using alternating active-phase algorithm." Computer Methods in Applied Mechanics and Engineering 377: 113674. https://doi.org/10.1016/j.cma.2021.113674.

Shen, Y., J. Li, Z. Ren, and C. Yang. 2025. "Geometrically nonlinear topology optimization of multi-material structures using the subdomain level set method." Journal of Mechanical Science and Technology 39 (6): 3309–3322. https://doi.org/10.1007/s12206-025-0525-6.

Sigmund, O., and S. Torquato. 1997. "Design of materials with extreme thermal expansion using a three-phase topology optimization method." Journal of the Mechanics and Physics of Solids 45 (6): 1037–1067. https://doi.org/10.1016/S0022-5096(96)00114-7.

Stegmann, J., and E. Lund. 2005. "Discrete material optimization of general composite shell structures." International Journal for Numerical Methods in Engineering 62 (14): 2009–2027. https://doi.org/10.1002/nme.1259.

Tavakoli, R., and S. M. Mohseni. 2014. "Alternating active-phase algorithm for multimaterial topology optimization problems: a 115-line MATLAB implementation." Structural and Multidisciplinary Optimization 49 (4): 621–642. https://doi.org/10.1007/s00158-013-0999-1.

Wan, C., H. Jiao, L. Lv, and C. Lu. 2024. "Multi-material topology optimization based on multiple simp of variable density method." Journal of Mechanical Science and Technology 38 (2): 749–759. https://doi.org/10.1007/s12206-024-0124-y.

Wan, Q., F. Dong, and S. Liu. 2025. "Multi-material isogeometric topology optimization with mapping-based interpolation." International Journal of Mechanical Sciences 305: 110767. https://doi.org/10.1016/j.ijmecsci.2025.110767.

Wang, H., L. Zhao, Q. Peng, Y. Li, X. Shao, and X. Huo. 2025. “Enhanced proportional topology optimization methods for the minimum compliance problem with volume constraints.” Journal of the Brazilian Society of Mechanical Sciences and Engineering 47 (12): 604. https://doi.org/10.1007/s40430-025-05897-z.

Wang, M. Y., and X. Wang. 2004. “‘Color’ level sets: a multi-phase method for structural topology optimization with multiple materials.” Computer Methods in Applied Mechanics and Engineering 193 (6–8): 469–496. https://doi.org/10.1016/j.cma.2003.10.008.

Wang, X., M. Cui, M. Gao, and Z. Peng. 2026. “Topology optimization of structures under thermo-mechanical coupling by the improved parameterized level set method.” Mechanics Based Design of Structures and Machines 54 (1): 2571736. https://doi.org/10.1080/15397734.2025.2571736.

Wang, Y., Z. Luo, Z. Kang, and N. Zhang. 2015. “A multi-material level set-based topology and shape optimization method.” Computer Methods in Applied Mechanics and Engineering 283: 1570–1586. https://doi.org/10.1016/j.cma.2014.11.002.

Yin, L., and G. K. Ananthasuresh. 2001. “Topology optimization of compliant mechanisms with multiple materials using a peak function material interpolation scheme.” Structural and Multidisciplinary Optimization 23 (1): 49–62. https://doi.org/10.1007/s00158-001-0165-z.

Zhang, X., L. Gao, M. Xiao, and J. Gao. 2025. “T-splines-based panel method for aerodynamic topology optimization of engineering shell structures using isogeometric analysis.” Computer Methods in Applied Mechanics and Engineering 444: 118154. https://doi.org/10.1016/j.cma.2025.118154.

Zhang, X., M. Xiao, L. Gao, and J. Gao. 2024. “A T-splines-oriented isogeometric topology optimization for plate and shell structures with arbitrary geometries using Bézier extraction.” Computer Methods in Applied Mechanics and Engineering 425: 116929. https://doi.org/10.1016/j.cma.2024.116929.

Zhou, Y., and I. Y. Kim. 2026. “Multi-scale multi-material topology optimization for lattice structures with interface connective microstructures.” Engineering Optimization 58 (2): 291–322. https://doi.org/10.1080/0305215X.2025.2465832.

Zuo, W., and K. Saitou. 2017. “Multi-material topology optimization using ordered SIMP interpolation.” Structural and Multidisciplinary Optimization 55 (2): 477–491. https://doi.org/10.1007/s00158-016-1513-3.